\documentclass[a4paper,11pt]{article}
\pdfoutput=1 

\usepackage{jheppub} 

\usepackage[T1]{fontenc} 
                     
\usepackage{microtype}
\usepackage[table,xcdraw]{xcolor}
\usepackage{multirow}
\usepackage{graphicx}
\usepackage{tikz}
\usetikzlibrary{decorations.pathmorphing}
\usetikzlibrary{shapes.misc}
\usetikzlibrary{arrows}

\usepackage{floatrow}
\usepackage{multicol} 
\usepackage{latexsym}

\def\circa#1{\,\raise.3ex\hbox{$#1$\kern-.75em\lower1ex\hbox{$\sim$}}\,}

\usepackage{amsmath,latexsym,amssymb,color,hyperref,graphicx}

\newcommand{\be}{\begin{equation}}
\newcommand{\ee}{\end{equation}}
\newcommand{\bea}{\begin{eqnarray}}
\newcommand{\ena}{\end{eqnarray}}
\newcommand{\no}{\noindent}
\newcommand{\nb}{\nonumber}

\renewcommand\l{\lambda}

\renewcommand\l{\ensuremath{\lambda}}

\newcommand\m{\ensuremath{\mu}}
\renewcommand\k{\ensuremath{\kappa}}

\newcommand\n{\ensuremath{\nu}}

\newcommand{\de}{\partial}

\renewcommand\l{\ensuremath{\lambda}}

\newcommand{\ba}{\begin{eqnarray}}
\newcommand{\ea}{\end{eqnarray}}
\newcommand{\plm}{M_{pl}} 
 
\makeatletter
\def\ps@mine{%
    \def\@oddfoot{\hfil\thepage\hfil}\let\@evenfoot\@oddfoot
    \let\@oddhead\@evenhead%
    \let\@mkboth\@gobbletwo
    \let\sectionmark\@gobble
    \let\subsectionmark\@gobble
    }
\renewcommand\section{\@startsection {section}{1}{\z@}%
                                   {-3.5ex \@plus -1ex \@minus -.2ex}%
                                   {2ex \@plus.2ex}%
                                   {\normalfont\large\sffamily\bfseries}}
\renewcommand\subsection{\@startsection {subsection}{1}{\z@}%
                                   {-3.5ex \@plus -1ex \@minus -.2ex}%
                                   {2ex \@plus.2ex}%
                                   {\normalfont\sffamily\bfseries}}
\makeatother

\numberwithin{equation}{section}

\tikzset{
    photon/.style={decorate, decoration={snake}, draw=black} 
}
\tikzset{cross/.style={cross out, draw=black, minimum size=2*(#1-\pgflinewidth), inner sep=0pt, outer sep=0pt},
cross/.default={1pt}}

\usepackage{subcaption}
\title{\boldmath Primordial Non-Gaussianity in Supersolid Inflation}
\author[a]{Marco Celoria, }
\author[b]{Denis Comelli,}
\author[c,d]{Luigi Pilo}
\author[e,f]{and Rocco Rollo}

\affiliation[a]{ICTP, International Centre for Theoretical Physics,\\ Strada Costiera 11, 34151, Trieste, Italy}
\affiliation[b]{INFN, Sezione di Ferrara,\\ I-44122 Ferrara, Italy}
\affiliation[c]{INFN, Laboratori Nazionali del Gran Sasso,\\ I-67010 Assergi, Italy}
\affiliation[d]{Dipartimento di Scienze Fisiche e Chimiche, Universit\`a degli Studi dell'Aquila,\\  I-67010 L'Aquila, Italy}
\affiliation[e]{Gran Sasso Science Institute (GSSI),\\Viale Francesco Crispi 7, I-67100 L'Aquila, Italy}
\affiliation[f]{Centro Nazionale INFN di Studi Avanzati GGI,\\ Largo Enrico Fermi 2,  I-50125 Firenze, Italy}

\emailAdd{mceloria@ictp.it}
\emailAdd{comelli@fe.infn.it}
\emailAdd{luigi.pilo@aquila.infn.it}
\emailAdd{rocco.rollo@gssi.it}

\abstract{
 We study primordial non-gaussianity in supersolid inflation. The dynamics of supersolid is formulated in terms of an effective field theory based on four scalar fields with a shift symmetric action minimally coupled with gravity. In the scalar sector, there are two phonon-like excitations with a kinetic mixing stemming from the completely spontaneous breaking of diffeomorphism.  
In a squeezed configuration, $f_{\text{NL}}$ of scalar perturbations is angle dependent and not proportional to slow-roll parameters showing a blunt violation of the Maldacena consistency relation. Contrary to solid inflation, the violation persists even after an angular average and generically the amount of non-gaussianity is significant. During inflation, non-gaussianity in the TSS and TTS sector is enhanced in the same region of the  parameters space where the secondary production of gravitational waves is sizeable enough to enter in the sensitivity region of LISA, while the scalar $f_{\text{NL}}$ is still within the current experimental limits.}

\begin{document}
\maketitle

\vspace{0.7cm}

\clearpage

\section{Introduction}
We have minimal knowledge of the Universe before radiation
domination. The most convincing solution of the horizon and flatness problems of the hot big bang model is to assume that  the  Universe had gone through an early phase of accelerated expansion driven by some sort of ``matter'' (inflaton). The only glimpse of the inflationary phase is the seed of primordial perturbations that gave rise to structure formation in the Universe via gravitational instability. Interestingly, many of
the inflationary phase features are determined by the spontaneous
symmetry breaking during inflation. In the simplest case of single
clock inflation~\cite{Cheung:2007st} 4-dimensional diffeomorphisms are
broken down to 3-dimensional diffeomorphisms of the hypersurface
$\phi=$constant, where $\phi$ is the inflaton field; the Weinberg
theorem holds~\cite{Weinberg:2003sw}, perturbations are adiabatic, and
the detailed theoretical predictions for primordial
non-gaussianity~\cite{Maldacena:2002vr, Acquaviva:2002ud} are not
affected by the reheating of the Universe. The downside is that
primordial non-gaussianity turns out to be very small, and the
amplitude of the stochastic background gravitational waves generated
during inflation is tiny and out of reach for LISA.

When more fields are present, symmetry breaking pattern can be
completely different, changing the predictions significantly for
primordial non-gaussianity. In the present work, we propose an
effective theory description based on up to four scalar fields that
allow us to study the symmetry breaking pattern during inflation
systematically~\cite{Rubakov:2004eb,Dubovsky:2004sg}. When diffeomorphisms are completely broken down to a global group required for the existence of de Sitter background space-time, the fluctuations of the scalar fields around their VEVs can be associated with the phonon-like excitations of a self-gravitating medium with the properties of a supersolid~\cite{Son:2005ak,Celoria:2017bbh,Celoria:2020diz}; the
breaking of spatial diffs by a solid-like medium was considered in~\cite{Endlich:2012pz}. Such a medium also provides a mass to the graviton~\cite{Rubakov:2004eb,Dubovsky:2004sg,ussgf,Pilo:2017fyg,Celoria:2017hfd} via gravitational Higgs mechanism. A similar breaking pattern during inflation was already considered in~\cite{Bartolo:2015qvr} with results similar to solid inflation. 
Our analysis shows that in fact the supersolid case is intrinsically rather different from solid inflation, as will be discussed in the rest of the paper. 
In the scalar sector, two independent modes are present and mix from
the beginning, in the Bunch-Davis vacuum, till the end of
inflation. Consequently, perturbations are not anymore purely
adiabatic.

The presence of non-adiabatic perturbations and anisotropic stress of the solid component of the medium leads to violation of the Weinberg theorem. In this sense, the option of inflating and forgetting is not available and reheating must be taken into account to determine the seed of primordial perturbations to be used at large scales as initial conditions for the standard radiation dominated phase of the Universe's evolution. The curvature of constant energy density hypersurface $\zeta$ does not coincide at superhorizon scales with ${\cal R}$ which represents the curvature of hypersurface orthogonal to
medium velocity, and both of them are not perfectly conserved at superhorizon scales; further, the Maldacena consistency relation~\cite{Maldacena:2002vr} is not satisfied. One can show that~\cite{Celoria:2020diz} in the instantaneous reheating approximation, the seed of primordial perturbations are almost adiabatic in agreement with the most recent CMB data~\cite{Akrami:2018odb}.
On a more phenomenological side, we point out the existence of a
region in the parameter space of supersolid inflation such that the
primordial tensor power spectrum (PS) and the related stochastic
background of gravitational waves could be significantly enhanced with
a blue-tilted spectral index via secondary production due to the
non-linear coupling among scalars and
gravitons~\cite{Celoria:2020diz}. Simultaneously, in the same region,
non-gaussianity related to tensor fields can be noticeably enhanced,
keeping the scalar prediction within the current experimental
constraints. Finally, in solid and probably also in supersolid inflation, as discussed in~\cite{Bartolo:2013msa} one should check how fast an anisotropic background gets to a Freedman-Robertson-Walker solution. In particular a statistical anisotropy can be generated~\cite{Bartolo:2014xfa} due to infrared modes that modify the background; such an effect is sizeable when inflation lasts for more than the minimal number of e-folds needed to generate the CMB anisotropies. In this work, we suppose a suitable duration of inflation that allows to transmit the correct amount of isotropy, and constraints on parameters will be obtained considering the current experimental limits on non-gaussianity only. We leave the detailed analysis of this interesting feature for future work.
  




The outline of the paper is the following. After a brief introduction to the effective description of a supersolid given in section \ref{supersolids}, the parameters entering in the quadratic and cubic action are discussed in \ref{mass-app}, while section \ref{PS} is devoted to the discussion of the power spectra, resuming the main results found in \cite{Celoria:2020diz}. Sections \ref{cubic}, \ref{effect},  \ref{NG_study} and \ref{ten} contain the analysis of primordial non-gaussianity. Finally, in section \ref{phen} we briefly analyze the phenomenological implications given by supersolid inflation. 
Our conclusions are drawn in section \ref{conc}.

\section{Supersolids}
\label{supersolids}
Single field inflation is the simplest choice in the vast menu of inflationary models. It successfully addresses all the drawbacks of the hot big bang model and predicts a tiny level of primordial non-gaussianity (PNG)  and very low tensor to scalar ratio in agreement with the lower bound from CMB~\cite{ade, Akrami:2018odb}.  Why then should one consider more complicated models?

The answer is related to the very different PNG predictions.
In single clock inflation, the inflaton background value $\varphi(t)$ spontaneously breaks the 4d-diffeomorphisms of general relativity (GR) down to 3d-diffeomorphisms of the hypersurface $\varphi=$const. The symmetry breaking pattern plays a crucial role in many aspects of inflation and primordial non-gaussianity. For instance, in the squeezed limit, the residual symmetry group determines the form of $f_{\text{NL}}$~\cite{Maldacena:2002vr,Creminelli:2011rh,Senatore:2012wy,Creminelli:2012ed,Hinterbichler:2012nm,Hinterbichler:2013dpa,Hui:2018cag}; see also~\cite{Matarrese:2020why} for a recent discussion. Things are different when 4d-diffeomorphisms are completely broken, leaving only a global symmetry as a leftover. The minimal number of scalar fields, that can implement such a scenario,  is four: ${\varphi^A\, , A=0,1,2,3}$ with a background value, during inflation, of the form
\be
\varphi^0= \bar \varphi(t), \, \qquad \varphi^l= x^l   \, , \quad l
=1,2,3 \, ,\qquad  \bar g_{\mu \nu} = a(t)^2 \,
\eta_{\mu \nu} \, .
\ee
To allow a FLRW (Freedman Lemaitre Robertson Walker)  background solution, the action describing the scalar fields dynamics needs to be symmetric under internal rotations and shift transformations
\be
\varphi^l \to {\cal R}^l_m \, \varphi^m \, , \qquad    \pmb{{\cal R}}^t
\pmb{{\cal R}}=\mathbf{1}\, ,   \qquad \varphi^A \to  \varphi^A + c^A
\qquad  \qquad A=0,1,2,3 \, .
\ee
The building block matrix
\be
C^{AB}=\partial_\alpha\varphi^A\,g^{\alpha\beta}\,\partial_\beta\varphi^B 
\ee
is used to write down the action
\be 
S=\plm^2\, \int dx^4\, \sqrt{-g}\,\left[ R+U(b,\,y,\,\chi,\,\tau_Y,\,\tau_Z,\, w_Y,\, w_Z)\right]\,,
\label{gact}
\ee 
where
\be
\begin{split}
& B^{lm}=C^{lm}\,, \qquad \qquad W^{lm} = B^{lm} - \frac{C^{0l} \,
  C^{0m}}{C^{00}} \, , \qquad  l,m=1,\,2,\, 3\, ; \\
& b=\sqrt{\text{Det}\left[\textbf{B}\right]}\, ,\qquad
\qquad y=u^\mu \, \partial_\mu \varphi^0\,,\qquad
\chi=\sqrt{-C^{00}}\, , \\
&\tau_X= \text{Tr}\left[\textbf{B}\right]\,,\qquad \tau_Y=
\frac{\text{Tr} \left[\textbf{B}^2\right]}{\tau_X{}^2}\,,\qquad
\tau_Z= \frac{\text{Tr} \left[\textbf{B}^3\right]}{\tau_X{}^3}\, ,  \\
&w_X= \text{Tr}\left[\textbf{W}\right]\,,\qquad w_Y= \frac{\text{Tr}
  \left[\textbf{W}^2\right]}{w_X{}^2}\,,\qquad w_Z= \frac{\text{Tr}
  \left[\textbf{W}^3\right]}{w_X{}^3}\, ;
\label{oper}
\end{split} 
\ee
and $u^\mu$ plays the role of  a normalized time-like four-velocity such that $u^\mu\partial_\mu\varphi^l=0$ 
\be
u^\mu=
-\frac{\epsilon^{\mu\nu\alpha\beta}}{6\;b\;\sqrt{-g}}\,\epsilon_{lmn}\;\partial_\nu\varphi^l\,\partial_\alpha\varphi^m\,\partial_\beta\varphi^n
\, , \qquad u^\mu\, u_\mu=-1 \, .
\ee
The action (\ref{gact}) can be interpreted as the relativistic generalization of the low-energy effective Lagrangian describing homogeneous and isotropic supersolids at zero-temperature~\cite{Son:2005ak, Landry:2019iel}.
Such an action is the most general at leading order in a derivative expansion compatible with the given internal symmetries, see~\cite{Celoria:2020diz} for more details. For the benefit of readers we point out that, besides the presence of $\varphi^0$, our choice for the independent operators is slightly different from~\cite{Endlich:2012pz}~\footnote{We use $b=\text{Det}\left[\text{\pmb{B}}\right]^{1/2}$ instead of $\tau_X=\text{Tr}\left[\text{\pmb{B}}\right]$.}.
From the energy momentum tensor (EMT) of the scalars' action, one can infer the energy density and pressure
\be
\rho = \plm^2 \left(-U + y \, U_y + \frac{y^2}{\chi} \, U_\chi \right)
 \, , \qquad p=\plm^2 \left(U -b \, U_b +\frac{\left(y^2- \chi^2\right)}{3 \, \chi }\, U_\chi \right) \, .
\label{prho}
\ee
The action (\ref{gact}) reduces to the one of a solid when the following operators are sent to zero:  $y, \, \chi, \, w_X, \, w_Y, \, w_Z$; that is  equivalent to remove the scalar $\varphi^0$.
 
\section{Mass Parameters and Cubic Vertices}
  \label{mass-app}

At the  background level, the EMT of the scalar fields is the one of a perfect fluid with energy density and pressure given by
\be
  \bar \rho=\plm^2\, \left[ -U+\frac{\bar \varphi'}{a} ( U_\chi+ U_y )\right]  \,,
\qquad   \bar p= \plm^2\, \left(U- \frac{1}{a^3}\, U_b \right) \,, \qquad {\cal H}^2= \left(\frac{a'}{a}\right)^2 = \frac{\bar \rho \; a^2}{6 \, \plm^2}\, ,
\ee
with $U$ and its derivatives evaluated at the background value of the independent operators
\be\label{back}
\bar b=\frac{1}{a^3}\, , \qquad \bar y= \bar \chi=\frac{\bar\varphi'}{a}\, ,
\qquad \bar \tau_Y=\bar w_Y=\frac{1}{3} \, ,\qquad \bar \tau_Z=\bar
w_Z=\frac{1}{9} \, .
\ee
The simplest way to identify the parameters entering in the quadratic and cubic expansion of (\ref{gact}) is to work in unitary gauge where the scalar fields are frozen to their background values and all perturbations are in the metric $g_{\m\n}=a^2(\eta_{\m\n}+h_{\mu\nu})$. Schematically, the expansion of the Lagrangian of the scalars has the following structure
\bea
  \sqrt{-g}\;U&\sim& f_1(U,\;U')\;h \hspace{1.6cm} \quad \; \; \, \Rightarrow
  \hspace{1.7cm} \; \bar \rho, \; \bar p 
  \\ &&\nonumber
+   f_2(U,\;U',\;U'')\;h^2  \hspace{1.1cm} \Rightarrow\hspace{1.6cm}\; M_{i=0,...4}
    \\&&\nonumber
   + f_3(U,\;U',U'',\;U^{(3)})\;h^3\hspace{0.35cm}
   \Rightarrow\hspace{1.5cm}\; \lambda_{i=1,...10} \, .
\ea
From SO(3) symmetry, we can define the following independent operators.
\begin{itemize}
\item Linear level: $h_{00},\;h_{ii}$ and  the associated parameters, the energy density $\bar \rho$ and the pressure $\bar p$.
\item
Quadratic level: $h_{00}^2,\;h_{0i}^2,\;h_{ii}^2,\;h_{ij}^2,\;h_{00}\,h_{ii}\;$ with 5 independent associated mass-like parameters (actually they have the dimension of a mass squared) $\{M_{i=0, \cdots, 4} \}$.
\item Cubic level: $h_{00}^3$, $h_{00}^2\;h_{ii}$, $h_{00}\;h_{ii}^2,$, $h_{00}\;h_{ij}^2$, $h_{00}\;h_{0i}^2$, $h_{0i} \;h_{ij} \;h_{0j}$, $h_{ii}^3, 
   \;h_{ii}\;h_{ij}^2$, $h_{ii}\;h_{0i}^2$, $h_{ij} \;h_{jk} \;h_{ki}$ with the corresponding ten parameters,  $ \lambda_{i=1,...,
      10 }$ .
\end{itemize}
In particular, the quadratic Lagrangian for scalars is given by
\be
    \begin{split}
{\cal L}_2 &=\frac{\plm^2\,a^4}{4}   \Big [
   \left(M_0+\frac{\bar \rho }{2\, \plm^2}\right)\,h_{00}^2-
   \left(M_2+\frac{\bar p}{ \plm^2}\right) \,h_{ij}^2-
   \left(2 \,M_4+\frac{\bar p}{ \plm^2}\right)\,h_{00}\, h_{ii}
    \\&
 \hspace{2.cm} +2\, \left(M_1+\bar p\right)
   \,h_{0i}^2+\left(\frac{M_3}{ \plm^2}+\frac{\bar p}{2}\right)\, h_{ii}^2 \Big ]  \, ;
\end{split}
\ee
with
\bea
&& M_0 = \frac{\bar\varphi'{}^2 \left(U_{\chi\chi }+2 \, 
    U_{y \chi}+U_{yy}\right)}{2 \,a^2} \, , \quad  M_1
=-\frac{\bar\varphi'{} \, U_{\chi }}{ a} \, ,\\
&& M_2 = -\frac{4}{9}\, \left(U_{w_Y}+U_{w_Z}+U_{\tau_Y}+U_{\tau_Z}\right) \, , \nb\\
&& M_3 =\frac{M_2}{3}+\frac{U_{bb}}{2 \,a^{6}\,} \, , \qquad M_4 = \frac{\bar\varphi ' \left[U_{b
       \chi}+U_{by} -a^3 \left(U_{\chi }+U_y\right)\right]}{2 \,a^4}\, ;
\ea
evaluated at the background values of the operators (\ref{back}). 
The EMT conservation at the background level is equivalent to
\be
\bar\varphi''-{\cal H}\,(1-3\, c_b^2)\;\bar\varphi'=0\, , \quad
\quad c_b^2 \equiv - \frac{M_4}{M_0}\, .
\label{tbkg}
\ee
Thanks to the above relations, it is easy to realize that for a generic $U$, at background level, the following quantity is conserved %
\be
\bar \sigma=\plm^2\,\frac{(U_\chi+U_y)}{\bar b} \, , \qquad \bar \sigma'=0 \, .
\ee
that we call background entropy per particle, see~\cite{Celoria:2017hfd,Celoria:2017bbh,Celoria:2020diz}.
We will consider inflation in a slow-roll (SR) regime for which
\be
\epsilon=1-\frac{{\cal H}'}{{\cal H}^2}= \frac{3}{2}\,(1+w) \ll 1
\,, \qquad \qquad \eta= \frac{\epsilon'}{\epsilon \, {\cal H}} \ll
1,\, \qquad \qquad w=\frac{\bar p}{\bar \rho}.
\label{epsdef}
\ee
It is useful to introduce the following parametrization for $\bar\sigma$ and  the masses $ M_i$ 
\be
\hat{\sigma}=\frac{\bar\sigma}{M_{\text{pl}}^2 \, H^2 \, \epsilon}\,,
\qquad M_i= 4\, H^2\, \epsilon\, c_i^2, 
\quad i=0,...,4
\label{mass_low}
\ee
in terms of the Hubble parameter $H=a^{-1} \, {\cal H}$ (almost constant during slow-roll), $\epsilon$ and $\{c_i\}$; with
\be
c_i' \propto \epsilon\;\;\;{\rm i.e.} \;\;\; M_i'\propto \epsilon^2 
\ee
slow varying in time. The $M_i$   proportionality to $\epsilon\sim(\rho+p)$ is required by dynamical stability (\ref{stab1}, \ref{stab2}), while the slow varying condition ensures the plane-wave solutions on subhorizon scales, as discussed in \cite{Celoria:2020diz}. 
We will often use $c_L^2$ defined by 
\be
c_2^2=\frac{3}{4}\,(1+c_L^2) \, ,
\label{cl_2} 
\ee
instead of $c_2^2$ and we will substitute $c^2_{3}$ in terms of the adiabatic sound speed $c_s^2$ that during SR is approximately equal to -1 (see also (\ref{stab2}))
\be
\begin{split}
\label{cs}
c_s^2\equiv \frac{\bar p'}{\bar\rho'} =&\frac{2}{3}\,(c_2^2-3\,c_3^2+3\,c_b^4\,c_0^2)=-1+{\cal O}(\epsilon,\,\eta)\\
& c_3^2=\frac{1}{2}+c_0^2 \, c_b^4+\frac{1}{3}\, c_2^2\,.
\end{split} 
\ee
The cubic order Lagrangian ${\cal L}_3$ in the unitary gauge, at the leading order in the SR expansion, results
\bea
&& \frac{\plm^2}{8}\, a^4 \,H^2\, \Big \{ 
h_{00} \,h_{0i}^2\, \left[2 \,\epsilon  \left(8 \,c_0^2 \,\left(c_b^2-1\right)+8\,
   c_1^2-\lambda _{10}+4\right)-12\right]+h_{00}^3\, \left[\epsilon  \,\left(8\,
   c_0^2+\frac{\lambda _2}{6}\right)+3\right]\nb\\
&&\nonumber
+   h_{0i}^2\, h_{jj} \,\left[\epsilon\,  \left(2 \,\hat{\sigma }\, a^{-3\,
   \left(c_b^2+1\right)}-16 \,c_0^2 \,c_b^2+8 \,c_1^2+16\, c_3^2-2\, \lambda _8+\frac{16\,
   \lambda _9}{3}+8\right)-12\right]\\
&&\nonumber
  + h_{00}^2 \,h_{ii}\, \left[\epsilon \, \left(4 \,c_0^2\,
   \left(c_b^2+1\right)-\frac{\lambda _3}{2}-2\right)+3\right]+ h_{00} \,h_{ij}^2\, \left[4\, \epsilon  \,\left(-2 \,c_0^2 \,c_b^2+c_2^2+\lambda
   _6+1\right)-6\right]\\
&&\nonumber +h_{00} \,h_{ii}^2\,
   \left[\epsilon\,  \left(4 \,c_0^2\, c_b^2-4 \,c_3^2+\frac{\lambda _4}{2}-\frac{4
   \lambda _6}{3}-2\right)+3\right]+  h_{ki} \,h_{ij}\,  h_{jk}\, \left[\frac{8}{3} \, \epsilon   \,\left(6  \,c_2^2-\lambda _7+2\right)-8\right] \\
&&\nonumber
+h_{ii}\,h_{ij}^2\, \left[\epsilon  \,\left(\frac{8\, \lambda _7}{3}-4\,
   \left(3 \,c_2^2+2 \,c_3^2+\lambda _5+1\right)\right)+6\right] \\
&&\nonumber   
+h_{ii}^3\, \left[\frac{\epsilon}{54} \,  \left(168  \,c_2^2-9 \,
   \lambda _1+72  \,\lambda _5-32  \,\lambda _7+36\right)-1\right]\\
&&
+   h_{0i}\,
  h_{ij}\, h_{j0} \,\left[24-16\, \epsilon  \,\left(c_1^2+c_2^2+\lambda _9+1\right)\right]
   \Big \} \, ;
   \ea
where the $\{\lambda_i \}$ parameters at the leading order in SR are defined by 
\be
\begin{split}
& \epsilon \, H^2 \, \lambda_1= a^{-9}\, U_{bbb}\,, \qquad
\epsilon \, H^2 \, \lambda_2= a^{-9\, c_b^2}\, \left(U_{\chi\chi\chi}+3\,U_{\chi\chi y}+3\,U_{\chi yy}+U_{yyy}\right)\,,\\
& \epsilon \,  H^2 \, \lambda_3=a^{-3-6\,
  c_b^2}\,\left(U_{byy}+2\,U_{by\chi}+U_{b\,\chi\chi}\right)\,, \qquad
 \epsilon \, H^2 \, \lambda_4=a^{-6-3\, c_b^2}\,\left(U_{bby}+U_{bb\chi}\right)\,,\\
& \epsilon \, H^2 \, \lambda_5= \frac{a^{-3}}{  9}\,\left(U_{b w_Y}+U_{b w_Z}+U_{b \tau_Y}+U_{b \tau_Z}\right)\,,\\
& \epsilon \, H^2 \, \lambda_6= \frac{a^{-3\, c_b^2}}{  9}\,\left(U_{y w_Y}+U_{y w_Z}+U_{y \tau_Y}+U_{y \tau_Z}+U_{\chi w_Y}+U_{\chi w_Z}+U_{\chi \tau_Y}+U_{\chi \tau_Z}\right)\,,\\
& \epsilon \, H^2 \, \lambda_7=\frac{1}{
  9}\,\left(U_{\tau_Z}+U_{w_Z}\right)\,, \qquad
 \epsilon \, H^2 \, \lambda_8=a^{-3-3\, c_b^2}\,\left(U_{by}\right)\,
 , \qquad
\epsilon \, H^2 \, \lambda_9=\frac{1}{
  9}\,\left(U_{w_Y}+U_{w_Z}\right)\, , \\
&\epsilon \, H^2 \, \lambda_{10}=-a^{-6\,c_b^2}\,\left(U_{\chi y}+U_{yy}\right)\,,
\end{split} 
\ee
and can be time dependent in general.  

\section{Power Spectrum in Supersolid Inflation}
\label{PS}
It is useful to recap some results obtained in \cite{Celoria:2020diz}, focusing on the elements relevant for the PNG computation done in the spatially flat gauge
\be
\begin{split}
& \varphi^0= \bar \varphi(t) + \pi_0, \, \qquad \varphi^l= x^l + \de_l \pi_L  \, , \quad l =1,2,3\, , \\
&ds^2{  =a^2\left[ (-1+2 \, \Psi)\; dt^2 + 2 \, dt \;dx^i \, \de_i F +
   \gamma_{ij}\;dx^i \;dx^j\right] \, ,} \\
& \gamma_{ij} = \delta_{ij}+ h_{ij}  \, , \qquad h_{ii}=\de_j h_{ij}=0 \, .
\end{split}
\ee
For the spin two field, in Fourier space,  we use the decomposition $h_{ij}(\textbf{k})=\sum_p\; \epsilon^p_{ij}( \hat{\textbf{k}})\,
h^p_{\textbf{k}}$ in terms of the circular polarization tensors $\epsilon^p_{ij}( \hat{\textbf{k}})$ ($p=+,\times$) that are traceless ( $\epsilon^p_{ii}( \hat{\textbf{k}})=0$ ) and transverse ( $\hat{\textbf{k}}^i\epsilon^p_{ij}( \hat{\textbf{k}})=0$ ).
The scalars~\footnote{$\Psi$ is related to the lapse and $F$ to the scalar part of the shift of the ADM formalism.} $\Psi$ and $F$ have algebraic equations of motion and can be integrated out in favor of the proper Goldstone modes $\pi_0$ and $\pi_L$. In particular, working in a quasi de Sitter background in SR regime, they result to be proportional to $\epsilon$ so that, at leading order, they can be neglected both in the quadratic and cubic action, and no operator containing such fields will be shown. 
Expanding (\ref{gact}) at the quadratic order in the fluctuations we get 
\bea\nonumber
{\cal L}_2&=&\frac{1}{2} \,a^2 \,\plm^2\,H^2\, \epsilon  \Big [
4 \,\left(c_1^2+1\right)\, (\partial_{i}\pi_L')^2+
\frac{8 \,c_0^2 }{\bar\varphi '{}^2}\, \pi _{0}'{}^2
-\frac{8\, \left(2 \,c_0^2\,c_b^2+c_1^2\right)}{\bar\varphi'}
    \,(\Delta\pi _L \,\pi _{0}' )
     +
     \\
     && \nonumber
     \frac{4 \,c_1^2 }{\bar\varphi'{}^2}\,(\partial_i\pi_{0})^2+
   4\,\left(1+2\,c_b^4\, c_0^2-\frac{4}{3}\,c_2^2\right) (\Delta\pi _L)^2
   -\frac{24\,a\,H\,(1+c_b^2)\,c_1^2}{\bar\varphi'}\;\pi_0\,\Delta\pi_L
   \Big ]+
   \\
   &&
   \frac{ a^2\, M_{\text{pl}}^2}{4}\,\left[
     (h_{ij}')^2-M_2\,a^2\,h_{ij}^2 + (\partial_{\k}h_{ij})^2 \right]
     \label{slowact}
  \ea
where $\Delta=\partial_i^2$. \\
In the scalar sector, there are two independent propagating scalar modes that, following~\cite{Celoria:2020diz}, can be quantized by diagonalizing the quadratic action and  implementing the corresponding Bunch-Davies { (BD)} vacuum.
It is important to point out that the parametrization (\ref{mass_low}) is more than a convenient choice but it is forced by stability and slow-roll. Stability derived from (\ref{slowact}) requires that~\cite{Celoria:2017hfd}
\be\label{stab1}
M_0 >0 \, , \qquad -(\bar\rho+\bar p)< M_1<0\, , \qquad M_2 >0 \, , \qquad
M_2>M_3 \, ;
\ee
on the other hand, in quasi de Sitter, $\bar p+\bar \rho \sim H^2 \, \epsilon$, thus $M_1 \sim \epsilon \, H^2$. Now, the adiabatic speed of sound (\ref{cs}) is explicitly  given by
\be\label{stab2}
\frac{\bar p'}{\bar \rho'}=\frac{a^2 \left(3 \, M_4  +M_2-3 \, M_3\right)}{6\;\epsilon \;\mathcal{H}^2}
\ee
and during slow-roll is very close to $-1$. Thus, unless $M_4$, $M_2$ and $M_3$ are precisely tuned, the choice  (\ref{mass_low}) is the only possible. Finally, notice that if $\pi_0$ should play the role of an inflaton field~\footnote{If $M_0 \sim  \, H^2$  then $\pi_0$ is a spectator field and physics is completely different.}, likewise $\pi_L$, then $M_0 \sim \epsilon \, H^2$.  

The quadratic action (\ref{slowact}) in the scalar sector describes two massless modes with a kinetic mixing. 
To remove the mixing, one should set $c_b^2=-1$ and $c_1^2 =2 \, c_0^2$; unfortunately stability imposes
that $M_1$ and then $c_1^2$ to be negative, while $M_0$ and then $c_0^2$ to be positive. Thus, the mixing between longitudinal phonons cannot be undone by a tuning of the parameters. The diagonalisation of  (\ref{slowact}) is non-trivial and it is discussed in~\cite{Celoria:2020diz}. The result is two independent longitudinal phonons with speed of sound $c_{s1}$ and $c_{s2}$.  

\begin{figure}[!tbp]
  \centering
   \begin{minipage}[b]{0.4\textwidth}
    \includegraphics[width=\textwidth]{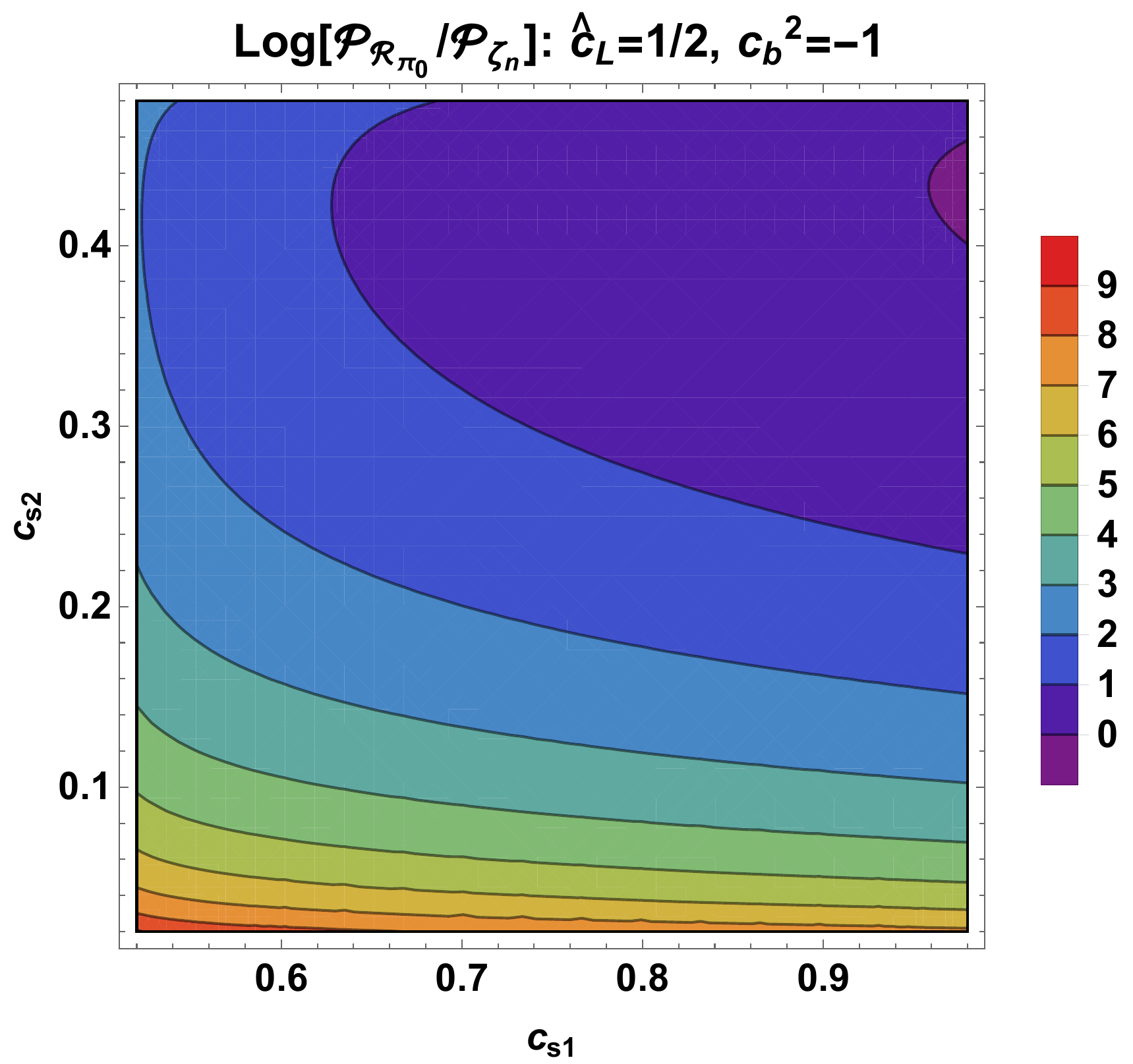}    
  \end{minipage}
  \hspace{1cm}
  \begin{minipage}[b]{0.41\textwidth}
    \includegraphics[width=\textwidth]{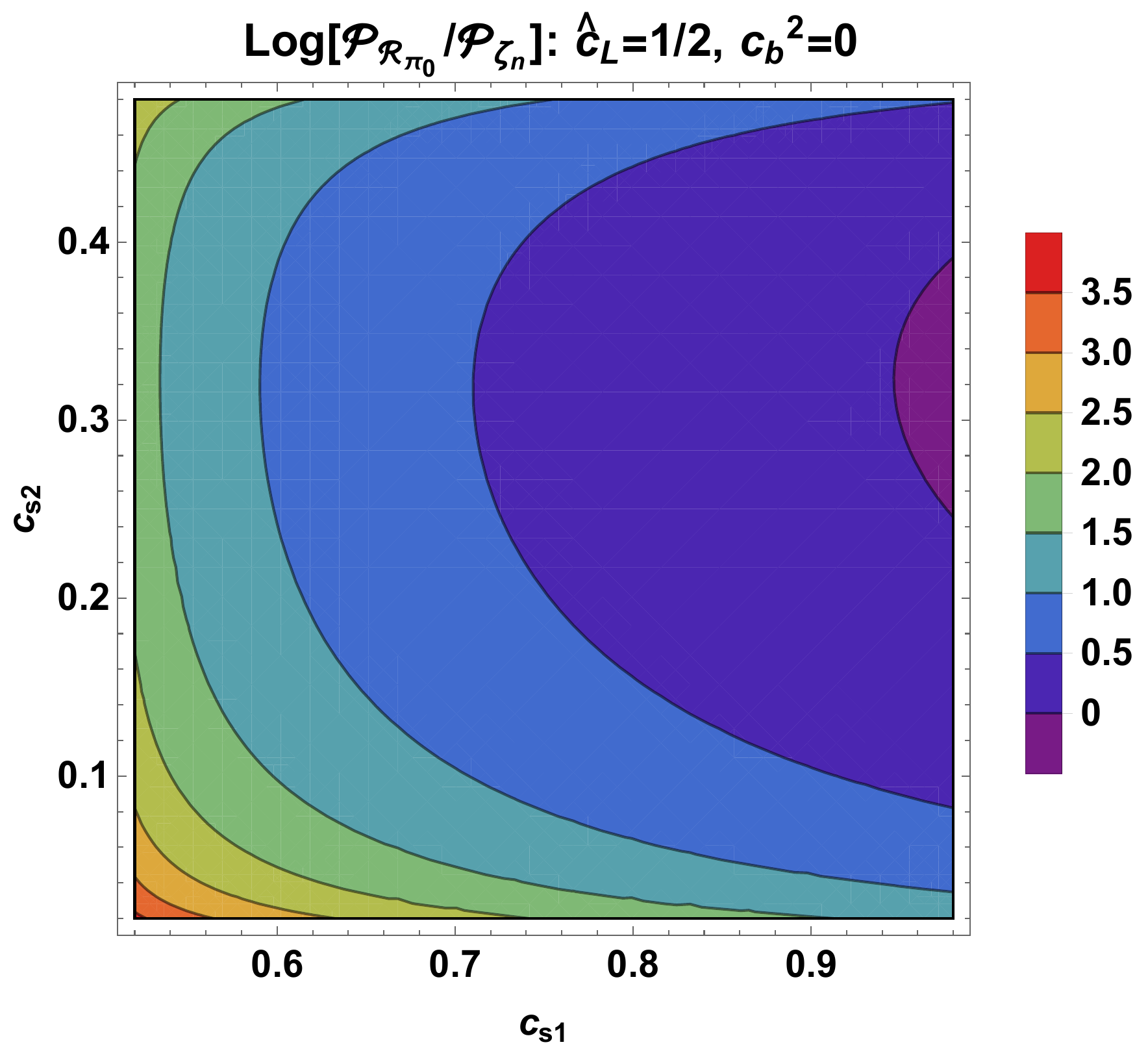}
  \end{minipage}
\caption{$Log_{{}_{10}}\left[{\cal P}_{{\cal R}_{\pi_0}}/{\cal P}_{\zeta_n}\right]$ plot once $\hat{c}_L$ is set to $1/2$. Note the ${\cal P}_{{\cal R}_{\pi_0}}$ enhancement particularly  important for $c_b^2=-1$ case.}
\label{P_ratio}
\end{figure}

An essential consequence of  (\ref{mass_low}) is that by consistency, imposing that the $\{ M_i \}$ are time independent at the leading order in slow-roll and $\bar \sigma$ is strictly constant, we get some useful relations among the parameters. Namely
\be
\begin{split}
\label{dep_par}
&  {M_0}'\sim O(\epsilon^2)\;\; \Rightarrow \;\;\lambda_3=-c_b^2 \, (16\, c_0^2+\lambda_2)\,;\\
& {\bar \sigma}''=0\;\hspace{0.9cm} \;\Rightarrow\;\; \lambda_4= 8\, c_0^2\, c_b^2 \,(1+3\, c_b^2)+c_b^4 \, \lambda_2 \,;\\
&  {M_2}' \sim O(\epsilon^2)\;\;\; \Rightarrow \;\; \lambda_5=-c_b^2\, \lambda_6 \,; \\
&  {M_3}' \sim O(\epsilon^2)\;\;\; \Rightarrow
\;\;\lambda_1=\,-8-24\,c_0^2\,c_b^4\;(1+c_b^2)-c_b^6\, \lambda_2\,;\\
& M_1' \sim O(\epsilon^2)\;\;\;\; \Rightarrow \;\;
 \lambda_8=c_b^2\;(\lambda_{10}-4\,c_1^2)+\hat\sigma\;a^{-3\,(1+c_b^2)}
 \, .
\end{split} 
\ee
Finally we can express $c_0^2$ and $c_1^2$ in terms of the diagonal sound speed parameters $c_{s1}$, $c_{s2}$ and $c_L$ according to~\cite{Celoria:2020diz}
\be 
\begin{split}
& c_0^2
=\frac{\left(c_L^2-c_{s1}^2\right)  \,\left(c_L^2-c_{s2}^2\right)}{2  \,c_b^4 \, \left(c_L^2-c_{s1}^2\right)-2 \, c_{s2}^2  \,\left(-2 \, c_b^2 \, c_{s1}^2+c_b^4+c_L^2 \,
   c_{s1}^2\right)}\,, \\
& c_1^2 
=\frac{\left(c_L^2-c_{s1}^2\right)
  \,\left(c_L^2-c_{s2}^2\right)}{-2\, c_b^2 \,c_L^2+c_b^4+c_L^2
  \left(c_{s1}^2+c_{s2}^2\right)-c_{s1}^2 \,c_{s2}^2}\,.
\label{expdia}
\end{split}
\ee
Thus, the number of free parameters is given by the two sound speeds ($c_{s1}$, $c_{s2}$), the ``graviton mass''~\footnote{ the parameter $c_L^2$ is  equivalently to $c_2^2$ by   eq.(\ref{PS_zeta_n}).} $c_2^2$, the mass ratio $c_b^2$ and the entropic parameter $\hat\sigma$ plus
the ten parameters $\{\lambda_i\}$ of the cubic interactions constrained by the five relations (\ref{dep_par}). Thus we end up with the following 10 independent  parameters: $\{ \hat\sigma,\, c_{s1}^2,\, c_{s2}^2, c_2^2, c_b^2,\lambda_{2,\,6,\,7,\,9,\,10} \}$.
An explicit example of a Lagrangian describing a supersolid where the constraints (\ref{dep_par}) are automatically incorporated is the following 
\be
\begin{split}
\label{toy-Lagrangian}
U = -6\, H^2 +\epsilon_0\,V(b)+\epsilon_0\, U_{\tilde\Lambda}({\cal X},\,{\cal Y}, \,\tau_{Y},\,\tau_{Z}, \,w_{Y},\,w_{Z}) \,, 
\end{split}
\ee
we have defined the combination of operators
\begin{equation*}
{\cal X}=b\,\chi\,,\qquad {\cal Y}= b\, y.
\end{equation*} 
The term $U_{\tilde\Lambda}$ in $U$ can be called a {\it background} $\Lambda$-medium 
due to the fact that, at the background level,  has the
equation of state of a cosmological constant
\be
\bar\rho_{\tilde\Lambda}=-\bar p_{\tilde\Lambda}\,,\qquad c_b^2=-1.
\ee
The perfect fluid $V(b)$ part has to be tuned to the specific functional form
\be
 V(b)=-4\, H^2 \, \log(b)\,,
 \ee
to ensure the presence of a quasi de Sitter phase with an almost constant $\epsilon$ factor
 \begin{equation*}
\epsilon(t)= \frac{3}{2}\,\frac{\rho+p}{\rho} = \epsilon_0 +O\left(\epsilon_0^2\right)\,. 
\end{equation*}
Thus, such a rather general Lagrangian automatically satisfies eqs.(\ref{dep_par}) and generates a slow-roll regime with an almost constant $\epsilon$. More details are given in Appendix \ref{toy_model}. 
Note that in \cite{Celoria:2017idi} and \cite{Celoria:2019oiu} we defined a bit different potentials named simply $\Lambda$-Media, characterised by the fact that $\rho_{\Lambda}+p_{\Lambda}=0$ worth exactly, not only on their  background value~\footnote{Another realization of an exact $\Lambda$-Medium is given in \cite{Ferrero:2020jts}.}. An example of a supersolid that it is also a $\Lambda$-Medium, in our notations, is $U_{\Lambda}(b\,y,\,\frac{\chi^2-y^2}{b^{2/3}},\,\tau_{Y},\,\tau_{Z}, \,w_{Y},\,w_{Z})$. Note that the constraint $M_1'\sim{\cal O}(\epsilon)$ is not satisfied for $U_{\Lambda}$. \\
Let us come back to the description of the perturbations PS.
One of the key features in (\ref{slowact}) is that $\pi_L-\pi_0$ mix
both at super and sub horizon scales~\footnote{ This is one of the
  important differences with the analysis in~\cite{Bartolo:2015qvr} is
  that the kinetic mixing was treated as small. In addition, the
  authors set $\bar {\varphi}'=a$ which  leads to a conserved
  background EMT only if $c_b^2=0$. Such a value of $c_b^2$  is rather
  peculiar, as we will see in what follows. The correct implementation
  of the St\"uckelberg trick at the background level requires a
  non-trivial background for $\varphi^0$ satisfying
  (\ref{tbkg}). In~\cite{Ricciardone_2017} the analysis is similar but
  a non-minimal coupling
  between the scalars and curvature is also considered.}.
As a result, any scalar field $\xi$, resulting from a combination of $\pi$ fields, can be written in Fourier space as a linear combination of two independent annihilation (creation) operators $a_k^{(i)}$ ( $a_k^{(i)}{}^\dagger$), $i=1,2$ according with
\be
\begin{split}
& \xi(\textbf{x})=\frac{1}{(2\,\pi)^3}\int\, d^3 k\, e^{i \,
  \textbf{k}\cdot \textbf{x}}\, \xi_{\textbf{k}}\,, \quad
\xi_{\textbf{k}}= \sum_{i=1}^2 \,\xi_{k}^{(i)}\,
a^{\dagger}_{\textbf{k}}{}^{(i)}+\xi_{k}^{(i)\,*}\,
a_{-\textbf{k}}^{(i)}  \,;
\label{modes}
\end{split}
\ee
where $\left[a^{(i)}_{\textbf{k}},\, a^{\dagger}_{\textbf{p}}{}^{(j)}\right]=(2\,\pi)^3 \, \delta^{(3)}(\textbf{k}-\textbf{p})\,\delta_{ij}$ and the modes $\{ \xi_{k}^{(i)} \, , \; i=1,2\}$ are two classical solutions of the linearized equations of motion for $\xi$ with initial conditions at early conformal time $t \to -\infty$ set by the choice of the BD vacuum. 
Thus, the two-point function and the related linear power spectrum for $\xi$ reads
\be
\begin{split}
& \langle \xi_\textbf{k}\, \xi_\textbf{p}\rangle = (2\,\pi)^3 \,
P_{\xi}(k)\, \delta^{(3)}(\textbf{k}+\textbf{p}) \, ;\\
&{\cal P}_{\xi}  = \frac{k^3}{(2\,\pi^2)} \,  P_{\xi}(k) \equiv
\frac{k^3}{(2\,\pi^2)}  \, \left(
  |\xi_{k}^{(1)}|^2+|\xi_{k}^{(2)}|^2\right) ={\cal P}^{(1)}_{\xi}+{\cal P}^{(2)}_{\xi}\, .\\
\end{split}
\ee 
As discussed in \cite{Celoria:2019oiu,Celoria:2020diz}, reheating\footnote{This is strictly valid in the approximation of an almost instantaneous reheating.} is likely to process primordial perturbations produced during inflation such that the initial conditions for the hot phase are determined by the curvature of constant number density of lattice sites $n_\ell$, given by the operator $b$. The corresponding curvature perturbation related to the $b$ operator is provided in the spatially-flat gauge at the linear order by
\be
\zeta_n\equiv \frac{k^2}{3} \pi_L \,.
\label{curv1}
\ee
Another important scalar field is given by the
curvature perturbation of the constant $\varphi^0$ hypersurface, namely
\be
{\cal R} _{\pi_0} \equiv\frac{\cal H} {\bar{\varphi}'} \pi_0 \,.
\label{curv2}
\ee
The two above perturbations are closely related to the Goldstone-like excitations $\pi_0$ and $\pi_L$, that are the fluctuations around the VEVs responsible for the complete breaking of four-dimensional diffeomorphisms during inflation. To figure out better their roles, notice that ${\cal R} _{\pi_0}$, when the single-field breaking pattern is considered (absence of $\varphi^a$ fields), tends to coincide with ${\cal R}$, the curvature of constant fluid velocity surface. Contrary, when only the solid breaking pattern is present (absence of $\varphi^0$), the $\zeta_n$ field tends to coincide with the $\zeta$ field, the curvature of constant energy density surfaces. \\
Finally, let us define  
\be
\label{redef}
\tilde {\cal R}_{\pi_0}'=\frac{{\cal H}}{\bar{\varphi}'} \pi_0' \equiv {\cal R}_{\pi_0}'-3\, c_b^2 \, {\cal H}\,{\cal R}_{\pi_0}\,. 
\ee
This last has no fundamental meaning, but it is useful to minimize the number of cubic interactions. \\
All scalar perturbations can be written in terms of $\zeta_n$, ${\cal R}_{\pi_0}$ and their time derivatives, whose linear PS are given by~\cite{Celoria:2020diz}
\be 
\begin{split}
&{\cal P}_{\zeta_n}^{(1)}= \frac{\left(c_b^2-c_{s1}^2\right){}^2 \left(
    c_L^2-c_{s2}^2\right)}{c_{s1}^5 \left(c_b^2-c_L^2\right){}^2
  \,\left( c_{s1}^2-c_{s2}^2\right)}\,{\cal P}\,, \qquad {\cal P}_{\zeta_n}^{(2)}={\cal P}_{\zeta_n}^{(1)} \, :\;\; c_{s1} \longleftrightarrow c_{s2}\,; 
\end{split}
\label{PSz}
\ee
\be
\begin{split}
& {\cal P}_{{\cal R}_{\pi_0}}^{(1)}=\frac{\left(c_b^2-c_{s2}^2\right){}^2}{c_{s 1} \, \left(c_L^2-c_{s2}^2\right) \left(c_{s1}^2-c_{s 2}^2\right)} \,{\cal P}\,, \qquad {\cal P}_{{\cal R}_{\pi_0}}^{(2)}={\cal P}_{{\cal R}_{\pi_0}}^{(1)}\, :\;\; c_{s1} \longleftrightarrow c_{s2}\,; 
\end{split}
\label{PS0}
\ee
together with the cross-correlation
\be
{\cal P}_{\zeta_n \, {\cal R}_{\pi_0}}^{(1)} ={\cal
    P} \, \frac{\left(c_{s1}^2-c_b^2\right)\left(c_{s\,2}^2-c_b^2\right)}{\left(c_L^2-c_b^2\right)\left(c_{s1}^2-c_{s2}^2\right) } \,\frac{1}{
    c_{s1}^{3}}\, \qquad {\cal P}_{\zeta_n \, {\cal R}_{\pi_0}}^{(2)}=-\frac{c_{s1}^3}{c_{s2}^3}\,{\cal P}_{\zeta_n \, {\cal R}_{\pi_0}}^{(1)}\,,
\label{cross}
\ee
where expressions (\ref{PSz}, \ref{PS0}, \ref{cross}) are manifestly symmetric under $c_{s1} \longleftrightarrow c_{s2}$ exchange. We have also introduced  the power spectrum ${\cal P}$ of canonical single field inflation given by
\be
\label{Psingle}
{\cal P} = \frac{H^2}{16\, \pi^2\, \plm^2 \, \epsilon} \, .
\ee
These relations are strictly valid for $c_b^2=0,\,-1$ only, where analytic solutions for the $\zeta_n$/${\cal R_{\pi_0}}$ modes can be found (see Appendix \ref{f_nl_sq}, eq. (\ref{an_sol})). Thus, from here below, we will consider the $c_b$ parameter fixed; while
the diagonal sound speeds are conventionally  ordered such that $c_{s2}^2< c_{s1}^2$; moreover stability requires that $ c_{s2}^2 < c_L^2< c_{s1}^2$. \\
In particular, in the rest of this work, we will focus on the $c_b^2=-1$ case. Our choice can be ascribed to the higher ${\cal R}_{\pi_0}$ sensitivity to low values of $c_{s2}$, as it can be easily inferred from fig. \ref{P_ratio}. As deeply argued in the next sections, this will enhance the tensor NG,\footnote{As well as the tensor power spectrum as argued in \cite{Celoria:2020diz}} in the limit of small $c_{s2}$. We would like to clarify that formal results exposed in Appendixes \ref{f_nl_sq}, \ref{TSS_total} and table \ref{Squeezed_shapes} will be still valid for $c_b^2=0$, and differences rise only once the $c_{s2}$ expansion is implemented.
In order to simplify the PNG notation, we define the following dimensionless ratios:
\be
 \label{r_par}
 r_{\zeta_n}^{(j)}=\frac{{\cal P}_{\zeta_n}^{(j)}}{{\cal P}_{\zeta_n}}
 \,, \qquad r_{L/0}^{(j)}=\frac{{\cal P}_{\zeta_n{\cal
       R}_{\pi_0}}^{(j)}}{{\cal P}_{\zeta_n}} \,,
\ee
specifying how far from ${\cal P}_{\zeta_n}$ the single-mode component ${\cal P}_{\zeta_n}^{(j)}$ or the cross-correlations $\zeta_n$-${\cal R}_{\pi_0}$ could be.\\
Considering that, during an almost instantaneous reheating, $\zeta_n$ is continuous during the transition and can be regarded as the seed of primordial adiabatic perturbations~\cite{Celoria:2020diz}, its PS is constrained by the CMB experiments and assumes the form:  
\be
\label{PS_zeta_n}
{\cal P}_{\zeta_n}={\cal P}_{\zeta_n}^{(1)}+{\cal P}^{(2)}_{\zeta_n}\equiv\frac{ {\cal P} }{\hat{c}_L^{5}} \simeq 10^{-9} \,,
\ee  
where $\hat{c}_L$ is an effective longitudinal sound speed~\footnote{We used a notation typical for the 2-point function in solid (and also in some single-field) inflationary models~\cite{Endlich:2012pz} where it is present an extra factor $c_L^{-5}$ in adiabatic power spectrum, in our case $c_L^{-5} \to \hat c_L^{-5}$.} which can be read off from (\ref{PSz}). \\
One can show~\cite{Celoria:2020diz} that when $10< \hat{c}_L^{-5}<100$,  the supersolid is dynamical stable, $c_{s1}$ and $c_{s2}$ are subluminal and the $\zeta_n$ power spectrum is well within the PLANCK constraints~\cite{Akrami:2018odb} by taking $10^{-11} \leq {\cal P} \leq 10^{-10}$.  For illustrative purposes we set $\hat{c}_L=1/2$  in our figures, which maximizes  the allowed $(c_{s1},\,c_{s2})$ region. \\
It is important to note that when $c_{s2}$ is much smaller than $c_{s1}$ and  $\hat c_L$, and ${\cal P}_{\zeta_n}$ is fixed, the power spectrum (\ref{PS0}) of ${\cal R}_{\pi_0}$ and the cross-correlation (\ref{cross}) become dominant due to the dependence on $c_{s2}$ (see fig.\ref{P_ratio}). This interesting subset of the parameter space, besides well reproducing PLANCK's data, has very distinctive features for secondary gravitational waves production~\cite{Celoria:2020diz} and for what concerns PNG as discussed in sections \ref{cubic} and \ref{NG_study}.  
We stress that the asymmetry in between the two $c_{s1}$ and $c_{s2}$ parameters starts once we imposed that only the PS of the $\zeta_n$ field is the adiabatic/observed one, as predicted by the matching conditions during an instantaneous reheating phase. It is this last condition that makes the ${\cal R}_{\pi_0}$ power spectrum sensitive to small $c_{s2}$ values.

\section{Cubic Lagrangian (spatially flat gauge)}
\label{cubic}

Primordial non-gaussianity is due to the interactions among the various dynamical fields during inflation and, at the leading order in perturbation theory,  originates from cubic terms that can be divided into interactions among scalars (SSS), gravitons (TTT), and mixed ones: (TSS),  (TTS). Thus the cubic Lagrangian can be split according to 
\be\label{L3}
{\cal L}_3 = {\cal L}^{(SSS)} +  {\cal L}^{(TTT)}+{\cal L}^{(TTS)}
+{\cal L}^{(TSS)}\, .
\ee
As we have discussed in section \ref{mass-app}, there are up to ten parameters in the cubic Lagrangian; we focus on the region of parameters where $c_{s2}^2 \ll 1$ that is probably the most interesting from a phenomenological point of view. The enhancement of the size of PNG in the operators containing $\pi_0$ is due to the extra negative powers of $c_{s2}$ in ${\cal R}_{\pi_0}$ and in the cross-correlation. It is convenient to associate at each cubic vertex a color code according to the power~\footnote{Such classification is strictly valid in the case $c_b^2=-1$.} of $c_{s2}{}^{-1}$ of the corresponding contribution to the 3-point function of $\zeta_n$ and then to $f_{\text{NL}}$ as shown in table \ref{default}. The color code goes from violet (no powers of $c_{s2}{}^{-1}$ are present) to the {\it red} (highest negative power $c_{s2}{}^{-8}$). Clearly, vertices labeled by a large wave-length color potentially give a large contribution to the 3-point function.\\
%
\begin{table}[!ht]
\begin{center}
\begin{tabular}{|c||c||c|c|c|c|}
\hline
Color&$c_{s2}^{-n}$& SSS& TSS &TTS& TTT\\
\hline
&&&&&
\\
{\color{violet} V}& $n=0$&$(\pi_L)^3,\;(\pi_L'^{\,2}\;\pi_L)$&   $(h\;\pi_L^2),\;(h\;\pi_L'^{\,2})$& $h^2\;\pi_L$ &$h^3$\\
&&&&&
\\
{\color{blue}B}& $n=3$&$ (\pi_0\;\pi_L\;\pi_L'),\;(\pi_0'\;\pi_L\;\pi_L'),\;(\pi_0'\;\pi_L^2)\;$&   $h\;\pi_0\;\pi_L',\;h\;\pi_0'\;\pi_L$ &$h^2\;\pi_0'$&\\
&&&&&
\\
{\color{green}G}&$n=5$& $\pi_0^2\;\pi_L $& $h\;\pi_0^2$&&  \\
&&&&&
\\
{\color{orange}O}&$n=6$& $\pi_0'^{\,3},\;(\pi_0'^{\,2}\;\pi_L),\;(\pi_0'\,\pi_0\,\pi_L')$& & 
   & \\
&&&&&
\\
{\color{red}R} & $n=8$& $\pi_0^2\;\pi_0'$   & & &\\
&&&&&
\\
\hline
\end{tabular}
\end{center}
\caption[Supersolid cubic operators classified in the small $c_{s2}$ limit]{Operators in the cubic Lagrangian (\ref{L3}) classified according the powers $c_{s2}^{-n},\;n=0,...,8$ of the corresponding contribution to $\left\langle\zeta_n^3\right\rangle$, in the limit $c_{s2}\ll 1$ and $c_b^2=-1$. When $c_b^2=0$ the $c_{s2}$ divergences are not as stiff as in the $c_b^2=-1$ case.}
\label{default}
\end{table}
\\
By expanding the action at the cubic order, we find in the scalar sector at leading order in $\epsilon$~\footnote{ The normalization of the vertices is chosen to simplify the Lagrangian (\ref{L3ZR}) written in terms of $\zeta_n$ and ${\cal R}_{\pi_0}$, characterized by an almost flat power spectrum.}:
\be 
\begin{split}
{\cal L}^{(SSS)}=&  \plm^2 \, {  H ^2}\,
  \epsilon \, a^4 \Big \{
  - \frac{1}{27}\,\de_{m n}
\pi_L\,\left[\de_{ij} \pi_L \,  \de_{k l}\pi_L \,   \left( V_{1} \, \delta_{ij}  \,\delta_{kl} \,
 \delta_{mn} + V_{2} \,
  \delta_{in}  \,\delta_{jk} \, \delta_{lm} \nb
\right.\right.\\
&\left.\left.
   +V_{3} \, \delta_{ij}  \,\delta_{kn}  \,\delta_{ml}\right) 
-\,  \de_i  \pi_L' \,  \de_l \pi_L' \,    \left( V_{4}  \,
  \delta_{il}  \,\delta_{mn} + V_{5} \,  {  \delta_{in}} \,
  \delta_{lm} \right)\right] \\
&+\frac{1}{9\;{\bar \varphi'}}\,\left[ \de_i \pi_0 \,    \de_{j l}
\pi_L  \, \de_n \pi_L' \, \left( B_{1} \, \delta_{lj}  \,\delta_{in} +
  B_{2} \, \delta_{ij} \, \delta_{ln}\right) \nb 
  -   \pi_0'\left( \,B_{3} \,  \, \de_i \pi_L'\, \de_i \pi_L' 
 \right. \right.  
\\
&
 \left.\left.
+ \de_{i j}
\pi_L \,   \de_{l n} \pi_L  \, \left( B_{4}\, \delta_{ij} \, \delta_{ln} +
  B_{5} \,  \delta_{il}  \,\delta_{jn} \right)\nb\right)\right] \\
  &+ \frac{1}{3\;\bar\varphi'{}^2}\, \de_i  \pi_0 \,  \de_l \pi_0 \,  \de_{m n}
\pi_L \,  \,\left(  G_{1} \, \delta_{il}  \,\delta_{mn} + G_{2} \, \delta_{im} \,
  \delta_{ln} \right)  \\
&
 +   { \frac{O_{1}}{\bar \varphi'{}^3}} \,
\pi_0'^3  -\frac{O_{2}}{3\;\bar \varphi'{}^2} \,  
    \pi_0'^2
\,\Delta \pi_L \nb
 -\frac{O_{3}}{3\;\bar \varphi'{}^2}\,  \pi_0' \,  \de_i
\pi_L' \de_i \pi_0 - \frac{R} { \bar \varphi'{}^3}\,\pi_0' \,
\de_i  \pi_0 \,  \de_i \pi_0 \Big \}\, .
\end{split}
\label{LSSS}
\ee
According to our classification scheme, for instance the vertex $O_{2}$ (orange) gives a larger contribution to the $f_{\text{NL}}$ than $G_{1}$ (green) but smaller than $R_{1}$ (red). The above operators are not all independent but can be related by total derivatives. For the violet operators we have the following spatial total derivative
\bea
 \det[\partial_{ij}\pi_{L}]=\frac{1}{6}\,(\Delta\pi_{L})^3-
  3\,\Delta\pi_{L} \,(\partial_{ij}\pi_{L})^2+
  2\,
\partial_{ij}\pi_{L}
\partial_{j\k}\pi_{L}
\partial_{\k i}\pi_{L},
\ea
so that the $V_i$ form factors will appear in the following combinations $V_2-2\,V_1$ and $V_3+3\,V_1$. 
For blue operators we get the following total derivatives
\be
\begin{split}
&\frac{1}{2}\,\partial_t\left(\pi_0\,((\Delta\pi_{L})^2-(\partial_{ij}\pi_{L})^2\right)+
\partial_{i}\,\left(\pi_0\,(-\Delta \pi_{L}\,\partial_{ i}\pi_{L}'+\partial_{ij}\pi_{L}\,\partial_{j} \pi_{L}')\right)
=\\
&\frac{1}{2}\,\pi_0'\,
\left((\Delta\pi_{L})^2-(\partial_{ij}\pi_{L}^2)\right) +
\partial_{i} \pi_0\,\left(-\Delta \pi_{L}\,\partial_{ i }\pi_{L}'+\partial_{ij}\pi_{L}\,\partial_{j} \pi_{L}'\right)
\label{totd}
\end{split}
\ee
that implies the presence in the observables of the following independent combinations $B_1+B_2$, $B_4-B_1/2$  and $B_5+B_1/2$. 
Cubic interactions among gravitons arise both from the Lagrangian $U$ and the Einstein-Hilbert (EH) action. We do not consider the TTT, TTS, and TSS interactions coming from EH action (see for instance \cite{Maldacena:2002vr}),  being subdominant with respect to the ones coming from $U$ during inflation.
Finally, the cubic TSS Lagrangian with a single graviton has the form
\bea
\label{TSS_lag}\nonumber
{\cal L}^{(TSS)} &=&
 \frac{1}{9}\,\plm^2 \, H^2 \,\epsilon \, a^4 \, h_{ij} \,
\left[ \left( V_{1}^{(tss)}\, \partial_{ij}\pi_L \, \Delta \pi_L+V_{2}^{(tss)}\, \partial_{il}\pi_L\, \partial_{jl}\pi_L+V_{3}^{(tss)}\,\partial_i \pi_L'\,\partial_j \pi_L'\right)
\right. \\
&& \left.-\frac{3}{ \bar\varphi'}\, \left( B_{1}^{(tss)}\, \partial_{i} \pi_0  \, \partial_j \pi_L'+B_{2}^{(tss)}\, \partial_{ij}\pi_L\, \pi_0'\right)
-\frac{9\, G^{(tss)}}{\bar\varphi'^{2}} \,   \partial_i  \pi_0 \, \partial_j  \pi_0 \right]\,.
\ea
The  TTS cubic Lagrangian with two gravitons reads
\be\label{LTTS}
\begin{split}
{\cal L}^{(TTS)}  &= \frac{1}{3}\;\plm^2 \, H^2\, \epsilon \,  a^4\;h_{ij} \,h_{mn} \,
\Big[ 
\delta_{im}\,\delta_{jn}\;
\left( B^{(tts)}\;\frac{\pi_0'  }{  \bar \varphi'} -   V_1^{(tts)}  \,\Delta \pi_L\right) 
-\;\delta_{in}\; V_2^{(tts)} \, \de_j \de_m
\pi_L \Big]\, .
\end{split}
\ee
The cubic interacting Lagrangian for gravitons  is given by
\be
{\cal L}^{(TTT)} 
 = \frac{\plm^2}{6}  H^2 \, \epsilon \, a^4  \; V_{T} \; h_{ij}
h_{jm} h_{mi}  \, .
\label{TTTc}
\ee
In Appendix \ref{C_action}, all vertices $X=V, B, G, O, R$ are given as functions of the derivatives of the $U$ Lagrangian. For the benefit of the reader interested in the details of the computation of PNG, the cubic Lagrangian is also written in Fourier space in terms of $\zeta_n$ and ${\cal R}_{\pi_0}$. By using (\ref{tbkg}) one has that $\bar{\varphi}'= a^{1-3\, c_b^2}$ and the vertices generically denoted by $X$  can be split into a time-independent part  $X_0$ and a part proportional to $ \hat\sigma$ whose time dependence is dictated by $c_b^2$ according to
 \be 
 \label{constant_structure}
 X= X_0+X_\sigma  
 \, \hat\sigma \, a^{-3\,(1+c_b^2)}\, 
 \ee
where $X_\sigma$ is an  $X$-dependent constant. 
The case $c_b^2=-1$ is special, being the full vertex
time-independent. Actually, all the terms proportional to $\hat\sigma$ in the cubic action reconstruct 
total spatial and time derivative terms, when the last relation in (\ref{dep_par}) is taken into account. Thus, effectively we can forget about $\hat\sigma$ as soon as we are interested in the bispectrum of our adiabatic-entropic modes, see Appendix \ref{tot_der}.

\section{Effective Theory Description}
\label{effect}

To compute PNG, we need to expand the action at least at the cubic order in the perturbations. Because of the action's shift symmetry, each $\pi^A$ field contains a least one derivative.  
Focusing on the scalar sector, schematically an interaction term with $n$ scalar fields~\footnote{Here we do not distinguish between $\pi_0$ and $k\,\pi_L$.} will be of the form
\be 
{\cal L}_n \sim   \plm^2 \, U_n  \sim M_{\text{pl}}^2\,H^2\,\lambda_n\, \left(\partial \pi \right)^n \,,
\ee
where the $n$-th derivative of $U$ with respect of its
arguments  $U_n$ is assumed to be proportional to $ U\sim  \, H^2$  times a dimensionless parameter $\lambda_n$;  typically  $\lambda_n$ will contain a combination of slow-roll parameters. The kinetic matrix of the $\pi$ fields has the general structure given in (\ref{slowact}) 
\begin{equation*}
{\cal K} \sim   \epsilon \,\, M_{\text{pl}}^2\,\k_i\; U \,\sim  \epsilon \,\, M_{\text{pl}}^2\,\k_i\; H^2 , 
\end{equation*}
where $\k_{i=0}\propto c_0^2$ for the $\pi_0$ field and
$\k_{i=L}\propto 1+c_1^2$ for $\pi_L$. By introducing the canonical normalized field $\Pi= (\epsilon\,\, M_{\text{pl}}^2\,\k_i\, U \,  )^{1/2} \, \pi $, we get
\be
{\cal L}_n \sim 
\left(M_{\text{pl}}\, H\right)^{2-n}\, \lambda_n \, (\k_i\,\epsilon)^{-\frac{n}{2}}\;(\partial\Pi)^n \, \equiv \, \frac{\left(\partial\Pi\right)^n}{\;\;\;\;\Lambda_n^{2\,(n-2)}}.
\ee
Consequently, we get for $\Lambda_n$
\be 
\Lambda_n=
\lambda_n{}^{-\frac{1}{2\,(n-2)}}\,\left(M_{\text{pl}}\, H\right)^{\frac{1}{2}}\, (\k_i\,\epsilon)^{\frac{n}{4\,(n-2)}}\,. 
\ee
Depending on the value of $\lambda_n$ there are at least two options: {\it democratic}
and {\it special}. In the democratic case $\lambda_n \sim  \lambda \, \epsilon \ll 1$ and all the derivatives of $U$ are of order $\epsilon$ while in the  special case $\lambda_n \sim \lambda \sim 
O(1)$. \\  
In both cases, the lowest cutoff $\Lambda$ is obtained for  $n=3$
\be 
\label{str_c}
\Lambda=\lambda_3^{-\frac{1}{2}}\,\left(M_{\text{pl}}\, H\right)^{\frac{1}{2}}\, (\k_i\,\epsilon)^{\frac{3}{4}} =
\begin{cases}
 \lambda^{-\frac{1}{2}}\,\left(M_{\text{pl}}\,
   H\right)^{\frac{1}{2}}\, (\k_i\,\epsilon)^{\frac{3}{4}} &
 \text{special} \\
\lambda ^{-\frac{1}{2}} \, \left(M_{\text{pl}}\,
  H\right)^{\frac{1}{2}}\, (\k_i^3\,\epsilon)^{\frac{1}{4}} &
\text{democratic} \\
\end{cases} \, .
\ee
As expected $\Lambda$ is proportional to the cutoff scale $\Lambda_2 =\sqrt{\plm\, m}$ of Lorentz breaking massive gravity in flat spacetime~\cite{Rubakov:2004eb,Dubovsky:2004sg,Rubakov:2008nh,Comelli:2013txa,Pilo:2017fyg,Celoria:2017hfd} with the graviton mass $m\sim H$.\\
Note that in the democratic approach, the cutoff scale is higher. In solid inflation~\cite{Endlich:2012pz} it was used the special approach and  only a unique combination of derivatives of $U$ is required to be small by consistency with  slow-roll, namely
\be
\epsilon = \frac{a^2 \varphi ' \left(U_\chi+U_y\right)-U_b}{4 \, a \, 
  \mathcal{H}^2} \ll 1\, .
\ee
In this paper we assume the democratic approach. By using (\ref{Psingle}) and (\ref{PS_zeta_n}), in order to avoid strong coupling during the inflationary period, we require that $\Lambda \gg H$ which gives  
\be
 \k_i \gg \,{\cal P}^{1/3} \,   
   \lambda^{\frac{2}{3}}=
  10^{-3}\;\hat c_{L}^{5/3}\;\lambda ^{\frac{2}{3}} \,.
\ee
Take the more stringent case $c_b^2=-1$.
In this case $\k_{0}\sim c_0^2$ and  $\k_L \sim 1+c_1^2 $; in the limit $c_{s2} \ll 1$ from (\ref{expdia}) we have  $c_0^2\sim c_{s2}^3$, $c_1^2 \sim c_{s2}^{-5} $. Thus, the most dangerous operator is a cubic interaction with three $\pi_0$; the absence of strong coupling implies
\be
\begin{split}
  c_{s2}  & \gg  
  10^{-1}\;\hat
  c_{L}^{5/9} \, {  \lambda^{\frac{2}{9}}}  \\
  & \gg 0.01 \; \lambda^{\frac{2}{9}} \,, \;\;\; \hat{c}_L= \frac{1}{2}\,,
\end{split}
\ee
that leaves enough space to explore the region of interest in the parameter space with a sufficiently small $c_{s2}$ sound speed. 

\section{Scalar Bispectrum}
\label{NG_study}
Once the cubic Lagrangian is provided, primordial non-gaussianity can be computed evaluating the various 3-point functions by using the in-in formalism. The computation is rather straightforward, so we include some of the details in Appendix \ref{in-in}.\\
We will focus on the $c_b^2=\,-1$ case, probably the most interesting case from a phenomenological point of view; the reason for that is two-folded. The relevant scalar fields $\zeta_n$ and ${\cal  R}_{\pi_0}$ have an almost flat power spectrum when $-1 \leq c_b^2\leq 0$ and for this particular case one can work out an analytic expression for the modes defined in (\ref{modes}) that are essential for the computation of the time integrals entering in the 3-point functions.
The Fourier transform of a generic 3-point function depends on the 3-momenta $\vec{k}_1$, $\vec{k}_2$ and $\vec{k}_3$ satisfying $\vec{k}_1 +\vec{k}_2 + \vec{k}_3 =0$. It is convenient to define the dimensionless parameter $f_{\text{NL}}$-local by~\footnote{The traditional choice of the factor $6/5$ comes from the following parametrization (local ansatz):\\
  $\zeta=\zeta_g+\frac{3}{5} f_{\text{NL}} \,\left(\zeta_g^2- \langle
    \zeta_g^2 \rangle \right)$ where $\zeta_g$ is the  gaussian part of the $\zeta$-field. As well known other shapes and related $f_{\text{NL}}$ can be defined, as the equilateral, orthogonal and folded ones.} 
\be
\begin{split}
& \left<\zeta_n(k_1)\,\zeta_n(k_2)\,\zeta_n(k_3)\right> \ \equiv \ (2\,\pi)^3\, \delta^{(3)}(\textbf{k}_1+\textbf{k}_2+\textbf{k}_3)\,{\cal B}(k_1,\,k_2,\,k_3) \,,\\
&{\rm with}\nonumber
\\
& {\cal B}(k_1,\,k_2,\,k_3) \equiv\frac{6}{5} \, f_{\text{NL}}\,
 \left[P_{\zeta_n}(k_1)\, P_{\zeta_n}(k_2)+P_{\zeta_n}(k_2)\,
 P_{\zeta_n}(k_3)+P_{\zeta_n}(k_3)\, P_{\zeta_n}(k_1)\right] \, .
\end{split}
\ee
In the computation of ${\cal B}$ we have retained the leading order
term in the slow-roll expansion as in~\cite{Endlich:2012pz}. However, 
there are cases~\cite{Akhshik_2015} where slow-roll corrections ($\pi_L$, $\pi_0$, lapse and shift next to leading terms) of the order
%
\begin{equation*}
\epsilon\,
\, N_{k_i}\,, \qquad \qquad N_{k_i}=-\log(-k_i \, c_{si}\,
t_e)\,, \qquad \qquad  i=1,2,3 \,,
\end{equation*} 
can give sizeable corrections,
where $N_{k_i}$ is the number of e-folds at the time $t_i=(c_{si}\,k_i)^{-1}$ of the sound horizon crossing of the $i$-th momentum, while $t_e$ is the end time of inflation. In this work, for each momentum, we consider $\epsilon$ sufficiently small to neglect those corrections. 

\subsection{Squeezed Configurations}
To avoid excessively long expressions, the main focus will be on the squeezed configurations, the most constrained one by observations, see for instance~\cite{Akrami:2019izv}. In a squeezed configuration we have $|\vec{k}_3| =k_L \ll |\vec{k}_1|=k_S$, and $\theta$ is angle between $\vec{k}_1$ and $\vec{k}_3$ then $|\vec{k}_2|= k_S+ k_L \, \cos \theta+ \cdots$.  
The bispectrum of $\zeta_n$ in the squeezed limit has the following
form
\be 
\label{viol_sq}
{\cal B}^{(SQ)} \equiv \frac{12}{5}\,f_{\text{NL}}^{(SQ)}\, P_{\zeta_n}(k_L)\,P_{\zeta_n}(k_S) \, ;
\ee
where $f_{\text{NL}}^{(SQ)}$ can be split into a monopole  ${f}_{\cal M }$ and a quadrupole  ${f}_{\cal Q }$ component
\be
f_{\text{NL}}^{(SQ)}=
{f}_{\cal M } +  {f}_{\cal Q }\,Y^0_2 
\, , 
\ee 
with
\begin{equation*}
 Y^0_2=-\frac{1}{4}\sqrt{\frac{5}{\pi}}(1-3\,\cos^2\theta) \, .
\end{equation*}
In our case the monopole term $ f_{\cal M } $ is not proportional to $n_s-1\sim O(\epsilon)$ as in single clock inflation, and actually $f_{\cal M }$ is in general of order $\epsilon^0$, if we consider the democratic approach.  Also, the quadrupole term $f_{\cal Q }$ is generally present and of order $\epsilon^0$.\\
As expected, the Maldacena consistency relation is explicitly violated being the symmetry breaking pattern different from the one of single-field inflation. 
The structure of different 3-point functions in a squeezed configuration is similar as discussed in Appendix \ref{f_nl_sq}.

The structure of a 3-point function in the squeezed limit of fields almost conserved on superhorizon scales and with a scale-free spectrum can be studied in general by extending the procedure exposed in \cite{Endlich:2013jia}. Such fields at large scales can be considered as classical stochastic variables. 
The long ``squeezed'' modes can be incorporated in the background modifying linear dynamics of the short modes. In other words, introducing a long-short $(k_L-k_S)$ modes splitting of the relevant fields, the effect of the squeezed components is to generate sort of {\it background fields} whose effect is to give rise to a non-linear and local correction of the 2-point correlation function. \\
In our case, in the scalar sector, the cubic Lagrangian is written in terms of $\zeta_n$ and ${\cal R}_{\pi_0}$ fields. The squeezing in the $j$-th vertex will be equivalent to
\begin{equation*}
{\cal L}_{k\, k'\,k''}^{(3)\,j} \to {\cal L}_{\textbf{k}_L\, \textbf{k}_s\,-(\textbf{k}_s+\textbf{k}_L)}^{(3)\,j} \approx \zeta_{n\,\textbf{k}_L}\left(/{\cal R}_{\pi_0\,\textbf{k}_L}\right)\,{\cal L}^{(2)\,j,\,\text{eff.}}_{\textbf{k}_s\,-(\textbf{k}_s+\textbf{k}_L)}
\end{equation*}     
which gives a contribution ${\cal L}^{(2)\,j,\,\text{eff.}}_{\textbf{k}_s\,-(\textbf{k}_s+\textbf{k}_L)}$ to the effective quadratic Lagrangian (a convolution over $k_L$ is understood). In the case of single-field inflation, there is only one scalar mode given by the comoving curvature ${\cal R}$ and the effective quadratic action coincides with the quadratic Lagrangian up to a rescaling of the
spatial coordinates~\cite{Creminelli:2011rh} which is a residual symmetry of single-clock inflation in the limit $k_L \to 0$
\begin{equation*}
{\cal L}^{(3)} \to {\cal L}^{(2)\,\text{eff.}}={\cal L}^{(2)}(\tilde x ,\, t)+ O\left(\partial_i {\cal R}_L\right) \,, \qquad \tilde x^i= (1+{\cal R}_{L})\,x^i\,.
\end{equation*}     
where ${\cal R}_L(x)$ is the long part of the comoving curvature. \\
The equivalence between ${\cal L}^{(2)\,\text{eff.}}$ and the rescaled quadratic action is the core of the Maldacena consistency relations~\cite{Maldacena:2002vr}, any deviation leads to a violation of such relations. By looking at the following ratio
\begin{equation*}
{\cal Q}^j=\frac{{\cal L}^{(2)\,j,\,\text{eff.}}_{\textbf{k}_s\,-(\textbf{k}_s+\textbf{k}_L)}}{{\cal L}^{(2)\,j}_{\textbf{k}_s\,-(\textbf{k}_s+\textbf{k}_L)}} \,.
\end{equation*}
we can predict the nature of the consistency violations. \\
Namely, an eventual time-dependence of ${\cal Q}^j$ can be related to the scale dependence of $f_{\text{NL}}$ in the squeezed limit producing possibly dangerous interactions. This is never the case for $c_b^2=-1$. \\ 
Still, the residual dependence on momenta directions, also related to the vertices' derivative structure, can give primordial anisotropies that can be very interesting from a phenomenological point of view. In general, $f_{\text{NL}}$ will be not suppressed by slow roll parameters and, as discussed above, with several features rather different from single-clock inflation.\\
Let us now focus on the region of the parameter space where $\hat{c}_L,\, c_{s1} \le 1$, while $c_{s2}\ll 1$. As shown in \cite{Celoria:2020diz}, in such a region, the secondary production of gravitational waves can significantly boost the tensor power spectrum enough to enter the sensitivity region of LISA. From (\ref{r_par}) and (\ref{expdia}), we have the following expansion 
\be \label{scs2}
\begin{split}
& r_{L/0}^{(1)}= \frac{\hat{c}_L^{5}}{ c_{s1}^5}+ O(c_{s2})\,, \qquad r_{L/0}^{(2)}=\frac{\hat{c}_L^{5}}{ c_{s1}^2}\, \frac{1}{c_{s2}^3} +O(c_{s2}^{-2})\,,\\
& r_{\zeta_n}^{(1)}= \frac{\hat{c}_L^{5}}{ c_{s1}^5}+ O(c_{s2})\,, \qquad r_{\zeta_n}^{(2)}=1-\frac{\hat{c}_L^{5}}{ c_{s1}^5} +O(c_{s2})\,;\\
& c_0^2 =-\frac{1}{2\, c_{s1}}\,
\left(1-\frac{c_{s1}^5}{\hat{c}_L^{5}}\,
\right)\,c_{s2}^3+O(c_{s2}^5)\,; \\
& c_1^2=\frac{1}{c_{s1}}\,
\left(1-\frac{c_{s1}^5}{\hat{c}_L^{5}}\,
\right)\,c_{s2}^5+O(c_{s2}^7)\,; \\
& c_2^2=\frac{3}{4}(1+c_{s1}^2)+{\cal O}(c_{s2}^5) \, ;
\end{split}
\ee
we have set $c_b^2=-1$, and $\hat{c}_L^{5}$ has
been defined in (\ref{PS_zeta_n}). By using the above
expressions, one can compute $f_{\text{NL}}$ for the bispectrum of $\zeta_n$; the result is given in table \ref{tab2}.\\ 
\vspace{1cm}
{\renewcommand\arraystretch{2.1}
\begin{table}[!ht]
\caption[Bispectrum of $\zeta_n$:
  $f_{\text{NL}}$ for a squeezed configuration and $c_{s2}\ll1$]{
Bispectrum of $\zeta_n$:
  $f_{\text{NL}}$ for a squeezed configuration and $c_{s2}\ll1$}
\centering
\begin{tabular}{|l|l|l|}
\hline
\rowcolor[HTML]{C798FA} 
$O\left(c_{s2}^0\right)$ &  \multicolumn{2}{l|}{$\zeta_n \, \zeta_n'^{2}$ : $-\frac{5}{144}\,\frac{\hat{c}_L^5}{c_{s1}^5}\, \left(V_{4}+V_{5} \, \cos(\theta)^2\right)$ }   \\ \hline 
\rowcolor[HTML]{C798FA} 
$O\left(c_{s2}^0\right)$ & \multicolumn{2}{l|}{ $\zeta_n^3$ : 
{\small $
\frac{\beta}{2}\,
\left[(V_{3}+3 \, V_{1} )+(3\, V_{2}+2\, V_{3})\,\cos(\theta)^2\right]$}  }   \\ \hline 
\rowcolor[HTML]{95AEF8} 
$O\left(c_{s2}^{-3}\right)$ & \multicolumn{2}{l|}{$\zeta_n'\,{\cal R}_{\pi_0}\,\zeta_n$ : $\frac{5}{72}\,\frac{\hat{c}_L^5}{c_{s1}^4} \,\left(B_{1}+B_{2} \, \cos(\theta)^2 \right)\, \frac{1}{c_{s2}^3}$ }                      \\ \hline
\rowcolor[HTML]{95AEF8} 
$O\left(c_{s2}^{-3}\right)$ &  $\tilde{\cal R}_{\pi_0}'\,\zeta_n'^2$ : $-\frac{5}{48}\,\frac{\hat{c}_L^{10}}{c_{s1}^7}\, B_{3}\, \frac{1}{c_{s2}^3}$     & $\tilde{\cal R}_{\pi_0}'\,\zeta_n^2$ : $\frac{5}{36}\frac{\hat{c}_L^5}{c_{s1}^4}\left(B_{\text{mix}}+B_{5}\, \cos\left(\theta\right)^2\right)\,\frac{1}{c_{s2}^3}$     \\ \hline
\rowcolor[HTML]{B4FDCD} 
$O\left(c_{s2}^{-5}\right)$ & \multicolumn{2}{l|}{$\zeta_n\,{\cal R}_{\pi_0}^2$                         : $-\frac{5}{16}\,\frac{\hat{c}_L^5}{c_{s1}^4}\, \left(G_{1}+G_{2} \, \cos(\theta)^2\right)\, \frac{1}{c_{s2}^5}$ }           \\ \hline
\rowcolor[HTML]{FAE198} 
$O\left(c_{s2}^{-6}\right)$ & $\tilde{\cal R}_{\pi_0}'\,{\cal R}_{\pi_0}\,\zeta_n'$ : $\frac{5}{24}\,\frac{\hat{c}_L^{10}}{c_{s1}^{6}} \,O_{3}\, \frac{1}{c_{s2}^6}$ 
& $\tilde{\cal R}_{\pi_0}'{}^2\,\zeta_n$ : $\frac{5}{12}\, \frac{\hat{c}_L^{10}}{c_{s1}^{6}} \,O_{2}\, \frac{1}{c_{s2}^6}$  \\ \hline
\rowcolor[HTML]{FAE198} 
$O\left(c_{s2}^{-6}\right)$ & \multicolumn{2}{l|}{ $\tilde{\cal R}_{\pi_0}'{}^3$ : $\frac{135}{16}\,\frac{\hat{c}_L^{10}}{c_{s1}^{6}} \,O_{1}\, \frac{1}{c_{s2}^6}$} \\ \hline
\rowcolor[HTML]{FAA1A1} 
$O\left(c_{s2}^{-8}\right)$ & \multicolumn{2}{l|}{$\tilde{\cal R}_{\pi_0}'\,{\cal R}_{\pi_0}{}^2$ : 
$-\frac{15}{16}\, \frac{\hat{c}_L^{10}}{c_{s1}^{6}} \,R\, \frac{1}{c_{s2}^8}$ }    \\ \hline
\end{tabular}
\label{tab2}
\end{table}
}
\\
We have introduced
\be
\beta= \frac{5}{216\,
  c_{s1}^2}\left(\frac{\hat{c}_L^{5}}{c_{s1}^5}+4\right) \, , \qquad 
B_{\text{mix}}=2\, B_{4}+B_{5}+\left(
  B_{4}+B_{5}\right)\,\frac{\hat{c}_L^5}{4\, c_{s1}^5} \, .
\ee
In each line of the table, we show the leading contribution in powers of $c_{s2}$ to $f_{\text{NL}}$ for the bispectrum of $\zeta_n$. The values of the vertices of the cubic Lagrangian are given in Appendix \ref{C_action}. The total value for $f_{\text{NL}}$ is obtained by summing up the various contributions.
Let us comment  the results in the table. Taking for instance $f_{\text{NL}}$ of the bispectrum of $\zeta_n$ and set $c_b^2=-1$. For what concerns the wave functions of $\zeta_n, {\cal R}_{\pi_0}$ and their time derivative, 
we can deduce the following leading order behaviour in the small parameter $c_{s2}$ 
\bea
&&\zeta_n,\,\zeta_n'\sim {\rm const},\quad {\cal R}_{\pi_0}\sim\frac{1}{c_{s2}^{3}},\qquad 
{\cal R}_{\pi_0}'\sim\frac{1}{c_{s2}^{2}}\,, \qquad
 \tilde{\cal R}'_{\pi_0} \sim(...)\frac{1}{c_{s2}^{2}}+
(...)\frac{1}{c_{s2}^{3}} \, . \label{as2}
\ea
Note that for the $\tilde{\cal R}_{\pi_0}'$ field  (as defined in (\ref{redef})) there are two competing contributions.
Given that the cross-correlation $r_{L/0}^{(2)}
\sim c_{s2}^{-3}$  (\ref{scs2}), the dominant contribution to $f_{\text{NL}}$ is given by the vertices with the greatest number of ${\cal R}_{\pi_0}$. A number of features follows.
\begin{enumerate}
\item
The contribution to $f_{\text{NL}}$ of {\it \color{violet} violet} vertices, typical of solid inflation, is little sensitive to $c_{s2}$;
indeed, in those  ${\cal R}_{\pi_0}$ is absent.
\item {\it\color{blue} Blue} vertices have a single ${\cal R}_{\pi_0}$ and contribute to $f_{\text{NL}}$ via the cross correlation term $r_{L/0}^{(2)} \sim c_{s2}^{-3}$. 
\item   {\it \color{green} Green} vertices and {\it \color{orange} orange} vertices, with at
  least two ${\cal R}_{\pi_0}$, contribute with a cross correlation of the form $r_{L/0}^{(2)}{}^2 \sim c_{s2}^{-6}$. 
\item
 A  vertex $\tilde{\cal R}_{\pi_0}'{}^3$, even if it has three $\tilde{\cal R}_{\pi_0}$ contributes as $ r_{L/0}^{(2)}{}^3\, c_{s\,2}{}^3\;\sim c_{s2}^{-6}$;
\item
  Similarly, the {\it \color{red} red} vertex can be rewritten as $\tilde {\cal R}_{\pi_0}' \, {\cal R}_{\pi_0}^2$ and  contributes as $r_{L/0}^{(2)}{}^3\, c_{s\,2}\;\sim \; c_{s\,2}^{-8}$ and is the dominant one.
\end{enumerate}
Clearly, in the deep $c_{s2}\ll 1$ region, the contributions from red, orange and green vertices tend to be very large compared with the experimental bounds. We come back to that in section \ref{phen}.\\
In Appendix \ref{f_nl_sq} the reader can find total squeezed expressions valid for any value of $c_{s2}$, where the contribution of each vertex ${\cal O}$ is factorized in an angular part and in an angle-independent part $ {\cal M}^{({\cal O})}$. 
Plots of the size of the above contributions are given in Appendix \ref{pics}. 

\subsection{Equilateral Configurations}
Proceeding as for the squeezed configuration, the expressions for $f_{\text{NL}}$ in the equilateral configuration read
\begin{equation*}
f_{\text{NL}}^{(EQ)}= \frac{5}{18} \, \frac{{\cal B}_{\zeta_n}}{P_{\zeta_n}(k)^2}\,,\qquad k_l=k\,,\; l=1,2,3\,.
\end{equation*}
Results in the case $c_{s2} \ll 1$ are given in table \ref{tab3}, where $\delta$ and $\gamma$ parameters are defined by 
\be
\delta=-\frac{5}{2592}\,\frac{\hat{c}_{L}^{10}}{c_{s1}^{12}}\,\left[243
  \,\frac{c_{s1}^5}{\hat{c}_L^5} \,\left(2\,
    \frac{c_{s1}^5}{\hat{c}_L^5}+1\right)-23\right] \, , \qquad
\gamma=-\frac{5 \, \left(243\,
    \frac{c_{s1}^5}{\hat{c}_L^5}+29\right)}{1296}\,
\frac{\hat{c}_L^{10}}{c_{s1}^{10}} \, .
\ee

{\renewcommand\arraystretch{2.1}
\begin{table}[!ht]
\centering
\begin{tabular}{|l|l|l|l|}
\rowcolor[HTML]{C798FA} 
$O\left(c_{s2}^0\right)$ & $\zeta_n^3$ : $\frac{\delta}{54}\,\left(-V_{2}+2\, V_{3} +8\,V_{1}\right)$   & \multicolumn{2}{l|}{$\zeta_n \, \zeta_n'^{2}$ : $\frac{\gamma}{54}\,\left(2\, V_{4}-V_{5}\right)$ }   \\ \hline
\rowcolor[HTML]{95AEF8} 
$O\left(c_{s2}^{-3}\right)$ & \multicolumn{3}{l|}{ $\zeta_n'\,{\cal R}_{\pi_0}\,\zeta_n$ : $\frac{5 \, \left(4 \,c_{s1}^5+\hat{c}_L^5\right)\,\hat{c}_L^{5}}{64\,c_{s1}^9}\,\frac{\left(2 B_{1}-B_{2}\right)}{9\,c_{s2}^{3}}$}    \\ \hline
\rowcolor[HTML]{95AEF8} 
$O\left(c_{s2}^{-3}\right)$ & $\tilde{\cal R}_{\pi_0}'\,\zeta_n'^2$ : $-\frac{5\, \hat{c}_L^{10}\,B_{3} }{48 \, c_{s1}^7 }\, c_{s2}^{-3}$ &  \multicolumn{2}{l|}{ $\tilde{\cal R}_{\pi_0}'\,\zeta_n^2$ : $\frac{5\,\left(4\,c_{\text{s1}}^5+\hat{c}_L^5\right)\,\hat{c}_L^5}{32 \,c_{\text{s1}}^{9}}\,\frac{\left(4\,B_{4}+\, B_{5}\right)}{9\,c_{s2}^{3}} $  }     \\ \hline
\rowcolor[HTML]{B4FDCD} 
$O\left(c_{s2}^{-6}\right)$ & \multicolumn{3}{l|}{$\zeta_n\,{\cal R}_{\pi_0}^2$                         :  $\frac{5 \,\hat{c}_L^{10}\,\left(G_{2}-2\, G_{1}\right)}{24\, c_{\text{s1}}^8 }\,\frac{1}{c_{\text{s2}}^{6}}$ }     \\ \hline
\rowcolor[HTML]{FAE198} 
$O\left(c_{s2}^{-6}\right)$ & $\tilde{\cal R}_{\pi_0}'\,{\cal R}_{\pi_0}\,\zeta_n'$ :  $\frac{5}{24}\, O_{3}\,\frac{\hat{c}_L^{10}}{c_{s1}^6}\, \frac{1}{c_{s2}^{6}}$ & \multicolumn{2}{l|}{ $\tilde{\cal R}_{\pi_0}'{}^2\,\zeta_n$ : $\frac{5}{12}\, O_{2}\,\frac{\hat{c}_L^{10}}{c_{s1}^6}\, \frac{1}{c_{s2}^{6}}$}\\ \hline
\rowcolor[HTML]{FAE198} 
$O\left(c_{s2}^{-6}\right)$ & \multicolumn{3}{l|}{ $\tilde{\cal R}_{\pi_0}'{}^3$ : $\frac{1490}{162}\, O_{1}\,\frac{\hat{c}_L^{10}}{c_{s1}^6}\, \frac{1}{c_{s2}^{6}}$ } \\ \hline
\rowcolor[HTML]{FAA1A1} 
$O\left(c_{s2}^{-8}\right)$ & \multicolumn{3}{l|}{$\tilde{\cal R}_{\pi_0}'\,{\cal R}_{\pi_0}{}^2$ :   $-\frac{170}{162}\,\frac{\hat{c}_L^{10}}{c_{s1}^6}\,R\, \frac{1}{c_{s2}^8}$}\\ \hline
\end{tabular}
\caption[Bispectrum of ${\zeta_n}$:
  $f_{\text{NL}}$ for an equilateral configuration and $c_{s2}\ll1$]{ Bispectrum of ${\zeta_n}$:
  $f_{\text{NL}}$ for an equilateral configuration and $c_{s2}\ll1$}
\label{tab3}
\end{table}
}
\vspace{1cm}
Note that the green interaction $\zeta_n\,{\cal R}_{\pi_0}^2$, in the equilateral configuration assumes an orange behavior being the $f_{\text{NL}}$ proportional to $c_{s2}^{-6}$. The same holds for the folded shape 
\begin{equation*}
k_1=2\, k_2=2\, k_3\,,
\end{equation*}
whose results are very similar and we do not report for brevity. In section \ref{phen} we check that experimental constraints on both equilateral and folded shapes can be satisfied.

\section{Tensor Bispectrum }
\label{ten} 
Let us now study 3-point function containing at least a tensor. We do not consider vertices coming from the Einstein-Hilbert of the action as discussed in section \ref{cubic}. Before proceeding let us recall the expression for the two-point tensor correlator
\begin{equation*}
\left\langle h^s_{\textbf{k}}\,h^p_{\textbf{p}}\right\rangle=(2\,\pi)^3 \, P_h(k)\, \delta_{ps}\, \delta(\textbf{k}+\textbf{p})\,, \qquad P_h(k)= (2\,\pi^2)\, \frac{{\cal P}_h(k)}{k^3}\,,
\end{equation*} 
with ${\cal P}_h$   the gaussian scale invariant tensor PS  whose expression is 
\be
{\cal P}_{h}^{(1)}=
\frac{H^2}{4\, \plm^2 \,\pi^2}\, \left(\frac{k}{H}\right)^{2 \, c_L^2\, \epsilon}\, (-H \, t)^{\frac{8}{3}\,c_2^2\,\epsilon}\, 
\simeq 4\,\epsilon \, {\cal P}. 
\ee
As shown in \cite{Celoria:2020diz}, when $c_{s2} \ll c_{s1}$, the secondary GWs production tends to overwhelm the primary one in the small $c_{s2}$ limit. 
The same mechanism gives rise to the TTT and TTS one-loop correction (work in progress). This last one will be naturally dominant in the same limit where the secondary PS dominates. 
In this work, we limit our analysis to the three-level contribution, whose expressions in the TTS and TSS cases can leave sizeable and very interesting signatures.
In the following sections we give explicit expressions for three tensor correlators
\be
\label{TTT_cor}
 \langle h_{k_1}^p\, h_{k_2}^q\, h_{k_3}^r\rangle = (2\, \pi)^3 \, {\cal B}_{TTT}^{p,q,r}\, \delta(\textbf{k}_1+\textbf{k}_2+\textbf{k}_3)\,;
\ee
two tensors and one scalars
\be 
\label{TTS_cor}
\langle \zeta_{n\,k_1}\, h_{k_2}^r\, h_{k_3}^s\rangle = (2\, \pi)^3 \, {\cal B}_{TTS}^{r,s}\, \delta(\textbf{k}_1+\textbf{k}_2+\textbf{k}_3)\,;
\ee
concluding with one tensor and two scalars 
\be 
\label{TSS_cor}
\langle \zeta_{n\,k_1}\, \zeta_{n\,k_2}\, h_{k_3}^s\rangle =(2\, \pi)^3 \, {\cal B}_{TSS}^{s}\, \delta(\textbf{k}_1+\textbf{k}_2+\textbf{k}_3)\,.
\ee
\subsection{\texorpdfstring{$\left<TTT\right>$}{Lg}}
Let us start with the 3-point function with only tensors. From the cubic Lagrangian (\ref{TTTc}), we have that the relevant vertex coupling reads
\be
\label{LTTT}
V_{T}=\frac{9}{2}\,(1+c_L^2)-2\,\lambda_7 \, .
\ee
Thus for $\langle h_{k_1}^p\, h_{k_2}^q\, h_{k_3}^r\rangle$, by using (\ref{TTT}) and (\ref{TTT_J}), we have 
\be
{\cal B}_{TTT}^{p,q,r}= 4\, \pi^4\,V_{T}\,\epsilon \,\frac{{\cal P}_h^3 }{{\cal P}}\, \frac{\varepsilon_{ij}^{p}\,\varepsilon_{li}^{q}\,\varepsilon_{lj}^{r}}{\Pi_i \, k_i^3}\, \delta_{p,q,r}\,{\cal I}_{h^3}\,,
\ee
where
\be
{\cal I}_{h^3}=\sum_{ij} k_i\, k_j^2-\Pi_i \,k_i+\sum_i \,k_i^3 \, \left[1-3 \, \gamma_e-\log\left(-k_T \, t\right)\right]\,, \qquad k_T= \sum_i k_i\,\, 
\ee
and 
\be
\delta_{p,q,r}=\frac{1}{3}\, \left(\delta_{pq}+\delta_{pr}+\delta_{rq}\right)\,. 
\ee
The corresponding $f_{\text{NL}}$ in squeezed limit is very suppressed and reads
\be
\begin{split}
f_{\text{NL}}& =\frac{10}{18\,\sqrt{2}}\,\delta_{p,q,r} \,V_{T}\, \epsilon \,\left(3 \,\gamma_e-7+3 \log(-2 \, k_s\, t)\right)\,\sin(\theta)^2
 \sim \epsilon
 \,\sin(\theta)^2 \ll 1\,.
\end{split}
\ee

\subsection{\texorpdfstring{$\left<TTS\right>$}{Lg}}
Consider the 3-point function of the form  $\langle \zeta_{n\,k_1}\,h_{k_2}^r\, h_{k_3}^s\rangle$. In the democratic approach, two cubic vertices are dominant: $h\, h \, \pi_L$ and $h\, h \, \pi_0'$; see (\ref{LTTS}). The computation of the bispectrum is done by using (\ref{TTS}) and (\ref{TTS_J}). 
\begin{itemize}
\item For the $h^2 \pi_L$ vertex we get
\be
{\cal B}_{TTS}^{(1),\,r,s}=\sum_{a=1}^2 \frac{8 \, \pi^4}{9} \,\epsilon\, \frac{{\cal P}_h\, {\cal P}_{\zeta_n}^{(a)}}{k_T\,
( k_1\,k_2\,k_3)^3}\, \varepsilon_{ik}^{r}\,\varepsilon_{lj}^{s}\, \left(\delta_{li}\,\delta_{kj}\, V^{(tts)}_{1}-\delta_{kl}\,\frac{\kappa_{1}^i\,\kappa_{1}^j}{\kappa_1^2}\, V^{(tts)}_{2}\right) \, {\cal I}_{h^2\,\pi_L}\,,
\ee
where we have rescaled the momentum of the scalar field and defined the total rescaled momentum $k_T$ 
\be
\kappa_1 =c_{s\,a}\, k_1\,,\qquad k_T=\kappa_1+k_2+k_3\,,   \qquad
a=1,2\, ;
\ee
and 
\be 
\begin{split}
{\cal I}_{h^2 \,\pi_L}=&\kappa_1^4 +\kappa_1^3 \, \sum_{i=2}^3 k_i+3\,\kappa_1^2  \,\sum_{i=2}^3 k_i^2+\kappa_1 \,\sum_{i=2}^3 k_i\,\left(7\, \sum_{i=2}^3 k_i^2-4\,\Pi_{i=2}^3 k_i\right)\\
&-\sum_{i=2}^3 k_i^2 \, \left(\Pi_{i=2}^3 k_i-4 \, \sum_{i=2}^3 k_i^2\right)-3\, k_T\, \sum_{i=2}^3 k_i^3 \left(\gamma_e+\log(-k_T\,t)\right)\,.
\end{split}
\ee
\item
In a similar way, one can compute the contribution from the vertex $h^2 \, \pi_0'$:
\be
{\cal B}^{(2),\,r,s}_{TTS}=\sum_{a=1}^2\frac{8\, \pi^4}{3}\, B^{(tts)}\,\epsilon\, \varepsilon_{ij}^{p}\,\varepsilon_{ij}^{s} \frac{ {\cal P}_h\,{\cal P}_{\zeta_n {\cal R}_{\pi_0}}{}^{(a)}}{k_T\,
( k_1\,k_2\,k_3)^3}\, {\cal I}_{h^2 \,\pi_0}\,,
\ee  
where
\bea 
{\cal I}_{h^2 \,\pi_0}&=&\kappa_1^4 +\kappa_1^3 \, \sum_{i=2}^3 k_i+3\,\kappa_1^2  \,\sum_{i=2}^3 k_i^2+3\, k_T\, \sum_{i=2}^3 k_i^3 \,\log(-k_T\,t)\\
&&\nonumber +\left(\sum_{i=2}^3 k_i\right)^2 \, \left(\Pi_{i \neq j=2}^3 k_i\,k_j (3\, \gamma_e-1)+(4-3\, \gamma_e)\, \sum_{i=2}^3 k_i^2\right)\\
&&\nonumber +\kappa_1 \,\left(3\, \Pi_{i\neq j=2}^3 k_i k_j^2+(7-3 \,\gamma_e)\,\sum_{i=2}^3 k_i^3\right)\,.
\ea
\end{itemize}
In the squeezed limit, when the scalar momentum $k_1$ (see eq. (\ref{TTS_cor})) is squeezed, the dominant contribution is the one from the $h^2 \, \pi_0'$ and has a characteristic {\it blue} operator behavior with the presence of a single scalar cross-correlation, while  the $h^2 \, \pi_L$ contribution is proportional to ${\cal P}_{\zeta_n}$; namely
\be
\begin{split}
&\left.\frac{{\cal B}_{TTS}^{(1)}}{P_{h}(k_S)\, P_{\zeta_n}(k_L)}\right |_{h^2\,\pi_L}\!\!\!\!\!\! \sim \frac{{\cal P}_{\zeta_n}}{{\cal P}} \propto c_{s2}^0\,, \qquad\qquad
\left.\frac{{\cal B}_{TTS}^{(2)}}{P_{h}(k_S)\, P_{\zeta_n}(k_L)}\right |_{h^2\,\pi_0'}\!\!\! \sim
\frac{{\cal P}_{\zeta_n {\cal R}_{\pi_0}} }{{\cal P}}
 \propto c_{s2}^{-3}\,.
\end{split}
\ee
Let us define $f_{\text{NL}}$ for the TTS bispectrum as
\be
f_{TTS}=\frac{1}{2}\, \sum_{r,s} \,\frac{{\cal
    B}_{TTS}^{r,s}(k_1,\,k_2,\,k_3)}{P_h(k_S)\,P_{\zeta_n}(k_L)}\,. 
\ee
By squeezing along the scalar direction and keeping only the dominant blue vertex, we get the following expression  
\be 
f_{TTS}=-\frac{8}{3}\,B^{(tts)}\, \left[3\, \gamma_e-7+3 \, \log(-2 \, k_S \,t)\right]\,
\epsilon\, r_{L/0}\,.
\ee
In figure \ref{Log_TTS}, we show the logarithmic plot of the dominant $\text{Abs}\left[f_{TTS}\right]$. In this extreme case, setting $ B^{(tts)}\, \epsilon$ to one, the $f_{TTS}$ can get as large as $10^{4\div 5}$ for $c_{s2}\sim 0.1$.

\subsection{\texorpdfstring{$\left<TSS\right>$}{Lg}}
Finally let us analyse the 3-point function $\langle \zeta_{n\, k_1}\,\zeta_{n\, k_2}\,h_{k_3} \rangle$.\\
As for the TTS case we can define a sort of $f_{\text{NL}}$-like parameter in order to understand what is the typical strength of the TSS Bispectrum. \\
Thus, with the $f_{TSS}$ symbol we mean 
\be
f_{TSS} \equiv \frac{{\cal B}}{(2\,\pi^2)^2 \,{\cal P}_h \, {\cal P}_{\zeta_n} }\, k_2^3\, k_3^3\,, 
\ee
where $k_3$ is the momentum related to the tensor field $h$, while $k_2$ is one of the momentum related with the scalar field $\zeta_n$ as presented in eq. (\ref{TSS_cor}). 
Consider a typical interacting TSS term
\begin{equation*}
{\cal L}_{TSS}^{i-j} \sim 
X\, \plm^2\, H^2 \, \epsilon\; \varepsilon^{s}(\vec k) \cdot \,D(k,\,k',\, k'')\,h_{ \vec k}\;\xi_{i,\, \vec k'}\,\xi_{j, \vec k''}\,; 
\end{equation*}
where the generic scalar fields $\xi_{i},\,\xi_{j}$ stand for any among $\zeta_n$, ${\cal R}_{\pi_0}$ and their time derivatives. Thus, the TSS ${i-j}$ interaction also selects the possible {\it color} related to the specific vertex\footnote{ If $i-j$ selects two ${\cal R}_{\pi_0}$ fields we may have an {\color{orange}orange} or {\color{green}green} vertex; only one ${\cal R}_{\pi_0}$ field implies a {\color{blue}blue} vertex; no ${\cal R}_{\pi_0}$ field implies a {\color{violet}violet} vertex.}. Manipulating    
 eq. (\ref{TSS_J}) one can demonstrate that such a parameter can be described by a very compact formula 
\be
\label{compact_fT}
 f_{TSS}^{i-j}= \sum_{l,m=1}^2\,X\, \frac{1}{4  \,\hat{c}_L^{5}\,k_1^3} \, 
\,r^{(l)}_{L/i}\,r^{(m)}_{L/j}\, S(k_1,\,k_2,\,k_3)\, \text{Re}\left[I^{i-j}_{l,m}\right]\,.
\ee
where
\be
r^{(l)}_{L/i}= \frac{{\cal P}_{\zeta_n-\xi_i}^{(l)} }{{{\cal P}_{\zeta_n}}}\,. 
\ee
The scale dependent factor $S$ describes the structure in polarization and spatial derivatives provided by the particular interaction  
\be
S=\varepsilon^{s}(k_1)\, D(k_1,\,k_2\,k_3)\,. 
\ee
For all our interactions, the $S$ function does not depend on the polarization $s$, thus the $f_{\text{NL}}$-parameter defined in eq. (\ref{compact_fT}) does not depend on $s$ and coincides with the average on the two polarizations.\\ 
Finally, a shape integral needs to be defined:
\be
\begin{split}
I^{i-j}_{l,m}=i\,\int_{-\infty}^{t_e} dt'\,a(t')^n\, f_{h}(-k_1\,t') \, \Big[ & f_i (-k_2\,c_{sl}\, t')\,f_j (-k_3\,c_{sm}\, t')
+(k_2\leftrightarrow k_3)
\Big]\,,
\end{split}
\ee
where the shape $f_i$ functions are given by
\be
f_i(-c_{sl}\,k\,t)=\frac{\xi_i(-c_{sl}\,k\,t)}{\text{lim}_{{}_{k\,t \to 0}}\,\xi_i(-c_{sl}\,k\,t)}\,. 
\ee
When $c_b^2=-1,\,0$, the $f$ functions have a simple Bessel-like form, in the other cases a numerical computation of $I_{l,m}$ is needed. Note that eq. (\ref{compact_fT}) is valid if and only if $S$ is symmetric under exchange of the scalar momenta, and this is always true in the squeezed, equilateral and folded shapes. In particular for the squeezed limit $k_1 \to k_T \equiv k_L$, $k_2 \to k_\zeta \equiv k_S$ and $k_3 \to k_S+ k_L \, \cos(\theta)$ 
\be 
S(k_1,\,k_2,\,k_3)-S(k_1,\,k_3,\,k_2) \propto k_L \, \cos(\theta) \to 0\,. 
\ee
We now need to compute five TSS vertices contributions: $h\, \pi_L^2$, $h\, \pi_L' \pi_0$, $h\, \pi_L'^2$ and $h\, \pi_0^2$.\\
Looking at vertices spatial derivative structure and contractions with polarization tensor, one can easily get that the folded limit always vanishes, i. e. $S(k,\frac{k}{2},\, \frac{k}{2})\equiv 0$.\\
Defining the following  2$\times$2 symmetric matrices
\be
\begin{split}
{\cal C}^{lm}_4=c_{sl}^4+c_{sl}^3\,c_{sm}+c_{sl}^2\,c_{sm}^2 &+c_{sl}\,c_{sm}^3+c_{sm}^4\,, \qquad  {\cal C}^{lm}_2=c_{sl}^2+c_{sl}\,c_{sm}+c_{sm}^2\,,\\
& {\cal C}^{lm}_1=c_{sl}+c_{sm}\,, \qquad l,m=1,2; 
\end{split}
\ee
we get  the  shapes  defined in table \ref{Squeezed_shapes}, while the equilateral formula are very similar and reported in Appendix \ref{TSS_total}. 
Note that each vertex is classified with our color-code, starting from {\it violet} (asymptotic $f_{TSS}$ of order $c_{s2}^0$) arriving to {\it green} vertices (asymptotic $f_{TSS}$ of order $c_{s2}^{-5}$). In tables \ref{Squeezed_TSS_T} and \ref{Equilateral_TSS_T} the reader can find the asymptotic $c_{s2}\ll 1$ of squeezed and equilateral $f_{TSS}$.
\vspace{2cm}
{\renewcommand\arraystretch{2.3}
{\Huge \begin{table}[htp]
\caption[$f_{TSS}$ parameters]{Squeezed shape  for $f_{TSS}$.}
\label{Squeezed_shapes}
\begin{center}
\begin{tabular}{|c||c|}
\hline
$h \,\zeta_n^2$ &  $\sum_{l,\,m=1}^2\, \left(V^{(tss)}_{1}-V^{(tss)}_{2}\right)\, \frac{1}{18\,\sqrt{2}}\, \frac{{\cal C}^{lm}_4}{{\cal C}^{lm}_1}\, \frac{1}{\hat{c}_L^{5}}\,r_{\zeta_n}^{(l)}\,r_{\zeta_n}^{(m)}\,\sin(\theta)^2$ \\\hline
 $h \,\zeta_n{}'{}^2$&  $-\sum_{l,\,m=1}^2\, V^{(tss)}_{3}\, \frac{c_{sl}^2\, c_{sm}^2}{18\,\sqrt{2}}\, \frac{{\cal C}^{lm}_2}{{\cal C}^{lm}_1}\, \frac{1}{\hat{c}_L^{5}}\,r_{\zeta_n}^{(l)}\,r_{\zeta_n}^{(m)}\,\sin(\theta)^2 $ \\\hline
 $h \,\zeta_n{}'\,{\cal R}_{\pi_0}$ &  $\sum_{l,\,m=1}^2\, B^{(tss)}_{1}\, \frac{c_{sl}^2}{6\,\sqrt{2}}\, \frac{{\cal C}^{lm}_2}{{\cal C}^{lm}_1}\, \frac{1}{\hat{c}_L^{5}}\,r_{\zeta_n}^{(l)}\,r_{L/0}^{(m)}\,\sin(\theta)^2$ \\\hline
$h\,\tilde{\cal R}_{\pi_0}'\, \zeta_n$   &  $-\sum_{l,\,m=1}^2\,  B^{(tss)}_{2}\, \frac{1}{6\,\sqrt{2}}\,\frac{{\cal C}^{lm}_4}{{\cal C}^{lm}_1}\,\frac{1}{\hat{c}_L^{5}}\, r_{L/0}^{(l)}\,r_{\zeta_n}^{(m)}\,\sin(\theta)^2$  \\\hline
   $h \,{\cal R}_{\pi_0}{}^2$ &  $-\sum_{l,\,m=1}^2\,  G^{(tss)}\, \frac{1}{2\,\sqrt{2}}\, \frac{{\cal C}^{lm}_2}{{\cal C}^{lm}_1}\, \frac{1}{\hat{c}_L^{5}}\,r_{L/0}^{(l)}\,r_{L/0}^{(m)}\,\sin(\theta)^2$ \\    \hline
\end{tabular}
\end{center}
\end{table}%
}
}
{\renewcommand\arraystretch{2.3}
\begin{table}[!ht]
\centering
\begin{tabular}{|l|l|l|l|}
\hline
\rowcolor[HTML]{C798FA} 
$O\left(c_{s2}^0\right)$ & $h\,\zeta_n^2$ : $V_{2-1}^{(tss)}\,\frac{\hat{c}_L^5}{c_{s1}^7}\,\left(1+4\, \frac{c_{s1}^5 }{\hat{c}_L^{5}}\right) \frac{\sin(\theta)^2}{36\, \sqrt{2}}$   & $h \,\zeta_n \, \zeta_n'$ : $-\frac{1}{12}\, V^{(tss)}_{3}\,\frac{\hat{c}_L^5}{c_{s1}^5}\frac{\sin(\theta)^2}{\sqrt{2}}$      \\ \hline
\rowcolor[HTML]{95AEF8} 
$O\left(c_{s2}^{-3}\right)$ & $h\,\zeta_n'\,{\cal R}_{\pi_0}$ : $B^{(tss)}_{1}\,\frac{\hat{c}_L^5}{c_{s1}^4}\frac{\sin(\theta)^2}{6\,\sqrt{2}}\, \frac{1}{c_{s2}^{3}}$                  & $h\,\tilde{\cal R}_{\pi_0}'\,\zeta_n$ : $-B^{(tss)}_{2}\,\frac{\hat{c}_L^5}{c_{s1}^4}\frac{\sin(\theta)^2}{6\,\sqrt{2}}\, 
\frac{1}{c_{s2}^3}$            \\ \hline
\rowcolor[HTML]{B4FDCD} 
$O\left(c_{s2}^{-5}\right)$ & $h\,{\cal R}_{\pi_0}^2$                         : $-\frac{3}{4}\,G^{(tss)}\,\frac{\hat{c}_L^5}{c_{s1}^4}\frac{\sin(\theta)^2}{\sqrt{2}}\, \frac{1}{c_{s2}^5}$    &          \\ \hline
\end{tabular}
\caption[Squeezed $f_{TSS}$ in supersolid inflation: $c_{s2} \ll 1$]{Squeezed $f_{TSS}$ for $c_{s2}\ll1$. We set  $V_{2-1}^{(tss)}=\left(V^{(tss)}_{1}-V^{(tss)}_{2}\right)$.}
\label{Squeezed_TSS_T}
\end{table}
}
{\renewcommand\arraystretch{2.3}
\begin{table}[!ht]
\centering
\begin{tabular}{|l|l|l|l|}
\hline
\rowcolor[HTML]{FFFFFF} 
\rowcolor[HTML]{C798FA} 
$O\left(c_{s2}^0\right)$ &  \multicolumn{2}{l|}{$h \,\zeta_n \, \zeta_n'$ : $-\frac{\hat{c}_L^5}{c_{s1}^6}\frac{\left[2\, c_{\text{s1}}\, \left(3 \,c_{\text{s1}} \,\left(c_{\text{s1}}+1\right)+2\right)+1\right]}{24 \,\sqrt{2}
   \, \left(2 c_{\text{s1}}+1\right){}^2}\, V^{(tss)}_{3}$ }      \\ \hline
\rowcolor[HTML]{C798FA} 
$O\left(c_{s2}^0\right)$ &  \multicolumn{2}{l|}{ $h\,\zeta_n^2$ : $\approx -\, \left(2 V^{(tss)}_{1}-V^{(tss)}_{2}\right)\,\frac{A_{\text{Log}}}{48\, \sqrt{2}\,\hat{c}_L^{5}}$}       \\ \hline
\rowcolor[HTML]{95AEF8} 
$O\left(c_{s2}^{-3}\right)$ & $h\,\zeta_n'\,{\cal R}_{\pi_0}$ : $ B^{(tss)}_{1}\,\frac{\hat{c}_L^5}{c_{s1}^5} \,\frac{\left(c_{\text{s1}}^2+c_{\text{s1}}+1\right) }{8\, \sqrt{2}\,\left(c_{\text{s1}}+1\right)}\,\frac{1}{c_{\text{s2}}^{3}}$  & $h\,\tilde{\cal R}_{\pi_0}'\,\zeta_n$ : $\approx B^{(tss)}_{2} \,\frac{A_{\text{Log}}}{8\, \sqrt{2}\, c_{\text{s1}}^2 }\,\frac{1}{c_{\text{s2}}^{3}}$      \\  \hline
\rowcolor[HTML]{B4FDCD} 
$O\left(c_{s2}^{-5}\right)$ & $h\,{\cal R}_{\pi_0}^2$                         : $-G^{(tss)}\,\frac{\hat{c}_L^5}{c_{s1}^4}\,\frac{3 }{8 \,\sqrt{2}}\, \frac{1}{c_{s2}^{6}}$    &              \\ \hline
\end{tabular}
\caption[Equilateral $f_{TSS}$ in supersolid inflation: $c_{s2} \ll 1$]{Equilateral $f_{TSS}$ for $c_{s2}\ll1$. We set $A_{\text{Log}}=\left(3\, \gamma _e+3 \,\log (-k\,t)-4\right)\,.$}
\label{Equilateral_TSS_T}
\end{table}
}

\clearpage

\section{Phenomenology}
\label{phen}

\subsection{Solid vs Supersolid}

Before giving an overview of primordial non-Gaussianity for a supersolid, it is interesting to briefly recap the main
differences with solid inflation introduced in~\cite{Endlich:2012pz}.
At the linear level, in the instantaneous reheating scenario, the
adiabatic power spectrum transmitted to the radiation phase
is pretty similar\footnote{At the first order in perturbation theory, the distinction between {\it special} and {\it democratic} is not very important: the dynamics is determined by the masses $\{M_l\} $ that are of order $\epsilon$ in both cases.}.
Indeed, we get~\cite{Celoria:2020diz}
\begin{eqnarray}
\text{Solid }:\;\;& {\cal P}_\Phi |_{Rad} \sim {\cal P}_{\zeta_n} \equiv {\cal P} \frac{1}{c_L^5}\\
\text{Supersolid }:\;\;& {\cal P}_\Phi |_{Rad} \sim {\cal P}_{\zeta_n}
                         \equiv {\cal P} \frac{1}{\hat{c}_L^5} \, ;
\end{eqnarray}
with $\hat{c}_L^5$ defined in (\ref{PS_zeta_n}). Isocurvature perturbations are very suppressed (of order $\epsilon\, \zeta_n$).
The crucial point is the fact that instantaneous reheating for a supersolid effectively almost filters out non-adiabatic modes.
Isocurvature modes have a much bigger amplitude but are mostly dissipated during the transition. Such a modes, particularly the ones related to $\pi_0$, are the main sources of the non linearities during inflation, affecting the scalar and tensor power spectra and bispectra.
In solid inflation there is a single scalar mode $\pi_L$, no isocurvature perturbation is present and all non linearities are due to the single mode whose amplitude is small as dictated by observed scalar power spectrum.
The main  difference between the two models resides in the presence in a supersolid of two independent scalar modes $\pi_L$ and $\pi_0$ with intrinsic mixing in the quadratic Lagrangian. Upon a non-trivial Hamiltonian diagonalization procedure, one gets the power spectra of $\zeta_n$ and ${\cal R}_{\pi_0}$ (trivially related to $\pi_L$ and $\pi_0$ in the flat gauge). While, given the reheating adopted, the power spectrum of $\zeta_n$ is strongly constrained  by observations, this is not the case for the one of ${\cal R}_{\pi_0}$ that can be significantly enhanced when $c_{s2}$ is small  triggering sizeable differences with solid inflation at the level of
non-Gaussianity and power spectra non linear corrections. 
The different non linear structure of solid and supersolids implies
that not only the size of non-Gaussianity is different but also its
form.
\begin{itemize}
\item The linear PS for tensor (gravitational waves) gets important corrections from the secondary production that can become of the same order or even dominant. As a consequence of the presence of $\pi_0$ in a supersolid not only can enhance the amplitude of the GW produced during inflation, but induces a blue tilt of the PS. A blue tilt makes the PS to grow as $k^{n_T-1}$ for large $k$ values and potentially can enter in the LISA, DECIGO and ET sensitivity region. The secondary production originates from the pure supersolid interaction $G^{(tss)}\, h \,\partial R_{\pi_0} \,\partial R_{\pi_0}$ of green type. In \cite{Celoria:2020diz}, the correction to the linear tensor PS  was estimated to be 
\be
{\cal P}_h |_{\text{One-loop}} \approx \frac{1}{2}\,G^{(tss)}{}^2 \, \pi^2\, \frac{\epsilon^2}{c_{s2}}\, {\cal P}_{{\cal R}_{\pi_0}}{}^2\,.  
\ee 
While for solids the tilt tends to be suppressed by
the slow-roll parameter $\epsilon$, for the supersolid we get
\be
n_T-1=12\,(1+c_b^2)+12\,c_b^2 \,\epsilon+2 \, \eta\,. 
\ee
When  $\epsilon,\,\eta \ll (1+c_b^2)$ we naturally get a conspicuous deviation from scale invariance toward the blue.   
\item For bispectra in the scalar sector there are important differences between solid and supersolid inflation. Consider a squeezed configuration.
While in solid inflation, if the special approach used
in \cite{Endlich:2012pz} is considered, a pure quadrupole term is
present, in a supersolid both a monopole and quadrupole term are found. The above can be interpreted as the indissoluble presence of a fluid component (responsible for the monopole term) and a solid component (that induces a quadrupole term). In addition, a monopole term is induced by the supersolid operators $w_{X/Z}$ and a quadrupole term by superfluid operator $\chi$.
It is worth to stress that the size of non linearities in supersolids can be enhanced up a factor $1/c_{s2}^8$ for small $c_{s2}$.
\item For a squeezed shape in the bispectra of the TSS sector, we get the same angular structure for both solids and supersolids 
\begin{equation*}
f_{TSS}^{(SQ)} \sim \left(\varepsilon_{\vec{k}_L} \cdot \vec{k}_S
\right)\cdot \vec{k}_S \, .
\end{equation*}
For a supersolid, the amplitude can be enhanced by a factor $1/c_{s2}^5$ for small $c_{s2}$.
\item Finally, for the bispectra (squeezed shape) in the sector TTS  both a monopole and a quadrupole are present; namely 
\begin{equation*}
f_{TTS}^{(SQ)} \sim \varepsilon_{\vec{k}_S} \cdot \varepsilon_{\vec{k}_S},\qquad \left(\varepsilon_{\vec{k}_S} \cdot \vec{k}_L \right)^2 \,.
\end{equation*}
While in solid inflation their size is similar, for a supersolid the monopole gets enhanced  by a factor  $1/c_{s2}^3$ for small $c_{s2}$. This leads a characteristic signature of supersolid inflation:  the non-linear and local corrections to the tensor PS that becomes rather different from solid inflation, as discussed in \cite{Malhotra:2020ket} where TTS predictions for the standard solid inflation \cite{Endlich:2012pz} scenario are compared with another ``{\it isotropic}'' (in the sense that the $f_{TTS}$ gives a monopole) model \cite{Adshead:2020bji}.
\end{itemize}


\subsection{PNG scenarios in supersolid inflation}

As it is evident from the previous analyses, the amount of non-gaussianity generated by supersolid inflation can be important. As already discussed, the presence of entropic perturbations with a sizeable amplitude (during inflation) generates not only noticeable non-gaussianity but also important non linear effects as secondary gravitational waves production. Clearly such enhancements can become leading for observables that start tiny at tree level, as it is the case for  the gravitational sector.
Related to this problem, as already discussed previously, it will be interesting to check the amount of non linear corrections that could in principle enhance or leave a peculiar angular and scale dependence to the primordial tensor PS. 
In particular the presence of important corrections to $f_{TTS}$ are expected to be related to sizeable non linear and local corrections to the tensor PS. Similarly, important corrections to $f_{SSS}$ and $f_{TSS}$ are expected to non linear correct the scalar PS.
Such a program is under study.
\\
So, let us come back to our ``tree level'' evaluation of the bispectra. Recalling that stability of the model gives the constraint $c_{s1}<\hat{c}_L<c_{s2}$ (we fixed for convenience $\hat{c}_L=1/2$), as one can infer from the figures in section \ref{pics}, we can discriminate two main regions in the $(c_{s1},\,c_{s2})$ space parameters. 
 \\
 An intermediate region where $c_{s2}$ is not too small if compared to $c_{s1}$ ($c_{s2}\lesssim c_{s1}$), and 
 an other region where one of the two diagonal sound speeds is very small, $c_{s2} \ll c_{s1}$. 
 \\
In the first region ($c_{s2}\lesssim  c_{s1}$), almost all the cubic interactions classified in this work equally participate in defining the total $f_{\text{NL}}$-parameters for SSS, TTS and TSS interactions.
In this case, some of the main sections' results need to be substituted with their total expressions given in Appendixes \ref{f_nl_sq} and \ref{TSS_total}. 
In particular, for the purely scalar sector, we have a summation of several terms that have to satisfy stringent constraints in the squeezed limit~\cite{Akrami:2019izv}, and this can be easily achieved by adjusting our free parameters. At the same time, setting the {\it coloured} X-couplings to one as shown in figures (\ref{Log_TTS},  \ref{TSS_dom}, \ref{TSS_cb1_squeezed}, \ref{cb01_EQ}) the TTS and TSS $f_{\text{NL}}$ can easily achieve values of order $10^2$ for $c_{s2}$ around $0.3$ making supersolid inflation an interesting model. A more quantitative and accurate analysis of this region is left for detailed future work.
\\
The main motivation to consider the region  $c_{s2} \ll c_{s1}$ is the possibility to enhance via secondary production the gravitational waves boost during inflation driven by a supersolid~\cite{Celoria:2020diz}. Following the same line of \cite{Celoria:2020diz} taking $c_{s2} \ll 1$ we will argue soon how to maximize the mixed scalar-tensor $f_{\text{NL}}$ parameters.\\
Let us start in the scalar sector for which $f_{\text{NL}}^{(SQ)}$ (squeezed configuration)  is better constrained by CMB observations, applying a vertex by vertex analysis and analyzing the figures (\ref{0t00}, \ref{V_plot}, \ref{B_G_O_plot}, \ref{O_plot}) we can desume the following features
\begin{itemize}
\item The contribution of violet operators to $f_{\text{NL}}^{(SQ)}$  has a minimum  in the region $c_{s1}\to 1$ and $c_{s2}\to 1/2$, while the maximum is attained for $c_{s1}\to 1/2$ and $c_{s2}\to 1/2$.
\item The  contribution of the remaining operators has only a minimum in the region $c_{s1}\to 1$ and $c_{s2}\to 1/2$ and diverges in the limit $c_{s2} \ll 1$. 
\end{itemize}
Once we impose that $|f_{\text{NL}}^{(SQ)}|<10$ for each vertex contribution, the results are the following absolute bounds for the couplings:
\bea
&&|V_{1,2,3}|\leq 214,\;\;\;
|V_{4,5}|\leq 9200,\;\;\;\\&&\nonumber
\\&&\nonumber
|B_{1,2}|\leq 36,\;\;\;
|B_{4,5}|\leq 7.2,\;\;\;
|B_{3}|\leq 144,\;\;\;
\\&&\nonumber
\\&&\nonumber
|G_{1,2}|\leq 4.6,\;\;\;
|O_{3}|\leq 18,\;\;\;
|O_{2}|\leq 3.6,\;\;\;
|O_{1}|\leq 1.8,\;\;\;
|R|\leq 2.2,\;\;\;
\ea
When $c_{s2}\leq 0.1$ the  constraints are much stronger:
\bea\nonumber
&& \hspace{-2.cm} |B_{1,2}|\leq 4.6 \;\hat c_{s2}^3,\;\; 
|B_{4,5}|\leq 4.6\;\hat c_{s2}^3,\;\; 
|B_{3}|\leq 98\;\hat c_{s2}^3,\;\; 
|G_{1,2}|\leq 10^{-2}\;\hat c_{s2}^5 \\&&\nonumber
\hspace{-2.cm} \\&&\nonumber
\hspace{-2.cm} |O_{3}|\leq 0.04\;\hat c_{s2}^6,\;\; 
|O_{2}|\leq 0.02\;\hat c_{s2}^6,\;\; 
|O_{1}|\leq 1.2\,10^{-3}\;\hat c_{s2}^6,\;\; 
|R|\leq 1.1\;10^{-3}\;\hat c_{s2}^8 \nonumber
\ea
where we normalized $c_{s2}$ as 
\begin{equation*}
\hat c_{s2}\equiv \left(\frac{c_{s2}}{0.1}\right)\,.
\end{equation*}
An interesting way to classify the vertices is to divide them into two main classes: operators compatible with a fluid or superfluid medium and operators compatible with a solid or supersolid medium, as shown in table \ref{class}.
{\renewcommand\arraystretch{1.8}
 \begin{table}[H]
\centering
\begin{tabular}{|l|c|c|l|}
\hline
\multicolumn{1}{|c|}{\textbf{$c_{b}^2=-1,\; \; c_{s2}=0.1$}} & \multicolumn{1}{l|}{\textit{\textbf{fluid/ superfluid}}} & \multicolumn{1}{l|}{\textit{\textbf{solid/ supersolid}}} & \textbf{$f_{\text{NL}}^{(SQ)} \le 10$}     \\ \hline
\textit{\textbf{Operators}}                                  & $b\,,\; y\,,\; \chi$                                     & $\tau_{Y/Z}\,,\;w_{Y/Z}$                                 &                                            \\ \hline
\multirow{2}{*}{\textbf{${\cal L}_3$}$\;\;\;\propto$ }                      & $\lambda_{2\,,\;10}$                                     & $\lambda_{9/7}$                                          &                                            \\ \cline{2-4} 
                                                             & \multicolumn{2}{c|}{$\lambda_6$}                                                                                    &                                            \\ \hline
\multirow{3}{*}{\textit{\textbf{Couplings}}}                 & $R$, $O_{1,\,2,\,3}$                             & $G_{1,\,2}$                                          & \multicolumn{1}{c|}{$\le 10^{-3\;\div \;-2}$} \\ \cline{2-4} 
                                                             & $B_{3}$                                                & $B_{1,\,2,\,4,\,5}$                                  & \multicolumn{1}{c|}{$\le 10^{0\;\div\; 1}$}       \\ \cline{2-4} 
                                                             & \multicolumn{1}{l|}{}                                    & $V_{1,\,2,\,3,\,4,\,5}$                              & \multicolumn{1}{c|}{$\le 10^{2\;\div\; 3}$}       \\ \hline
\end{tabular}
\caption[Supersolid operators and medium classification]{Operators in the cubic SSS Lagrangian classified medium-wise.}
 \label{class}
\end{table}}
\no
The operators $b,\,y,\,\chi$ pertain to a fluid or a superfluid, $\tau_{Y/Z}$ characterize solids while $w_{Y/Z}$ supersolids, see \cite{Celoria:2020diz}. The couplings in  ${\cal L}_{3}$ are derivatives of the Lagrangian $U$ with respect to the operators (\ref{oper}) and can also be divided according to their presence in the cubic Lagrangian of a fluid/superfluid and or of a solid/supersolid. Last but not least, only the coupling  $\l_{6}$ depends on second derivatives with respect to operators characterizing both fluid/superfluids and solid/supersolids. As a result, a generic vertex can be ascribed to a specific class of medium. 
In the table \ref{class} it is shown the contribution of a vertex to  $f_{\text{NL}}^{(SQ)}$ when $f_{\text{NL}}^{(SQ)}$ is set to be $\le 10$ and $c_{s2}\sim 0.1$. The result is that $R$, $O$, and $G$ operators are rather constrained and very sensitive to the ``portion'' of fluid/superfluid of the medium. \\
Thus the result is clear: if we want to enter a region with small enough $c_{s\,2}$ values, we have to mitigate the squeezed red, orange vertices typical of the fluid/superfluid component of the medium to avoid too large contribution to $f_{\text{NL}}^{(SQ)}$~\footnote{This is not a case. Indeed, these are the elements with a high number of $\pi_0$-fields typical of fluids only.}. \\
In this way, we can have the non-gaussianity in the scalar sector completely under control and entirely constrained by PLANCK results~\cite{Akrami:2019izv}, and at the same time to be able to get a sizeable secondary tensor production (of the same order of the primary one or even dominant) as described in \cite{Celoria:2020diz} and to maximize the TSS/ TTS primordial non-gaussianity.

\begin{figure}[!ht]
  \centering
    \includegraphics[width=8.cm]{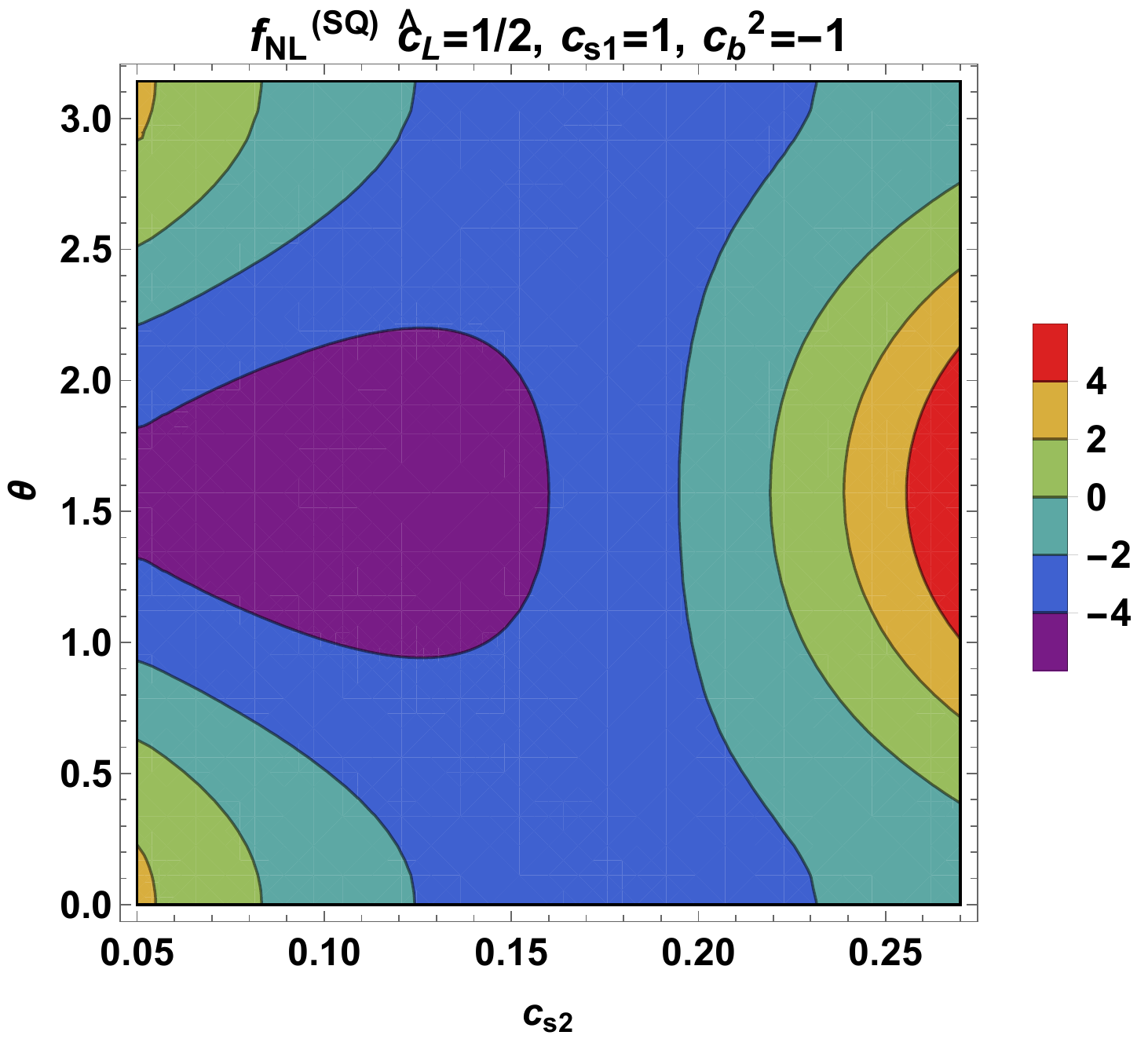}
\caption[Plot of the total squeezed $f_{\text{NL}}$ in supersolid inflation]{Plot of the total squeezed $f_{\text{NL}}$ with $\lambda_9=0.05$.}
\label{sq_plot}
\end{figure}

\begin{figure}[!ht]
  \centering
    \includegraphics[width=11.cm]{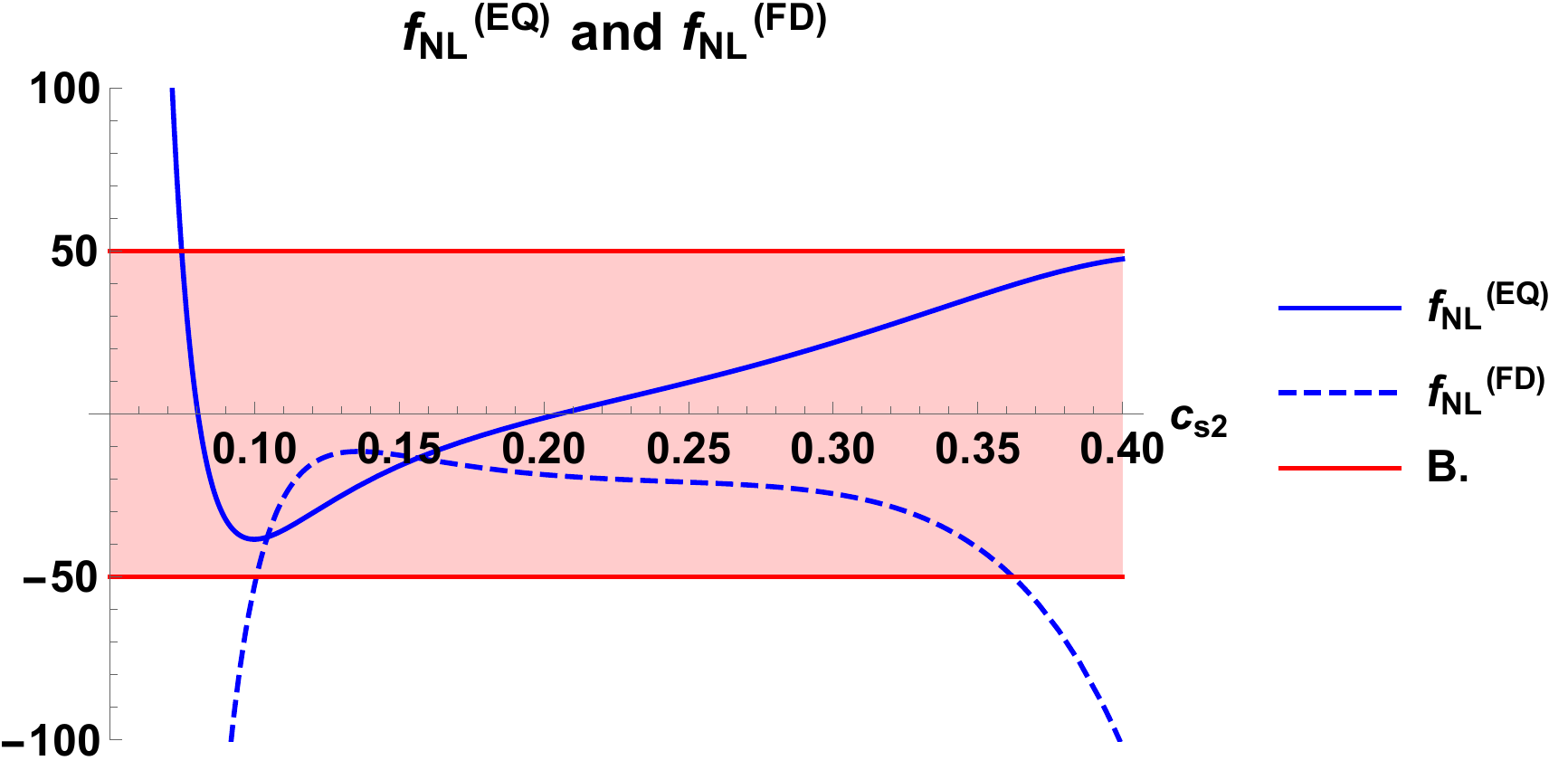}
\caption[Plot of the total equilateral and folded $f_{\text{NL}}$ in supersolid inflation]{Plot of the total equilateral $(k_i=k)$ and folded $(k_1=k,\; k_{2/3}=k/2)$ $f_{\text{NL}}$ with $\lambda_9=0.05$. The red zone is the region $ |f_{\text{NL}}| < 50$, considering the constraints $f_{\text{NL}}^{(EQ)}\sim -4 \pm 43$ and $f_{\text{NL}}^{(ORTHO)}\sim -26 \pm 21$ at the $68\%$ CL~\cite{ade}.}
\label{eq_plot}
\end{figure}

\begin{figure}[!ht]
  \centering
   \begin{minipage}[b]{0.45\textwidth}
    \includegraphics[width=\textwidth]{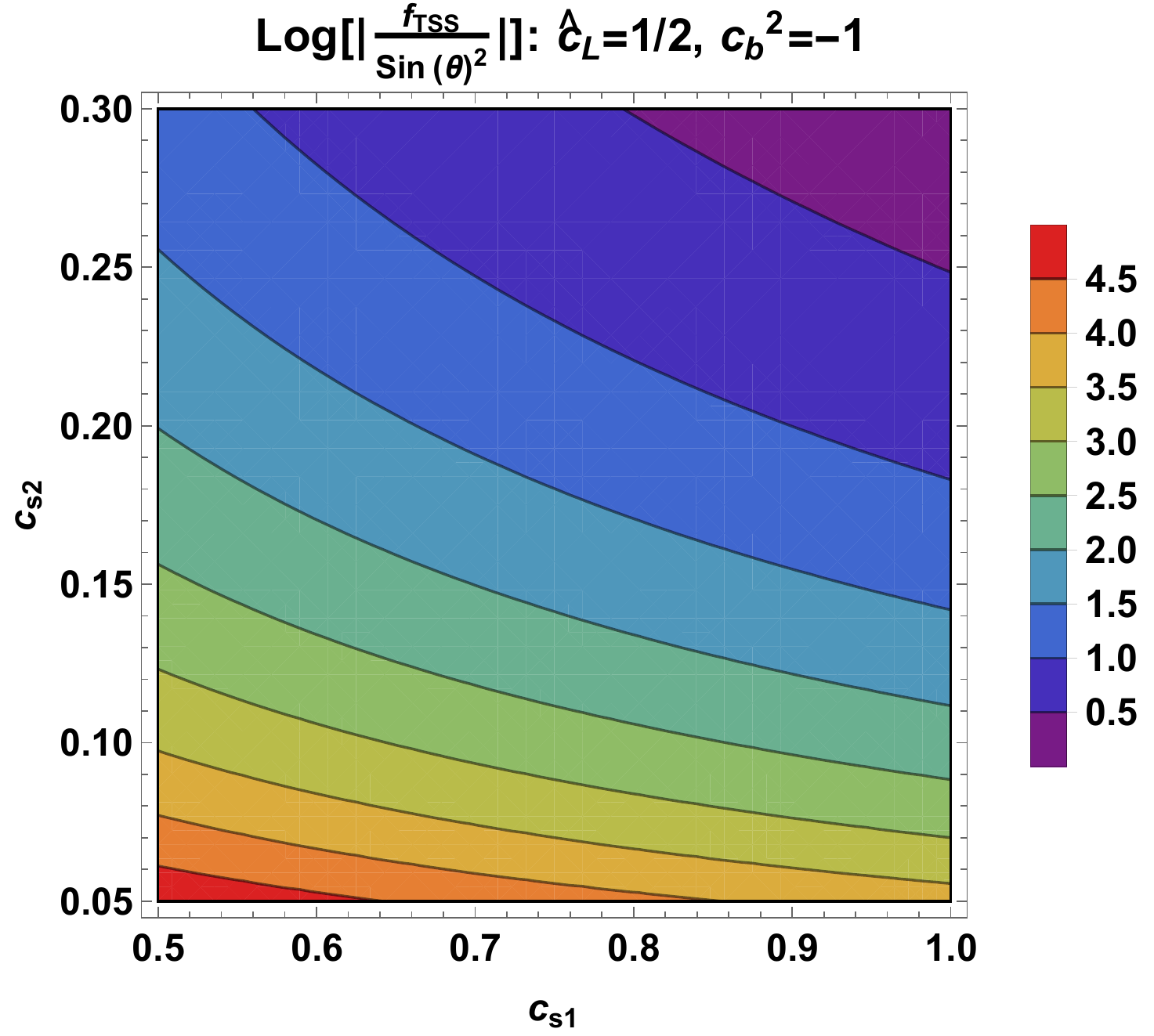}    
  \end{minipage}
  \begin{minipage}[b]{0.45\textwidth}
    \includegraphics[width=\textwidth]{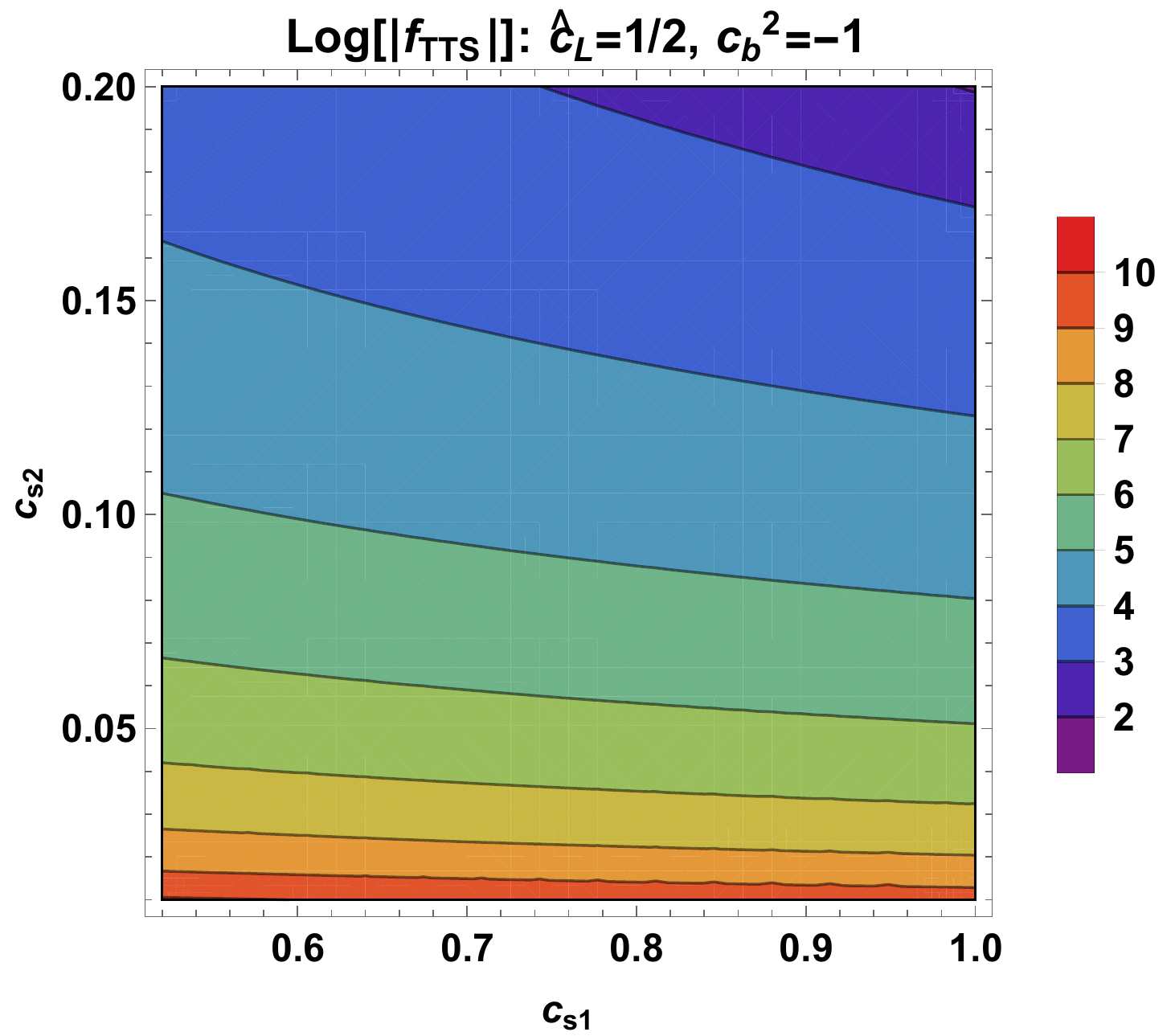}
  \end{minipage}
\caption[Dominant $f_{TSS}$ and $f_{TTS}$ in supersolid inflation]{On the left the dominant $h\, {\cal R}_{\pi_0}^2$ $f_{\text{NL}}$-like parameter with $\sin(\theta)^2$ set to one and $\lambda_9=0.05$. On the right the dominant $h^2\, {\cal R}_{\pi_0}$ $f_{\text{NL}}$-like parameter with $\epsilon$ set to 1, and $k_{S}\,t \sim 10^{-5}$. For instance, considering $\epsilon \sim 10^{-2}$ we get $f^{TTS}$ of order $10^{3}/10^{4}$ in the region $c_{s2}\sim 0.1$.}
\label{dom_T}
\end{figure}

Let us give one among several descriptive examples. 
Consider $\lambda_{2,6,10}$ free parameters as slightly $c_{s2}$-dependent. The remaining $\lambda_9$ and $\lambda_7$ as purely constant parameters. These last ones are defined by $w$-operators, playing a crucial role in the tensor sector.
By setting the $\lambda_{2,6,10}$ parameters as in figure \ref{tune_par} (explicit expressions are given in Appendix \ref{boost}), and expanding the total $f_{\text{NL}}$ squeezed in $c_{s2}$ we get 
\be
\label{tota_f}
f_{\text{NL}}^{(SQ)}= f_{\cal M}(c_{s1},\, \lambda_{9}, \,\lambda_{7})+ f_{\cal Q}(c_{s1},\, \lambda_{9},\,\lambda_{7})\; Y_2^0+O(c_{s2})\,.
\ee
\begin{figure}[!ht]
  \centering
    \includegraphics[width=9.cm]{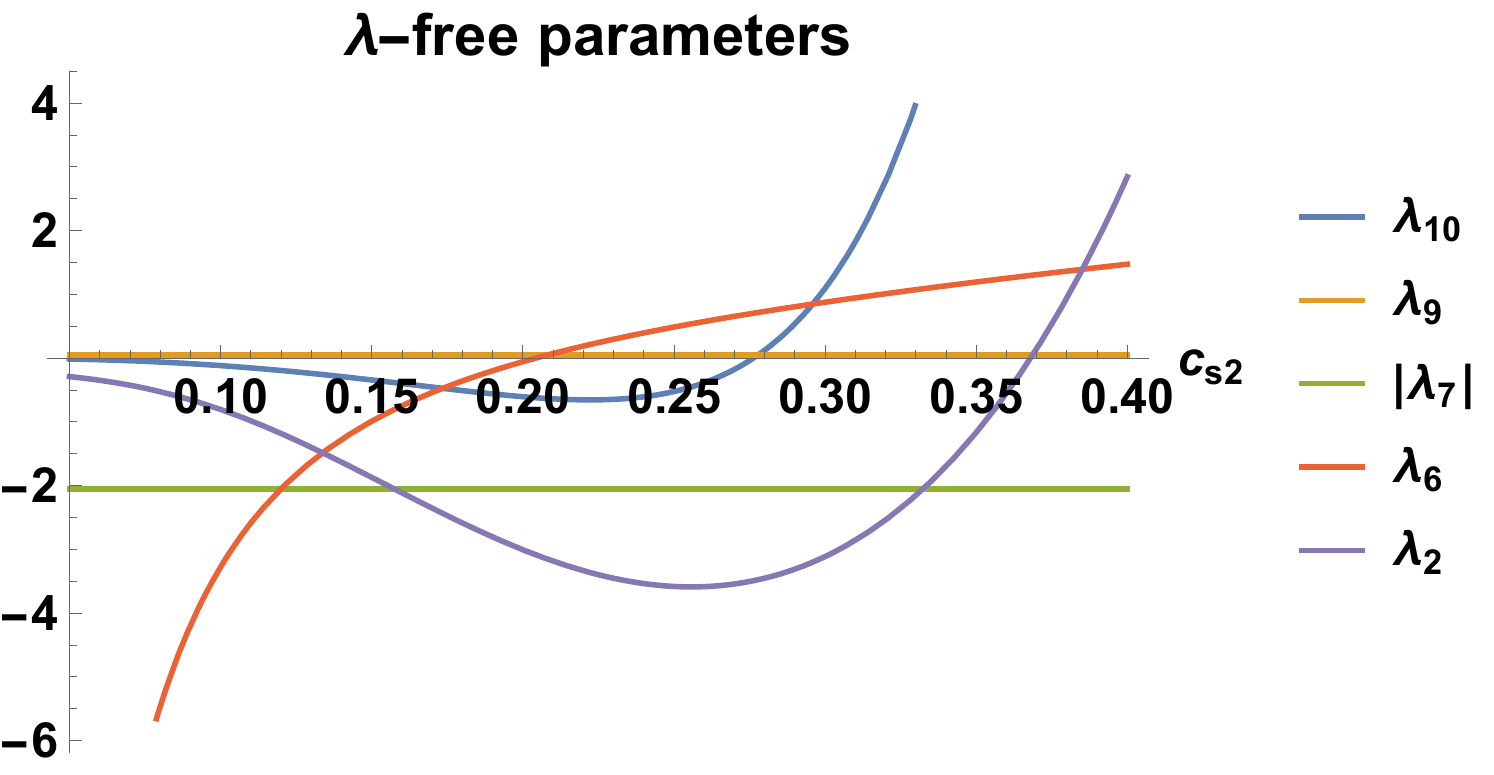}
\caption[$\lambda_{2,\,6,\,10}$ tuning]{Plot of the particular $\lambda_{2,6,7,9, 10}$ functions used to make the total squeezed $f_{\text{NL}}$ slightly dependent on $c_{s2}$ in a region $c_{s2} \leq 0.3$ .}
\label{tune_par}
\end{figure}
\\
For instance, in this particular example (see Appendix \ref{boost}) we reproduce a solid-like SSS scenario where $f_{\cal Q}$ is set to an optimal value around $10$ for small $c_{s2}$ values as it can be inferred from figure \ref{sq_plot}. \\
\\
At the same time, with the same parameters, we verified that the total equilateral and folded $f_{\text{NL}}$ assume values between $0$ to $\pm 50$ in a large range of the $c_{s2}$ parameter as shown in figure \ref{eq_plot}, and the situation is almost unchanged varying $c_{s1}$ \footnote{We used the $\lambda$ parameters appearing in red and orange monopole to compensate the squeezed {\it green} vertex. Consequently, the equilateral/folded shape is of green type (${\cal O}(c_{s2}^{-6})$) at maximum.}. In order to not have an excessively high equilateral/folded shape, we are forced to take $\lambda_9$ between $0.1$ and $0.01$.\\
\\
With this particular choice, we have the possibility to access to an interestingly small-$c_{s2}$ region whose effect is to enhance the $f_{TTS}$ and the $f_{TSS}$ parameters as it can be deduced by figures \ref{dom_T}. \\
\\
Note how, taking $\lambda_9$ relatively small, the dominant $f_{TSS}$ contribution is always between $10\div 10^3$ in the region $\left( c_{s2} \sim 0.1-0.2,\, c_{s1}\sim 1\right)$ in agreement with the current WMAP constraints~\cite{Shiraishi_2019}, while the $f_{TTS}$ parameter is even more enhanced thanks to the $\lambda_6$ parameters ($\sim 10^5\, \epsilon$ if $\lambda_9 \sim 10^{-2\;\div\;-1}$). Finally, the TTT $f_{\text{NL}}$ is well within the PLANCK constraints, being the $V_T$ parameter almost untouched by this parameterization and defined by $\lambda_7$ and $c_L$ only (see (\ref{LTTT})).

\section{Conclusions}
\label{conc}

A considerable effort is devoted to search for signatures of inflation by studying the CMB and the large scale structure of the Universe. The progress in detecting GWs and the forthcoming space interferometer LISA
give new opportunities to test the physics of inflation, and more in general, the early Universe on scales very
different from the ones of the CMB~\cite{Bartolo2_2019,  Bartolo_2019, Bartolo:2019yeu}. In this context, primordial non-gaussianity can be used to distinguish single clock inflation from multi-field models. Specifically, in single-clock inflation the 4D-diffeomorphism of GR are broken down to the 3D-diffeomorphisms of the constant inflaton field hypersurface and  non-gaussianity is tiny~\cite{Maldacena:2002vr}, the production of gravitational waves tends to be
suppressed for modes that re-enter the horizon during radiation domination~\footnote{This is a consequence of the red-tilted spectral index for tensors in single-field inflation.}, making almost impossible the direct detection by  LISA~\cite{Boyle_2008, Smith_2019, Bartolo:2016ami, Guzzetti:2016mkm}.
We have studied the physical impact of a change of the symmetry breaking pattern during inflation. By using effective field theory based on four St\"uckelberg-like scalar fields, the complete breaking of diffeomorphisms down to a global group can be described in terms of a
self-gravitating supersolid computing $f_{\text{NL}}$ for  squeezed and equilateral configurations extending the results of~\cite{Celoria:2020diz}.
Let us schematically highlight the main differences among single-field inflation, solid inflation (democratic approach) and our supersolid inflation model.
\begin{enumerate}
\item Symmetry breaking pattern: broken time reparametrization in single field inflation (one scalar DoF),
 broken spatial diffs.  in solid inflation (one scalar and two vector
 DoFs), completely broken time and space diffs. in  supersolid inflation (two scalar and two vector DoFs).
\item Graviton ``mass term''~\footnote{The presence of the
    term  $\epsilon\,a^4\,\plm^2\,H^2\,c_2^2\;h_{ij}^2$ in the gravitational action (\ref{slowact}), becomes a  ``mass term'' once we reduce to Minkowski space (this is the origin of the denomination ``mass''). In dS space its presence  introduces a mild time dependence ($a^{-\frac{8}{3}\,\epsilon\,c_2^2 }$) for the superhorizon power spectrum of the gravitational waves.}: massless in single field inflation, 
massive in solid  and supersolid inflation.
\item Perturbations  
\begin{itemize}
\item Single field inflation: $\zeta$ and ${\cal R}$ equivalent at
  superhorizon scales and  reheating independent. 
\item Solid inflation:  $\zeta$ and ${\cal R}$ not equivalent at
  superhorizon scales, weakly time dependent
  $a^{-\frac{4}{3}\,c_2^2\,\epsilon}$ and  reheating dependent.
\item Supersolid inflation: two independent scalar perturbations ${\cal
    R}_{\pi_0}$ and $\zeta_n$. The former is the analog of the
  comoving curvature perturbation in single field inflation, while
  $\zeta_n$ is the analog of $\zeta$ in solid inflation. In the
  instantaneous reheating approximation, $\zeta_n$ determines the
  primordial seed of adiabatic perturbations, while ${\cal R}_{\pi_0}$
  is mostly related to  entropic perturbations.
\end{itemize}
\item  Bispectrum of $\zeta_n$ in a squeezed configuration.
\begin{itemize}
\item 
  Single field inflation: a tiny monopole term,
  $f_{\text{NL}}^{SQ}\sim {\cal O}(\epsilon)$.
\item Solid inflation: a sizeable  quadrupole term, $f_{\text{NL}}^{SQ}\sim {\cal O}(1)\;Y^0_2(\theta) $
\item  Supersolid inflation:   sizeable  monopole and quadrupole
  terms, $f_{\text{NL}}^{SQ}\sim {\cal O}(1)\;(\#_1+\#_2\,Y^0_2(\theta))$.
\end{itemize}
It is evident that the violation of the consistency relations for
solid and supersolid inflation is related both to the magnitude and
the angular structure. While in solid inflation, an angular average
restores the consistency relation~\cite{Bordin:2017ozj}, this does not happen in supersolid inflation.
\item When the sound speed of scalar DoF is small, the corresponding power spectrum and $f_{\text{NL}}^{(SQ)}$ in a squeezed configuration scale as
\begin{itemize}
\item Single field inflation\footnote{In this case our prototype  is the $P(X)$ model~\cite{Chen_2007}}: $P_\zeta\sim\frac{1}{c_s}$, $\quad f_{\text{NL}}^{(SQ)}\sim\frac{1}{c_s^2}$.
\item Solid inflation: $P_\zeta\sim\frac{1}{c_s^5}$, $\quad f_{\text{NL}}^{(SQ)}\sim\frac{1}{c_s^2}$.
\item Supersolid inflation: $P_{{\cal R}_{\pi_0}}\sim\frac{1}{c_{s2}^5},\; P_{\zeta_n}\sim\frac{1}{c_{s2}^0},\;
P_{\zeta_n{\cal R}_{\pi_0}}\sim\frac{1}{c_{s2}^3}$, $\quad f_{\text{NL}}^{(SQ)}\sim\frac{1}{c_{s2}^8},...,\frac{1}{c_{s2}^0}$.
\end{itemize}
\end{enumerate}
As it is evident, the main difference between supersolid and  the
other two alternative models (single filed and solid inflation) is the
presence of two independent modes with phonon-like dispersion
relations and a kinetic mixing present throughout the inflationary period.
By studying the relevant 3-point functions, we have shown that the form and size of primordial non-gaussianity are very different from single-clock inflationary models. 
The most general supersolid produces a considerable enhancement for
all sectors (SSS, TTS, and TSS)\footnote{The same is true for the TTT
  sector if one considers one-loop corrections that  will be studied in a
  future work.} in the small $c_{s2}$ limit. Thus we have essentially two main possibilities.
\begin{itemize}
\item Do not tune the free parameters. In this case, we
  cannot consider values of  $c_{s2}$ too small, effectively reducing
  the allowed  region of the parameters $c_{s1}$ and
  $c_{s2})$. Contrary to solid inflation, monopole and quadrupole
  $f_{\text{NL}}^{SQ}$ terms arise, violating the Maldacena
  consistency relation in the SSS,TTS and TSS sectors. Getting the
  enhancement of GWs during inflation is more difficult.  
\item We have enough free-parameters, to mitigate the {\it
    too-large} contribution to $f_{\text{NL}}$ of  SSS green $\propto
  {\cal O}(c_{s2}^{-5})$, orange $\propto {\cal O}(c_{s2}^{-6})$ and
  red $\propto {\cal O}(c_{s2}^{-8})$ vertices in the small $c_{s2}$
  limit. In this case, we can have an interesting
  enhancement of the TTS and TSS sector; indeed the presence of the $w$-operators, specific of a supersolid still gives
\begin{equation*}
f_{TSS}   \propto {\cal O}(c_{s2}^{-5}) \,, \qquad f_{TTS}   \propto {\cal O}(c_{s2}^{-3})\, .\end{equation*}  
The former could leave a signature in LSS next-generation experiments,
giving a specific imprint~\footnote{The same mechanism is used to
  compute anisotropic corrections due to non-Gaussianities in the SSS
  sector; for  solid inflation see~\cite{Endlich:2013jia, Bartolo:2014xfa}.} on the distribution of galaxies~\cite{Akhshik_2015, Dimastrogiovanni:2014ina, Dimastrogiovanni_2016}, while the latter could play a crucial role in the generation of anisotropies of the gravitational waves background~\cite{Malhotra:2020ket, Iacconi:2020yxn, Adshead:2020bji, Iacconi:2019vgc}.
\end{itemize}  
We also mention that in this paper we suppose a suitable number of e-folds needed to solve the horizon problem in order to avoid the issue with statistical anisotropy discussed in~\cite{Bartolo:2014xfa}. Regardless of the choice made, supersolid inflation is a very interesting model whose features could be tested in next generation experiments probing the early Universe. 
\section*{Note Added}
After the completion of the present work the preprint~\cite{Cabass:2021iii}
was posted with
some overlap with our results.
\section*{Acknowledgments}

The work of DC and LP was supported in part by Grant No.2017X7X85K ``The dark Universe: A synergic multimessenger approach'' under the program PRIN 2017 funded by Ministero dell'Istruzione, Universit\'a e della Ricerca (MIUR).

\appendix

\section{Cubic Action}
\label{C_action}
In this appendix we give the expression for the cubic Lagrangian.
Consider the case where momenta are not too squeezed. The lapse and the shift fields are higher order in the slow-roll parameters and enter in the cubic Lagrangian at the leading order in slow-roll.

\subsection{Scalar Sector}

 In Fourier space, it is convenient to   replace $\pi_0$ and $\pi_L$ in terms of  ${\cal R}_{\pi_0}$ and
$\zeta_n$  (characterised by a flat power spectrum)
\be
\zeta_n= \frac{k^2}{3}\pi_L,\qquad {\cal R}_{\pi_0}=\frac{{\cal H}}{\bar\varphi'} \pi_0\,. 
\ee
Each cubic vertex can be parametrized  in the form
\be
{\cal L}^{(SSS)} = M_p^2\, H^2\,\epsilon \, X
\,a^n \, D\left(k,\,k',\,k''\right)\,\xi_{k}\,\xi_{k'}\,\xi_{k''}\,, 
\ee
  where $X$  is a dimensionless constant, $D$ is a dimensionless combination of momenta and the fields entering the vertex are denoted by  $\xi=\{\,\zeta_n ,\, \zeta_n{}',\,H^{-1}\, {\cal R}_{\pi_0},\, H^{-1}\,{\cal
    R}_{\pi_0}{}' \,\}$; integration over momenta and the
  delta functions enforcing momentum conservation in each vertex is understood. 
 Thus, the cubic Lagrangian in the scalar sector can be
  written as 
\be\label{L3ZR}
\begin{split}
\frac{{\cal L}^{(SSS)} }{\epsilon  \, H^2 \, M_p^2}=& a^4\left[V_{1}+V_{2}\,
(\hat k\cdot \hat k')(\hat k\cdot \hat k'')(\hat k'\cdot \hat k'')+V_{3}\,(\hat k'\cdot \hat k'')^2
\right]\, \zeta_{n\,\vec k}\, \zeta_{n\,\vec k'}\,\zeta_{n\,\vec k''}\,+ \\
& \frac{a^4}{k'\;k''}\,\left[V_{4}\,(\hat k'\cdot \hat k'') +V_{5}\,(\hat k\cdot \hat k')(\hat k\cdot \hat k'') \right]\, \zeta_{n\,\vec k}\,  \zeta_{n\,\vec k'}' \,\zeta_{n\,\vec k''}' \\
&+a^3\,\frac{k'}{k''}\;\left(B_{1}\,(\hat k'\cdot \hat k'')+
B_{2}\, (\hat k\cdot \hat k')(\hat k\cdot \hat k'') \right)\, 
\zeta_{n\,\vec k}\,\frac{{\cal R}_{\pi_0\,\vec k'}}{H}\, \zeta_{n\,\vec k''}'\\
&+\frac{a^3}{k'\;k''}\,B_{3}\,(\hat k'\cdot \hat k'') \,\frac{ \tilde {\cal R}_{\pi_0\,\vec k}'}{H}\, \zeta_{n\,\vec k'}' \, \zeta_{n\,\vec k''}'  \\
&+a^3\,\left(B_{5}\,(\hat k'\cdot \hat k'')^2+B_{4} \right)\,\frac{\tilde {\cal R}_{\pi_0\,\vec k}'}{H}\,\zeta_{n\, \vec k'}\,\zeta_{n\, \vec k''}\\
&+a^2\,k'\;k''\;\left( G_{2}\,(\hat k\cdot \hat k'')(\hat k\cdot \hat k')+G_{1}\,(\hat k'\cdot \hat k'') \right) \, \zeta_{n\,\vec k}{}\,\frac{{\cal R}_{\pi_0\, \vec k'}}{H}\,\frac{{\cal R}_{\pi_0\,\vec  k''}}{H}\\
&+ a\,O_{1}\,\frac{\tilde {\cal R}_{\pi_0\,\vec k}'}{H}\,\frac{\tilde {\cal R}_{\pi_0\,\vec k'}'}{H}\,\frac{\tilde {\cal R}_{\pi_0\,\vec k''}'}{H}+ a^{2}\,O_{2}\,\frac{\tilde {\cal R}_{\pi_0\,\vec k}'}{H}\,\frac{\tilde {\cal R}_{\pi_0\,\vec k'}'}{H}\,\zeta_{n\, \vec k''} \\
&+ a^2\,O_{3}\,\frac{\tilde {\cal R}_{\pi_0\,\vec k}'}{H}\,\zeta_{n\,\vec  k'}'\,
\frac{{\cal R}_{\pi_0\,\vec k''}}{H}+a\,R\,k'\;k''\;(\hat k'\cdot \hat k'')\,
\frac{\tilde {\cal R}_{\pi_0\,\vec k}'}{H}\,\frac{{\cal
    R}_{\pi_0\,\vec k'}}{H}\,\frac{{\cal R}_{\pi_0\,\vec k''}}{H}
\,;
\end{split} 
\ee
{  where} $k=|\vec k| $ and $\hat k =\frac{ \vec k}{k}$.
 The values of dimensionless couplings  of the vertices
  are given in the following table. 
  \begin{table}[H]
  \begin{center}
\begin{tabular}{|c|c|}
    \hline
  Vertex & Value \\
  \hline
${\color{violet} V_{1}}$ & $ 36  \,c_L^2-108 \,c_0^2  \,c_b^4-\frac{9 \, \lambda _1}{2}+36 \, \lambda_5-16 \,
              \lambda_7$   \\
    \hline
   {\color{violet} $V_{2}$}&   $ 81 \, c_L^2-72 \, \lambda _7+117$ \\
    \hline
{\color{violet} $V_{3}$} &$ 108  \,c_0^2  \,c_b^4-135 \,
 c_L^2-108 \, \lambda _5+72 \, \lambda _7-135$ \\
\hline
{\color{violet} $V_{4} $}&  $-108  \,c_0^2  \,c_b^4-27  \,c_L^2+\frac{27 \, \lambda _8}{2}-36
\,\lambda _9-27-\frac{27}{2}\,a^{-3\,(1+c_b^2)}\,\hat{\sigma}$\\
    \hline
{\color{violet} $V_{5} $} & $-27    \, \left(4  \,c_1^2-3  \,c_L^2-4 \, \lambda
             _9+1\right)$ \\
 \hline
\end{tabular}
 \end{center}
\end{table}
\begin{table}[H]
\begin{center}
 \begin{tabular}{|c|c|}
        \hline
  Vertex & Value \\
  \hline
     {\color{blue} $B_{1}$} &  $24\, \lambda _9-9 \,\lambda_8$ \\
     \hline 
     {\color{blue} $B_{2}$} & $36 \,c_1^2-72\, \lambda _9+9 \,\hat{\sigma}\,a^{-3(1+c_b^2)}$ \\
     \hline
      {\color{blue} $B_{3}$} & $ \frac{9}{2} \,  \left(-4 \,c_1^2-8 \,c_0^2 \,c_b^2+\lambda _{10}\right)$\\
     \hline
    {\color{blue}  $B_{4} $} & $-36 \,c_0^2 \,c_b^2+\frac{9\,\lambda _4}{2}-12 \,\lambda _6+\frac{9}{2} \,\hat{\sigma}\,a^{-3(1+c_b^2)}$\\
     \hline
   {\color{blue} $B_{5}$} & $36 \,c_0^2\, c_b^2+36 \,\lambda_6-\frac{9}{2} \,\hat{\sigma}\,a^{-3(1+c_b^2)}$\\
   \hline
 {\color{green}  $G_{1}$} &   $12 \,c_0^2\, c_b^2+\frac{3\, \lambda _8}{2}-4 \,\lambda_9-\frac{3}{2} \,\hat{\sigma}\,a^{-3(1+c_b^2)}$\\
   \hline
   {\color{green} $G_{2}$} &$ 12 \, \lambda _9$\\ \hline
 \end{tabular}
 \end{center}
\end{table}

\vspace{-1.cm}

\begin{table}[H]
  \begin{center}
\begin{tabular}{|c|c|}
    \hline
  Vertex & Value \\
  \hline
{\color{orange}  $O_{1}$ } & $\frac{ \, \lambda _2}{6} $ \\
\hline
{\color{orange} $O_{2}$} &$ \frac{3}{2} \,\lambda_3 $ \\ 
\hline
  {\color{orange} $O_{3}$} &$ -3 \,  \,\lambda _{10} $\\
  \hline
{\color{red} $ R$} & $\frac{1}{2}\,  \, \left(8 \,c_0^2+4
                           \,c_1^2+\lambda _{10}\right) $\\
  \hline
\end{tabular}
 \end{center}
\end{table}

\subsection{TSS Sector}
In the TSS sector we have
\be
\begin{split}
\frac{{\cal L}^{(TSS)} }{\epsilon \, M_p^2\,H^2}  = & a^4\,\,
  h_{ij\,\vec k}\,\left(G^{(tss)}\,  k'^i\,k''^j \, 
\frac{{\cal R}_{\pi_0\,\vec k'}}{a\,H}\,\frac{{\cal R}_{\pi_0\,\vec k''}}{a\,H}
\right.\\
 & \left.
  +B^{(tss)}_{1}\, \frac{k''}{k'}\;\hat k'^i\,\hat k''^j  \,\zeta_{n\,\vec k'}'\,
  \frac{{\cal R}_{\pi_0\,\vec k''}}{a\,H} + 
   \, B^{(tss)}_{2}\, \hat k''^i\,\hat k''^j  \,\frac{\tilde{\cal R}_{\pi_0\,\vec k}'}{a\,H}\,\zeta_{n\,\vec k''} 
  \right.\\ 
  &\left.
  +   V^{(tss)}_{3}\,\frac{ \hat k'^i\,\hat k''^j}{k'\;k''}\,  \zeta_{n\,\vec k'}' \, \zeta_{n\,\vec k''}' +
   \left[V^{(tss)}_{1}\, \hat k'^i\, \hat k'^j +V^{(tss)}_{2}\,
  (\hat k'\cdot k'') \, \hat k''^j \,\hat k'^i \right]\,\zeta_{n\,\vec k'}\,\zeta_{n\,\vec k''}\right)\,.
\end{split}
\ee
The expressions for the vertices are given in the following table
\begin{table}[!ht]
  \begin{center}
\begin{tabular}{|c|c|}
\hline
  Vertex & Value \\
  \hline
 {\color{violet} $V_{1}^{(tss)}$} & $-54\, (1+c_L^2)+36\, c_b^2 \, \lambda
                                    _5+24\, \lambda _7 $\\
  \hline
  {\color{violet} $V_{2}^{(tss)}$} &$ \frac{135 }{2}\,(\,c_L^2+1)-36\, \lambda
   _7 $ \\
  \hline
 {\color{violet} $V_{3}^{(tss)}$} & $18 \,c_1^2-\frac{27  }{2}\,(c_L^2+1)-18 \,\lambda
                                    _9+18 $\\
  \hline
  {\color{blue} $B_{1}^{(tss)}$} & $ -\frac{4}{3} \, \lambda _9$ \\
  \hline
  {\color{blue} $B_{2}^{(tss)}$} & $\frac{4 \, \lambda _6}{3} $\\
\hline
 {\color{green} $ G^{(tss)}$} &   $\frac{2 \,c_1^2}{9}+\frac{2 \lambda _9}{9}$
  \\
\hline
\end{tabular}
 \end{center}
\end{table}

\subsection{TTS Sector}
For the TTS sector we have
\begin{equation*}
\begin{split}
\frac{{\cal L}^{(TTS)} }{\epsilon \, M_p^2\,H^2}= \,
a^4\, & h_{ij\,\vec k}\,h_{mn\,\vec k'} \\
&\left[
\delta_{im}\,\delta_{jn}\;\left( B^{(tts)}\;\frac{\tilde{\cal R}_{\pi_0\,\vec k''}'}{a\,H}+
 V^{(tts)}_{1}\;\zeta_{n\, \vec k''}\right)+
 V^{(tts)}_{2}\;\delta_{in}\;\left(\hat k^{j ''}\;\hat k^{m ''}\right)\;\zeta_{n\, \vec k''}
\right]\,; 
\end{split}
\end{equation*}
the vertices are only of violet and green type given following table.
\begin{table}[!ht]
  \begin{center}
\begin{tabular}{|c|c|}
\hline
  Vertex & Value \\
  \hline
{\color{violet} $V_1^{(tts)}$} & $-\frac{9}{2}
                                   \,\left(c_L^2+1\right)+3\, c_b^2
  \,  \lambda _6+2\, \lambda _7$ \\
\hline
{\color{violet}  $V_2^{(tts)}$}  & $\frac{27}{2}\,
   \left(c_L^2+1\right)-6 \,\lambda _7$ \\
  \hline
  {\color{blue}  $B^{(tts)}$}  &$ 3 \,\lambda _6$ \\
  \hline
\end{tabular}
 \end{center}
\end{table}

\section{Entropy Term and Total Derivatives}
\label{tot_der}
In several points of the paper we stated that $\hat \sigma$ appears only as a coefficient multiplying a total derivative term.  Let us show
how it comes about.
Let us consider  the  operators of interest as a  background
(\ref{back}) plus a  perturbation (all orders)
\be
b=\bar b+\delta b,\quad Y=\bar Y+\delta Y,\quad \chi=\bar \chi+\delta
\chi \,; 
\ee
and let us expand the Lagrangian for the scalar fields in (\ref{gact}) in the following way
\be\label{relU}
\delta U=(U_b\;\delta b+U_Y\;\delta Y+ U_{bY} \;\delta b\,\delta Y+ U_{b\chi} \;\delta b\,\delta \chi+...)
\ee
Now, from this expansion we can identify the derivatives of $U$ that
contains $\bar\sigma$.
To do that, we have to identify  where the $\bar\sigma$  appears.
For $c_b^2=-1$, we  have 
\bea\nonumber
&& \bar \sigma=\frac{(U_\chi+U_Y)}{\bar{b}},\quad 
\bar p+\bar \rho=\plm^2(\bar Y\,\bar b\,\bar\sigma-\bar b\;U_b)\simeq 0, \\ 
&& M_4=4\,H^2\,\epsilon\,c_4^2=\frac{\bar Y\,\bar b}{2}(U_{b\chi}+U_{bY}-\sigma), \quad \epsilon\;H^2\,\lambda_8=U_{bY}\;\underbrace{\to}_{(\ref{dep_par})}\;\bar \sigma+...
\ea
Thus, we see that  $U_Y=\bar b\,\bar\sigma+ \cdots$, 
$U_b= \bar Y \,\bar\sigma+ \cdots$ and finally $U_{b\chi}= \bar \sigma+\cdots$.
So the coefficient  $\delta U|_{\bar\sigma} $ in the expansion of $U$
that multiplies $\bar\sigma$ is given by 
\be
\delta U|_{\bar\sigma}=(\bar Y \;\delta b+\bar b\;\delta Y +
\;\delta b\,\delta \chi) \, .
\ee
Note that when  we implement the constraints in (\ref{dep_par}), we have to impose
$U_Y=\bar b\,\bar\sigma+...$,
$U_b= \bar Y \,\bar\sigma+...$, $U_{b\chi}= 0+...$ and $U_{bY}= \bar\sigma+...$
that implies
\be
\delta U|_{\bar\sigma}=(\bar Y \;\delta b+\bar b\;\delta Y +  \;\delta b\,\delta Y) =\bar\sigma\;(\bar Y\;\delta b +\bar b\;\delta Y+\delta b \,\delta Y)=\delta(b\;Y)
\ee
Last step is to show that the term $b\,Y$ form a total derivative.\\
With the field content of our model it is easy to build the following  total derivative term \cite{Ballesteros:2016kdx}:
\be
\de_\mu Q^\mu \equiv 
\sqrt{g}\,b\, y = 
  \epsilon^{\mu \nu \alpha \beta} \, \epsilon_{A
  B C D} \, \de_\mu \varphi^A \, \de_\nu \varphi^B \, \de_\alpha
\varphi^C \, \de_\beta \varphi^D \, .
\label{TD}
\ee
Thus, at cubic order, we have that  all the terms in the expansion
of the Lagrangian for the scalar fields in (\ref{gact}) 
proportional to $\bar\sigma$  can be collected in ${\cal
  L}_{\bar\sigma}^{(3)} $ given by
\bea
{\cal L}_{\bar\sigma}^{(3)}\!\!\! &=&\!\!\! \bar\sigma \big \{\pi_0'+\Delta\pi_L\;\bar\varphi'+\partial_t(\pi_0\;\Delta\pi_L)-\partial_i(\pi_0\,\partial_i\pi_L')+
\\ \nonumber
&&\frac{1}{2}\,\partial_t \left[\pi_0\,((\Delta\pi_{L})^2-(\partial_{ij}\pi_{L})^2)\right]+
\partial_{i}\,\left[\pi_0\,(-\Delta \pi_{L}\,\partial_{ i}\pi_{L}'+\partial_{ij}\pi_{L}\,\partial_{j} \pi_{L}')\right]
+\det(\partial_{ij}\pi_L)\,\bar\varphi' \big \} \,.
\ea 
It is well known that spatial derivative terms never contribute to the
primordial bispectrum. Surprisingly, in general this is not the case
of a time derivative term~\cite{Garcia_Saenz_2020, Arroja_2011, Rigopoulos_2011, Burrage_2011}.
We can write the cubic total derivative expression proportional to $\bar\sigma$ as
\be
\begin{split}
  {\cal L}_\sigma^{(3)} = \partial_{t} Q \,,\qquad Q =\;\frac{\bar\sigma}{2}\,\pi_0\, \left[\left(\partial^2 \pi_L\right)^2-\left(\partial_{ij}\pi_L\right)^2\right]\,.
\end{split}
\ee
Using perturbation theory, such a local term gives  a contribution
local in time of the form
\be
\left\langle \zeta_n^3(t) \right\rangle \sim i \int d^3x \,
\left\langle 0| \left[\zeta_n (t)^3, Q (t) \right] |0 \right\rangle\,.
\label{comm}
\ee
Such terms are commonly estimated considering a quadratic field
redefinition and absorbing the time derivative terms in the cubic
Lagrangian~\cite{Maldacena:2002vr,Seery_2005,Chen_2007}. In our particular case, $Q $ does depend by $\pi_0$ and $\pi_L$
conjugate momenta and then it  commutes with $\zeta_n \sim
\nabla^2\pi_L$ and the contribution to the bispectrum of $\zeta_n$
(\ref{comm}) is zero.
Notice that, it is the choice of an instantaneous reheating
that selects $\zeta_n$ as the reference scalar perturbation that sets
the initial conditions for the radiation domination phase of the Universe.
Quite different scenarios can be considered computing $\zeta$, or ${\cal R}$ bispectra where the conjugate momenta appear~\cite{Celoria:2020diz}. 
As final remark we point out that the vertices $V_{4/5}$ and $G_{1/2}$
do not contribute to  (\ref{TD}); indeed $\bar{\sigma}$ cancels due to the condition $M_{1}{}'\approx 0$.

\section{PNG and the In-In Formalism}
\label{in-in}

In this section a general description of the in-in formalism and  the main formula to get our results are given.
The in-in formalism is based on the  interacting picture, each quantum
field~\footnote{To make clear that we are dealing with quantum fields
  an $\hat{}$ is used.} $\hat \xi$ evolves following
\be
\frac{d \hat \xi}{dt}=-i \left[\hat \xi, \hat H_2 \right]\,, 
\ee
where $\hat H_2$ is the quadratic(free) Hamiltonian. The leading
contribution to the 3-point function of f$\hat \xi^3$ operator, by expanding the evolution operator, will be given by
\be
 \langle \hat \xi(\textbf{x}_1)\,\hat \xi(\textbf{x}_2)\,\hat \xi(\textbf{x}_3) \rangle = i\,\int_{-\infty}^t \, dt'\,\langle 0 \mid \left[ \hat \xi(\textbf{x}_1)\,\hat \xi(\textbf{x}_2)\,\hat \xi(\textbf{x}_3), \hat H_{in}^I (t')\right] \mid 0 \rangle\,.
\ee
By definition of the interacting Hamiltonian, the reader can easily verify that supersolid inflation can be included in the class of theories such that  
\be
 \hat H_I =- \int d^3 x \;\hat {\cal L}^{(3)}(\textbf{x}\,, t)\,,
\ee 
where $\hat {\cal L}^{(3)}$ is the cubic Lagrangian density. Thus, in
Fourier space the bispectrum reads 
\be
\label{BS_general}
\begin{split}
{\cal B}(k_1,\,k_2,\,k_3) \equiv & \frac{2}{(2\,\pi)^9} \, 
\int  \Pi_{i=1}^3 \,d^3 q_i \,\cdot \\
&\;\;\int_{-\infty}^t dt' \,a(t')\,\delta^{(3)}\left(\sum_j\,\textbf{q}_j\right)\, \text{Re} \left[i\,\langle 0 \mid \hat \xi_{k_1}\,\hat \xi_{k_2}\,\hat \xi_{k_3}\mid_t \, \hat {\cal L}_{q_1\,q_2\,q_3} \mid_{t'} \mid 0 \rangle\right] ;
\end{split}
\ee
where ${\cal L}_{q_1\,q_2\,q_3}$ is the Fourier transform of ${\cal
  L}^{(3)}$. Equation (\ref{BS_general}) is completely general and can
be specialized to the case scalar, tensor or mixed case. 
\begin{enumerate}
\item \textbf{SSS}:\\
vertex of the form 
\be 
{\cal L}_{k,\,k'\,k''}= X\, \epsilon \, \plm^2 \,H^2\, a^n \, D(k,\,k',\,k'') \, \xi_{\omega}{}_k\,\xi_{\delta}{}_{k'}\,\xi_{\rho}{}_{k''}\,,
\ee
where $\xi_{\text{Greek letter}}$ can be any scalar field among
$\zeta_n$, $\,\zeta_n'$, $H^{-1}{\cal R}_{\pi_0}$,  $H^{-1}\,{\cal
  R}_{\pi_0}'$. For the bispectrum of $\zeta_n$ we have
\be
\label{SSS} 
{\cal B}_{\zeta_n^3}= \int dk^3\,dk'^3\,dk''^3 \, {\cal I}\,, \qquad {\cal I}= 2 \, \text{Re} \left[{\cal J}\right]\,; 
\ee
\be
\begin{split}
\label{SSS_J}
&{\cal J}= i\,X\, \epsilon \, \plm^2 \,H^2\,\, D(k,\,k',\,k'') \, \\
&\;\;\sum_{i,\,j,\,l=1}^2 \zeta_n{}_{k_1}^{(i)}\,\zeta_n{}_{k_2}^{(j)}\, \zeta_n{}_{k_3}^{(l)} \mid_{t_e} \int_{-\infty}^{t_e}\, dt'\, a^{n}\,\xi_{\omega}{}_k^{(m)\,*}\,\xi_{\delta}{}_{k'}^{(n)\,*}\,\xi_{\rho}{}_{k''}^{(o)\,*}\, \cdot \\
 &\Big\{ \delta(\textbf k_3+\textbf k'')\,\left[\delta^{l,i,j}_{o,n,m} \, \delta(\textbf k_1+\textbf k')\,\delta(\textbf k_2+\textbf k)+ \delta^{l,i,j}_{o,m,n}\, \delta(\textbf k_1+\textbf k)\delta(\textbf k_2+\textbf k') \right] \\
&\;\;+\delta(\textbf k_3+\textbf k')\,\left[\delta^{l,i,j}_{n,o,m} \, \delta(\textbf k_1+\textbf k'')\delta(\textbf k_2+\textbf k)+ \delta^{l,i,j}_{n,m,o} \delta(\textbf k_1+\textbf k)\delta(\textbf k_2+\textbf k'') \right] \\
&\;\;+\delta(\textbf k_3+\textbf k)\,\left[\delta^{l,i,j}_{m,o,n}\, \delta(\textbf k_1+\textbf k'')\delta(\textbf k_2+\textbf k')+ \delta^{l,i,j}_{m,n,o}\, \delta(\textbf k_1+\textbf k')\delta(\textbf k_2+\textbf k'') \right]\Big\}
\end{split}
\ee
where $\delta^{l,i,...,j}_{m,n,...,0}=\delta^l_m\, \delta^i_n\, ...\,\delta^j_o$. The dimensionless  function $D$ encodes the structure of  spatial derivatives in the vertex structure, while the $n$ and $X$ are vertex dependent constants.
\item \textbf{TSS}:\\
 vertex of the form 
 \be 
 {\cal L}_{k,\,k'\,k''}= X\, \epsilon \, \plm^2 \,H^2\, a^n \, D_{ij}(k,\,k',\,k'') h_{ij \, k}\,\xi_{\omega}{}_{k'}\,\xi_{\delta}{}_{k''}\,,
\ee 
which leads to
\be 
\label{TSS}
{\cal B}_{\zeta_n^2\,h^p}= \int dk^3 dk'^3 dk''^3\,{\cal I}\,, \qquad {\cal I}= 2 \, \text{Re}\left[{\cal J}\right]\,;
\ee
\be
\label{TSS_J}
\begin{split}
{\cal J}=&i\,X\, \plm^2\, H^2 \, \epsilon \, D_{ij}(k,\,k',\,k'')\\
&\; \sum_{l,m=1}^2 h_{k_1} \,\zeta_n{}_{k_2}^{(l)}\, \zeta_n{}_{k_3}^{(m)}|_{t_e} \,\sum_{r}\,\varepsilon_{ij}^{(r)}{}_k\;\int_{-\infty}^{t_e} dt' \, a^{n}\,  h^{*}_{k}\,\xi_{\omega}{}^{(n)\,*}_{k'}\, \xi_{\delta}{}^{(o)\,*}_{k''}|_{t'} \,\cdot \\
&\delta(\textbf k_1+\textbf k)\,\left[ \delta^{p,l,m}_{r,n,o} \,\delta(\textbf k_2+\textbf k')\delta(\textbf k_3+\textbf k'')+\delta^{p,l,m}_{r,o,n} \, \delta(\textbf k_2+\textbf k'')\delta(\textbf k_3+\textbf k') \right]\,.
\end{split} 
\ee
\item \textbf{TTS}:\\
vertex of the form 
\be 
{\cal L}_{k,\,k'\,k''}=X\,\plm^2 \,H^2 \, \epsilon\,  a^n \, D_{lm}(k,\,k',\,k'')\, h_{im \, k} \,h_{il \, k'}\, \xi_{\omega}{}_{k''}\,,
\ee
which leads to:
\be
\label{TTS}
{\cal B}_{\zeta_n\, h^p h^q}= \int dk^3 dk'^3 dk''^3 \,{\cal I}\,, \qquad {\cal I}= 2 \, \text{Re}\left[{\cal J}\right]\,;
\ee
\be 
\begin{split}
\label{TTS_J}
{\cal J}=& i \,X\, \plm^2\, H^2 \, \epsilon \,D_{lm}(k,k',k'')\,\\
&\sum_{m=1}^2 h_{k_1}h_{k_2} \zeta_n{}_{k_3}^{(m)}\mid_{t_e}\, \sum_{r,s}\,\varepsilon_{im}^{(r)}{}_k \;\varepsilon_{il}^{(s)}{}_{k'}\, \delta_j^{m}\;\int_{-\infty}^{t_e} dt' \, a^{n}\,  h^{*}_{k}\,h^{*}_{k'}\, \xi_{\omega}{}^{(j)\,*}_{k''}|_{t'}\\
& \delta(\textbf k_3+\textbf k'')\,\left[\delta_{s,r}^{p,q} \, \delta(\textbf k_1+\textbf k')\delta(\textbf k_2+\textbf k)+ \delta_{r,s}^{p,q}\, \delta(\textbf k_1+\textbf k)\delta(\textbf k_2+\textbf k') \right]\,.
\end{split} 
\ee

\item \textbf{TTT}:\\
vertex of the form 
\be
{\cal L}_{k\,k'\,k''}= X\, \,\plm^2\,H^2\,\epsilon\, a^3 \,  h_{ij \, k} \,h_{jl \, k'} \,h_{li \, k''}\,,
\ee
which leads to:
\be
\label{TTT}
{\cal B}_{h^0 h^p h^q}= \int dk^3 dk'^3 dk''^3 \,{\cal I}\,, \qquad {\cal I}= 2 \, \text{Re}\left[{\cal J}\right]\,;
\ee
\be
\begin{split}
\label{TTT_J}
{\cal J}=&i \, X \, \plm^2 \, H^2\,\epsilon\,\\
&\sum_{i=1}^2 h_{k_1}\,h_{k_2}\,h_{k_3}\mid_{t_e} \sum_{r,s,u}\,\varepsilon_{ij}^{(r)}{}_k\,\varepsilon_{jl}^{(s)}{}_{k'}\,\varepsilon_{li}^{(u)}{}_{k''} \,\int_{-\infty}^{t_e} dt'  \, a^{n}\,  h^{*}_{k}\,h^{*}_{k'}\, h^{*}_{k''}|_{t'} \, \cdot\\
&\;\; \cdot \,\Big\{ \delta(\textbf k_3+\textbf k'')\,\left[\delta_{u,s,r}^{o,p,q}\, \delta(\textbf k_1+\textbf k')\delta(\textbf k_2+\textbf k)+ \delta_{u,r,s}^{o,p,q}\, \delta(\textbf k_1+\textbf k)\delta(\textbf k_2+\textbf k') \right] \\
&\;\;+\delta(\textbf k_3+\textbf k')\,\left[\delta_{s,u,r}^{o,p,q}\, \delta(\textbf k_1+\textbf k'')\delta(\textbf k_2+\textbf k)+ \delta_{s,r,u}^{o,p,q}\, \delta(\textbf k_1+\textbf k)\delta(\textbf k_2+\textbf k'') \right] \\
&\;\;+\delta(\textbf k_3+\textbf k)\,\left[\delta_{r,u,s}^{o,p,q}\, \delta(\textbf k_1+\textbf k'')\delta(\textbf k_2+\textbf k')+ \delta_{r,s,u}^{o,p,q}\, \delta(\textbf k_1+\textbf k')\delta(\textbf k_2+\textbf k'') \right]\Big\}\,.
\end{split} 
\ee
\end{enumerate}
The results presented can be obtained by evaluating the above integrals.
In the  cases $c_b^2=-1,\,0$, the integral ${\cal J}$  can be
computed analytically, while when $c_b^2 \in (0,\,1)$ a numerical approach is needed.

\section{Squeezed SSS \texorpdfstring{$f_{\text{NL}}$}{Lg}}
\label{f_nl_sq}

In this appendix we give the full expressions for
  $f_{\text{NL}}$ in a
  squeezed configuration.
Among ten vertices, only four give an independent contribution to $f_{\text{NL}}$.
The contribution from each  independent operator ${\cal O}$ of (\ref{L3ZR}) to
$f_{\text{NL}}$ can be split in an  overall amplitude ${\cal M}$ times
a momentum-independent part wich gives
the monopole and the quadrupole structure according with
\bea\label{decfnl}
f_{\text{NL}}^{({\cal O})}= ({X}_{1}+{X}_{2}\, \cos^2\theta)\;{\cal M}^{({\cal O})}
=(X_{\cal M}+X_{\cal Q}\,Y_2^0)\;{\cal M}^{({\cal O})} \, .
\ea
It is useful to define the following recurrent combinations  in
$f_{\text{NL}}$ of the diagonal sound speeds
\be
{\cal C}_1= c_{s 1}+c_{s2}\,, 
\ee
\be
{\cal C}_2= c_{s1}^2+c_{s1}c_{s2}+c_{s2}^2\,, 
\ee
\be
{\cal C}_4= c_{s1}^4+c_{s1}^3 c_{s2}+c_{s1}^2 c_{s2}^2+c_{s1} c_{s2}^3+c_{s2}^4\,.
\ee
In the case of $c_b^2=-1$, at the leading order in SR, $\zeta_n$ and ${\cal R}_{\pi_0}$ fields have analytic solutions of the form 
\begin{equation}
\label{an_sol}
\begin{split}
& \zeta_{n}^{(l)} \sim k^{-\frac{3}{2}}\,\left(k\, t\right)^{\frac{5}{2}}\, H_{\frac{5}{2}}^{(1)}(-k\,t\, c_{sl})\,, \\
& {\cal R}_{\pi_0}^{(l)} \sim k^{-\frac{3}{2}}\,\left(k\, t\right)^{\frac{3}{2}}\, H_{\frac{3}{2}}^{(1)}(-k\,t\, c_{sl})\,, \qquad l=1,2\,; \\ 
\end{split}
\end{equation}
used to evaluate the in-in time integrals.\\
We choose the following four independent vertices for their simplicity
\begin{itemize}
\item $\zeta_n^3$ vertex
\be
f_{\text{NL}}^{(\zeta_n^3)}=
\left[V_{3}+3\,V_{1}+(3 \, V_{2}+2\,V_{3})\, \cos\left(\theta\right)^2\right]\, {\cal M}^{(\zeta_n^3)}\,,
\ee
\be
\begin{split}
& {\cal M}^{(\zeta_n^3)}=
\frac{5}{16}\,\frac{\left(r_{\zeta_n}^{(1)}+r_{\zeta_n}^{(2)}\right)}{27\,{\cal
    C}_1\,\hat{c}_L^{5}}\left(5\,{\cal C}_1\,\left( c_{s 1}^3\,
    r_{\zeta_n}^{(1)}{}^2+c_{s 2}^3\, r_{\zeta_n}^{(2)}{}^2\right)+4
  \,r_{\zeta_n}^{(1)}\,r_{\zeta_n}^{(2)}\,{\cal C}_4\right) \, .
\end{split}
\ee
\item $\zeta_n'{}^2\, \zeta_n$ vertex
\be
f_{\text{NL}}^{(\zeta_n'{}^2\, \zeta_n)}=  \,\left[V_{4}+V_{5}\, \cos\left(\theta\right)^2  \right]\, {\cal M}^{(\zeta_n'{}^2\, \zeta_n)}\,, 
\ee
\be
\begin{split}
&{\cal M}^{(\zeta_n'{}^2\, \zeta_n)}=-\frac{5}{16}\,
\frac{\left(r_{\zeta_n}^{(1)}+r_{\zeta_n}^{(2)}\right)}{27\,{\cal
    C}_1\,\hat{c}_L^{5}}\left(3\,{\cal C}_1\,\left( c_{s 1}^5\,
    r_{\zeta_n}^{(1)}{}^2+c_{s 2}^5\, r_{\zeta_n}^{(2)}{}^2\right)+4
  \,r_{\zeta_n}^{(1)}\,r_{\zeta_n}^{(2)}\,{\cal C}_2 \, c_{s 1}^2\,
  c_{s 2}^2\right) \, .
\end{split}
\ee
\item $\zeta_n \,\zeta_n'\, {\cal R}_{\pi_0}$ vertex
\be
f_{\text{NL}}^{(\zeta_n \,\zeta_n'\, {\cal R}_{\pi_0})}= \,\left[B_{1}+B_{2}\, \cos\left(\theta\right)^2\right]\, {\cal M}^{(\zeta_n \,\zeta_n'\, {\cal R}_{\pi_0})}\,, 
\ee
\be
\begin{split}
&{\cal M}^{(\zeta_n \,\zeta_n'\, {\cal R}_{\pi_0})}=\frac{5}{16}\,\frac{\left(r_{\zeta_n}^{(1)}+r_{\zeta_n}^{(2)}\right)}{9\,{\cal C}_1\,\hat{c}_L^{5}}\left(3\, c_{s,\,1}^3\,{\cal C}_1\, r_{L/0}^{(1)}\,r_{\zeta_n}^{(1)}+2 \,r_{L/0}^{(1)}\,r_{\zeta_n}^{(2)}\,c_{s\,2}^2\,{\cal C}_2 \right)+\left[ c_{s\,1} \to c_{s\,2} \right]\,.
\end{split} 
\ee
\item $\zeta_n \, {\cal R}_{\pi_0}^2$ vertex
\be
f_{\text{NL}}^{(\zeta_n \, {\cal R}_{\pi_0}^2)}= \,\left[G_{1}+G_{2}\, \cos\left(\theta\right)^2\right]\, {\cal M}^{(\zeta_n \, {\cal R}_{\pi_0}^2)}\,, 
\ee
\be
\begin{split}
&{\cal M}^{(\zeta_n \, {\cal R}_{\pi_0}^2)}=-\frac{5}{16}\,\frac{\left(r_{\zeta_n}^{(1)}+r_{\zeta_n}^{(2)}\right)}{3\,{\cal C}_1\,\hat{c}_L^{5}}\left(3\, c_{s,\,1}\,{\cal C}_1\, r_{L/0}^{(1)}{}^2+2 \,r_{L/0}^{(1)}\,r_{L/0}^{(2)}\,{\cal C}_2 \right)+\left[ c_{s\,1} \to c_{s\,2} \right]\,.
\end{split}
\ee
\end{itemize}
The remaining vertices can be written as follows
\begin{itemize}
\item $\tilde {\cal R}_{\pi_0}' \, \zeta_n'^2$ vertex
\be
f_{\text{NL}}^{(\tilde {\cal R}_{\pi_0}'   \zeta_n'^2)}=B_{3} \,
{\cal M}^{(\tilde {\cal R}_{\pi_0}' \, \zeta_n'^2)} \,, \qquad {\cal
  M}^{(\tilde {\cal R}_{\pi_0}'   \zeta_n'^2)}=3\,\,
\frac{\left(r_{L/0}^{(1)}+r_{L/0}^{(2)}\right)}{\left(r_{\zeta_n}^{(1)}+r_{\zeta_n}^{(2)}\right)}\;{\cal
  M}^{(\zeta_n'{}^2  \zeta_n)} \,.
\ee
\item $\tilde {\cal R}_{\pi_0}'{}^3$ vertex
\be
f_{\text{NL}}^{(\tilde {\cal R}_{\pi_0}'{}^3)}= O_{1} \,  {\cal M}^{(\tilde {\cal R}_{\pi_0}'{}^3)} \,, \qquad {\cal M}^{(\tilde {\cal R}_{\pi_0}'{}^3)}=81\, {\cal M}^{(\zeta_n^3)} \;\; \left[r_{\zeta_n} \to r_{L/0}\right]\,. 
\ee
\item $\tilde {\cal R}_{\pi_0}'{}^2 \,\zeta_n$ vertex
\be
\begin{split}
& f_{\text{NL}}^{(\tilde {\cal R}_{\pi_0}'{}^2 \,\zeta_n)} =   \,O_{2} \,   {\cal M}^{(\tilde {\cal R}_{\pi_0}'{}^2 \,\zeta_n)} \,,\\
&{\cal M}^{(\tilde {\cal R}_{\pi_0}'{}^2 \,\zeta_n)} =27\,{\cal M}^{(\zeta_n^3)} \\
& \qquad\qquad\qquad \left[r_{\zeta_n}^{(i)}{}^3 \to r_{L/0}^{(i)}{}^2 \, r_{\zeta_n}^{(i)}\,,\;\; r_{\zeta_n}^{(i)}{}^2\,r_{\zeta_n}^{(j)} \to \frac{1}{3} \,r_{L/0}^{(i)}{}^2 \, r_{\zeta_n}^{(j)}+\frac{2}{3}\, r_{L/0}^{(i)}\, r_{L/0}^{(j)} \, r_{\zeta_n}^{(i)}\right]\,.
\end{split}
\ee
\item $\tilde {\cal R}_{\pi_0}'\, \zeta_n' \,{\cal R}_{\pi_0}$ vertex
\be
f_{\text{NL}}^{(\tilde {\cal R}_{\pi_0}'\, \zeta_n' \,{\cal R}_{\pi_0})}=O_{3} \,   {\cal M}^{(\tilde {\cal R}_{\pi_0}'\, \zeta_n' \,{\cal R}_{\pi_0})} \,, \qquad {\cal M}^{(\tilde {\cal R}_{\pi_0}'\, \zeta_n' \,{\cal R}_{\pi_0})}=3\,\, \frac{\left(r_{L/0}^{(1)}+r_{L/0}^{(2)}\right)}{\left(r_{\zeta_n}^{(1)}+r_{\zeta_n}^{(2)}\right)}\;{\cal M}^{(\zeta_n \,\zeta_n'\, {\cal R}_{\pi_0})}.
\ee
\item $\tilde {\cal R}_{\pi_0}'\, {\cal R}_{\pi_0}^2$ vertex
\be
f_{\text{NL}}^{(\tilde {\cal R}_{\pi_0}'\, {\cal R}_{\pi_0}^2)}=R \,
{\cal M}^{(\tilde {\cal R}_{\pi_0}'\, {\cal R}_{\pi_0}^2)}\,, \qquad
{\cal M}^{(\tilde {\cal R}_{\pi_0}'\, {\cal R}_{\pi_0}^2)}=3\,\,
\frac{\left(r_{L/0}^{(1)}+r_{L/0}^{(2)}\right)}{\left(r_{\zeta_n}^{(1)}+r_{\zeta_n}^{(2)}\right)}\;{\cal
  M}^{(\zeta_n \, {\cal R}_{\pi_0}^2)} \, .
\ee
\end{itemize}

\section{Equilateral TSS \texorpdfstring{$f_{\text{NL}}$}{Lg}}
\label{TSS_total}

In this appendix we give the complete expressions for  $f_{TSS}$  in
the TSS sector for an equilateral configuration.
\begin{itemize}
\item $h\,\zeta_n^2$ vertex
\be
f_{TSS}= \sum_{l\,m} \hat{c}_L^{-5}\,r_{\zeta_n}^{(l)}\,r_{\zeta_n}^{(m)} \, {\cal I}_{\zeta_n^2}\,,
\ee
\be 
\begin{split}
{\cal I}_{\zeta_n^2}=&\frac{2\,V^{(tss)}_{1}-V^{(tss)}_{2}}{48\, \sqrt{2}\,(1+c_{sl}+c_{sm})^2}\, \left[ c_{sl}^5+2\, c_{sl}^4\, (1+c_{sm})+2 \, c_{sl}^3\,(2+c_{sm}+c_{sm}^2) \right.\\ 
  &\left.  \qquad\qquad+c_{sl}\,(1+c_{sm})(11+6\,c_{sm}+2\,c_{sm}^3-6\, \gamma_e) \right.\\ 
  &\left. \qquad \qquad+(1+c_{sm})^2 (4+3\,c_{sm}+c_{sm}^3-3\, \gamma_e)\right.\\ 
  &\left. \qquad \qquad+c_{sl}^2 \, \left[10+2\, c_{sm} (3+c_{sm}+c_{sm}^2)-3 \, \gamma_e \right]\right.\\ 
  &\left. \qquad \qquad-3 \, (1+c_{sl}+c_{sm})^2 \,\log\left( -(1+c_{sl}+c_{sm})\,k\, t \right)\right] \,.
\end{split}
\ee
\item $h\,\zeta_n'^2$ vertex
\be
f_{TSS}= \sum_{l\,m} \hat{c}_L^{-5}\,r_{\zeta_n}^{(l)}\,r_{\zeta_n}^{(m)} \, {\cal I}_{\zeta_n'^2}\,,
\ee
\be 
\begin{split}
{\cal I}_{\zeta_n'^2}=-V^{(tss)}_{3}& \frac{c_{sl}^2\,c_{sm}^2}{24\,\sqrt{2}\,(1+c_{sl}+c_{sm})^2}\,\left[c_{sl}^3+2\, c_{sl}^2\,(1+c_{sm})\right.\\ 
  &\left. +2\, c_{sl}\,(1+c_{sm}+c_{sm}^2)+(1+c_{sm})(1+c_{sm}+c_{sm}^2)\right]\,.
\end{split}
\ee

\item $h\,\zeta_n'\,{\cal R}_{\pi_0}$ vertex
\be
f_{TSS}= \sum_{l\,m} \hat{c}_L^{-5}\,r_{\zeta_n}^{(l)}\,r_{L/0}^{(m)} \, {\cal I}_{\zeta_n' {\cal R}_{\pi_0}}\,,
\ee
\be 
\begin{split}
{\cal I}_{\zeta_n' {\cal R}_{\pi_0}}=B^{(tss)}_1& \frac{c_{sm}^2}{8\,\sqrt{2}\,(1+c_{sl}+c_{sm})^2}\,\left[c_{sl}^3+2\, c_{sl}^2\,(1+c_{sm})\right.\\ 
  &\left. +2\, c_{sl}\,(1+c_{sm}+c_{sm}^2)+(1+c_{sm})(1+c_{sm}+c_{sm}^2)\right]\,.
\end{split}
\ee

\item $h\,{\cal R}_{\pi_0}'\,\zeta_n$ vertex
\be
f_{TSS}= \sum_{l\,m} \hat{c}_L^{-5}\,r_{\zeta_n}^{(l)}\,r_{L/0}^{(m)} \, {\cal I}_{{\cal R}_{\pi_0}'\,\zeta_n}\,,
\ee
\be 
\begin{split}
{\cal I}_{{\cal R}_{\pi_0}'\,\zeta_n}=&\frac{-B^{(tss)}_2}{8\, \sqrt{2}\,(1+c_{sl}+c_{sm})^2}\, \left[ c_{sl}^5+2\, c_{sl}^4\, (1+c_{sm})+2 \, c_{sl}^3\,(2+c_{sm}+c_{sm}^2) \right.\\ 
  &\left.  \qquad\qquad+c_{sl}\,(1+c_{sm})(11+6\,c_{sm}+2\,c_{sm}^3-6\, \gamma_e) \right.\\ 
  &\left. \qquad \qquad+(1+c_{sm})^2 (4+3\,c_{sm}+c_{sm}^3-3\, \gamma_e)\right.\\ 
  &\left. \qquad \qquad+c_{sl}^2 \, \left[10+2\, c_{sm} (3+c_{sm}+c_{sm}^2)-3 \, \gamma_e \right]\right.\\ 
  &\left. \qquad \qquad-3 \, (1+c_{sl}+c_{sm})^2 \,\log\left( -(1+c_{sl}+c_{sm})\,k\, t \right)\right] \,.
\end{split}
\ee
\item $h\,{\cal R}_{\pi_0}{}^2$ vertex
\be
f_{TSS}= \sum_{l\,m} \hat{c}_L^{-5}\,r_{L/0}^{(l)}\,r_{L/0}^{(m)} \, {\cal I}_{{\cal R}_{\pi_0}^2}\,,
\ee
\be 
\begin{split}
{\cal I}_{{\cal R}_{\pi_0}^2}= -3\,&\frac{G^{(tss)}}{8\,\sqrt{2}\,(1+c_{sl}+c_{sm})^2}\,\left[c_{sl}^3+2\, c_{sl}^2\,(1+c_{sm})\right.\\ 
  &\left. +2\, c_{sl}\,(1+c_{sm}+c_{sm}^2)+(1+c_{sm})(1+c_{sm}+c_{sm}^2)\right]\,.
\end{split}
\ee
\end{itemize}

\section{Supersolid Lagrangian: an example}
\label{toy_model}
 A simple way to implement the consistency relations ${M}'_i=0$ is to
use the Lagrangian (\ref{toy-Lagrangian})
\be
U= -6\,H^2+\epsilon_0\;V+\epsilon_0\; U_{\text{$\tilde\Lambda$}} \,;
\label{Lmod}
\ee   
$V=-4\;H^2\,\log(b)$ is the Lagrangian of a perfect fluid needed to have $\epsilon$ approximately constant, while $U_{\text{$\tilde\Lambda$}}$ is a background-$\Lambda$-Media (\ref{toy-Lagrangian}) that automatically satisfies all the constraints (\ref{dep_par}).
The reason of this choice is quite simple. The parameters entering in
the quadratic Lagrangian are given by
\be
\begin{split}
& c_0^2=c_4^2=\frac{1}{8}\,\left(
U_{\tilde \Lambda,\,{\cal X}{\cal X}}+2\,
U_{\tilde\Lambda,\,{\cal X}{\cal Y}}+
U_{\tilde\Lambda,\,{\cal Y}{\cal Y}}\right)\,,\\
& c_1^2=-\frac{1}{4}\,U_{\tilde\Lambda,\,{\cal X}}\,,\\
& c_2^2=-\frac{1}{9}\,\left(U_{\tilde\Lambda,\,\tau_Y}+U_{\tilde\Lambda,\,\tau_Z}+
U_{\tilde\Lambda,\,w_Y}+U_{\tilde\Lambda,\,w_Z}\right)\,,\\
& c_3^2=\frac{1}{2}+\frac{2}{3}c_2^2+c_0^2\,,\\
&c_b^2=-1\,,\qquad 
\hat\sigma= ( U_{\tilde\Lambda,\,{\cal Y}}+U_{\tilde\Lambda,\,{\cal X}} )\,.
\end{split} 
\ee
For the independent parameters in the cubic Lagrangian we get 
\be
\begin{split}
&\lambda_2=U_{\tilde\Lambda,\,{\cal X}{\cal X}{\cal X}}+3\, U_{\tilde\Lambda,\,{\cal X}{\cal Y}{\cal Y}}+3\, U_{\tilde\Lambda,\,{\cal X}{\cal X}{\cal Y}}+U_{\tilde\Lambda,\,{\cal Y}{\cal Y}{\cal Y}} \,, \\
&\lambda_6= -\partial_{\cal X} c_2^2-\partial_{\cal Y} c_2^2\,, \\
&\lambda_7= \frac{1}{9}\, \left( U_{\tilde\Lambda,\,\tau_Z}+U_{\tilde\Lambda,\,w_Z}\right)\,, \\
&\lambda_9= \frac{1}{9}\, \left( U_{\tilde\Lambda,\,w_Y}+U_{\tilde\Lambda,\,w_Z}\right)\,, \\
&\lambda_{10}=-U_{\tilde\Lambda,\,{\cal Y}{\cal Y}}-U_{\tilde\Lambda,\,{\cal X}{\cal Y}} \,.
\end{split} 
\ee
Thus, (\ref{Lmod}) indeed satisfies all the requirements for a
consistent slow-roll dynamics.

\section{Boosting the tensor sector}
\label{boost}
\begin{figure}[H]
  \centering
    \includegraphics[width=10.cm]{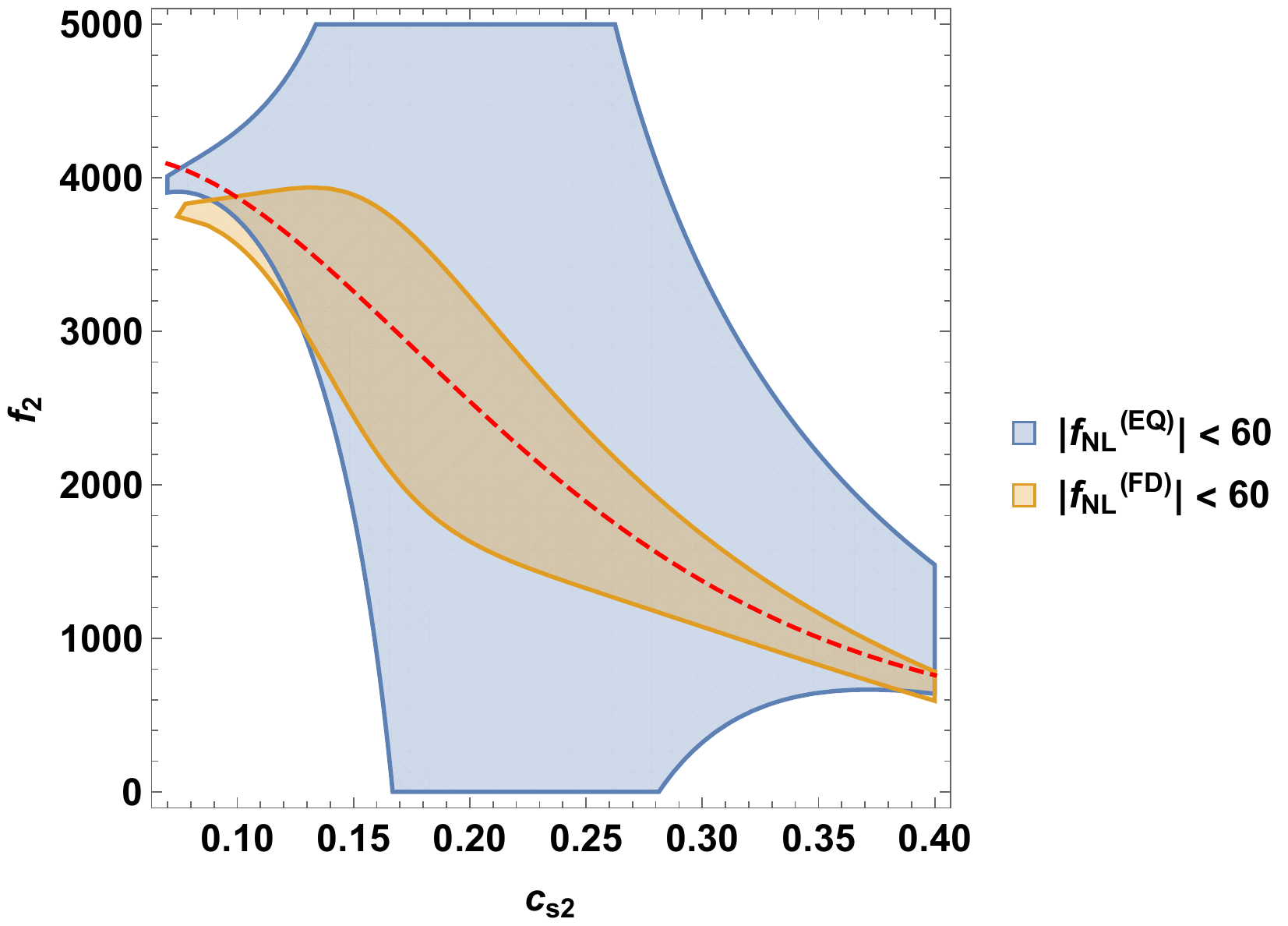}
\caption[Intersection region: $f_{\text{NL}}^{(EQ)},\, f_{\text{NL}}^{(FD)} \le 60$]{A plot showing the existence of a common region of small sound speed $c_{s2}$ with equilateral and folded shapes both within the CMB constrains. The red-dashed line represents the $f_2$ function in eq. (\ref{eqf}).}
\label{par_con}
\end{figure}

In this appendix we discuss the parameterization used in section
\ref{phen} to keep  the SSS sector under control. 
Figure \ref{tune_par} was obtained by the following choice for the five independent parameters $\{\lambda_2,
\lambda_6, \lambda_7, \lambda_9,\lambda_{10}\}$  in the cubic Lagrangian 
\be
\begin{split}
\label{lambda_5}
& \lambda_2 \approx \left(-8.91+2\,10^{-3}\, f_2\right)\,\hat{c}_{s2}+\left(0.26+7\,10^{-5}\, f_2\right)\,\hat{c}_{s2}^3+ 10^{-4}\,f_2 \,\hat{c}_{s2}^4\,,\\
& \\
& \lambda_6 \approx -3.75\,\hat{c}_{s2}^{-2}+.06+0.41\,\hat{c}_{s2}+10^{-3}\,\hat{c}_{s2}^2\,,\\
& \\
& \lambda_{10}=\left(-0.04-2\,
  10^{-5}\,f_2\right)\,\hat{c}_{s2}^3+\left(1.2 \, 10^{-3}+2\,
  10^{-7}\,f_2\right)\,\hat{c}_{s2}^5+\left(4\,10^{-4}-3.4\,
  10^{-8}\,f_2\right)\,\hat{c}_{s2}^8\, ,
\end{split} 
\ee 
while 
\be
\lambda_7\approx-2.05\,, \qquad \lambda_9=0.05\, .
\ee
The function  $f_2$ corresponds to the dashed-red line shown in figure \ref{par_con} 
\begin{equation}
\log(f_2)\approx 8.23+0.28\, \hat c_{s2}-0.32\, \hat{c}_{s2}^2+0.04 \,
c_{s2}^3\,.
\label{f2form}
\end{equation}
One could have expanded $f_2$ in powers, rewriting (\ref{lambda_5}) as
a Laurent series, however (\ref{f2form}) is more compact. We chose this function, considering the intersections of the regions $|f_{\text{NL}}^{(FD)}|,\, |f_{\text{NL}}^{(EQ)}|\leq 60$ in order to get an almost violet squeezed $f_{\text{NL}}$ without exceeding the equilateral/folded experimental constraints, see figure \ref{par_con}. This is only one possibility to get the squeezed $f_{\text{NL}}$ of order $c_{s2}^0$ and within the experimental constraints obtaining
\be\label{eqf}
f_{\text{NL}}^{(SQ)}\approx 2.3+ 9.5\, \cos(2\,\theta)-\left(6.9+5.5\,\cos(2\,\theta)\right)\, \hat c_{s2}+ -11.4\,\hat c_{s2}^2 +O(\hat c_{s2}^3)\,,
\ee
whose total plot is shown in figure \ref{sq_plot}, and allowing us to enhance the tensor PNG having access to a sufficiently small $c_{s2}$ values. 

\clearpage
\section{Bispectrum figures}
\label{pics}
%
In this section we give the amplitude ${\cal M}^{({\cal O})}$ (where ${\cal O}$ are the operators of the interactions terms) defined in (\ref{decfnl}):
\be
f_{\text{NL}}^{({\cal O})}= ({X}_{1}+{X}_{2}\, \cos^2\theta)\;{\cal M}^{({\cal O})}
=(X_{\cal M}+X_{\cal Q}\,Y_2^0)\;{\cal M}^{({\cal O})}\,.
\ee
Thus, for each $f_{\text{NL}}$ we plot ${\cal M}$ functions by setting $({X}_{1}+{X}_{2}\, \cos^2\theta)$ to one, and $c_b^2=-1$.
\begin{figure}[ht!]\centering
\includegraphics[width=7.3cm]{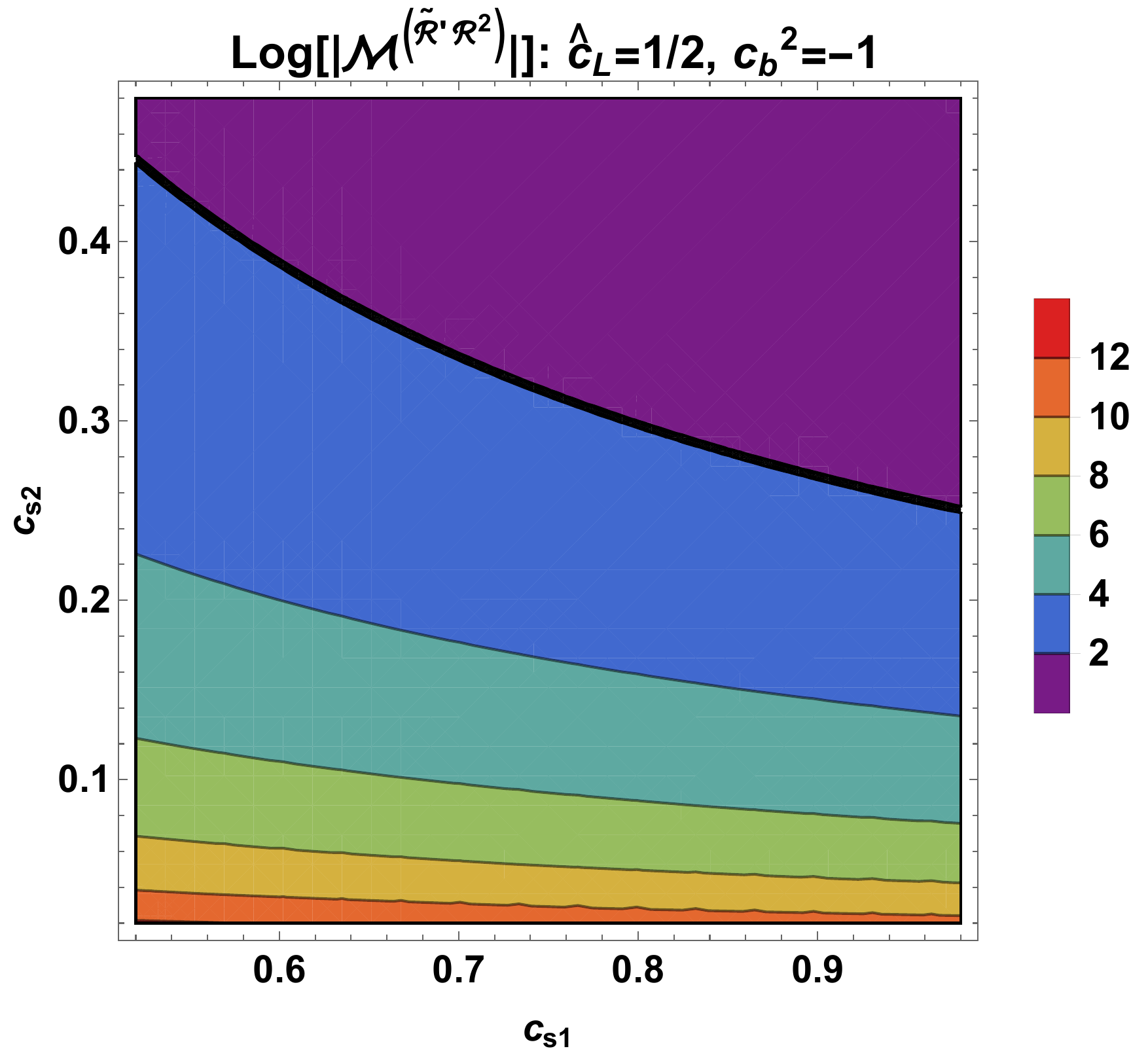}
\caption[Dominant SSS vertex: $f_{\text{NL}}$ squeezed]{Dominant squeezed SSS vertex: $Log_{{}_{10}}\left[\mid {\cal M}^{\tilde{{\cal R}}'{\cal R}^2} \mid \right]$. The black line delimits the region where ${\cal M}^{\tilde{{\cal R}}'{\cal R}^2} \sim 10^2$}
\label{0t00}
\end{figure}
\begin{figure}[!ht]
  \centering
   \begin{minipage}[b]{0.457\textwidth}
    \includegraphics[width=\textwidth]{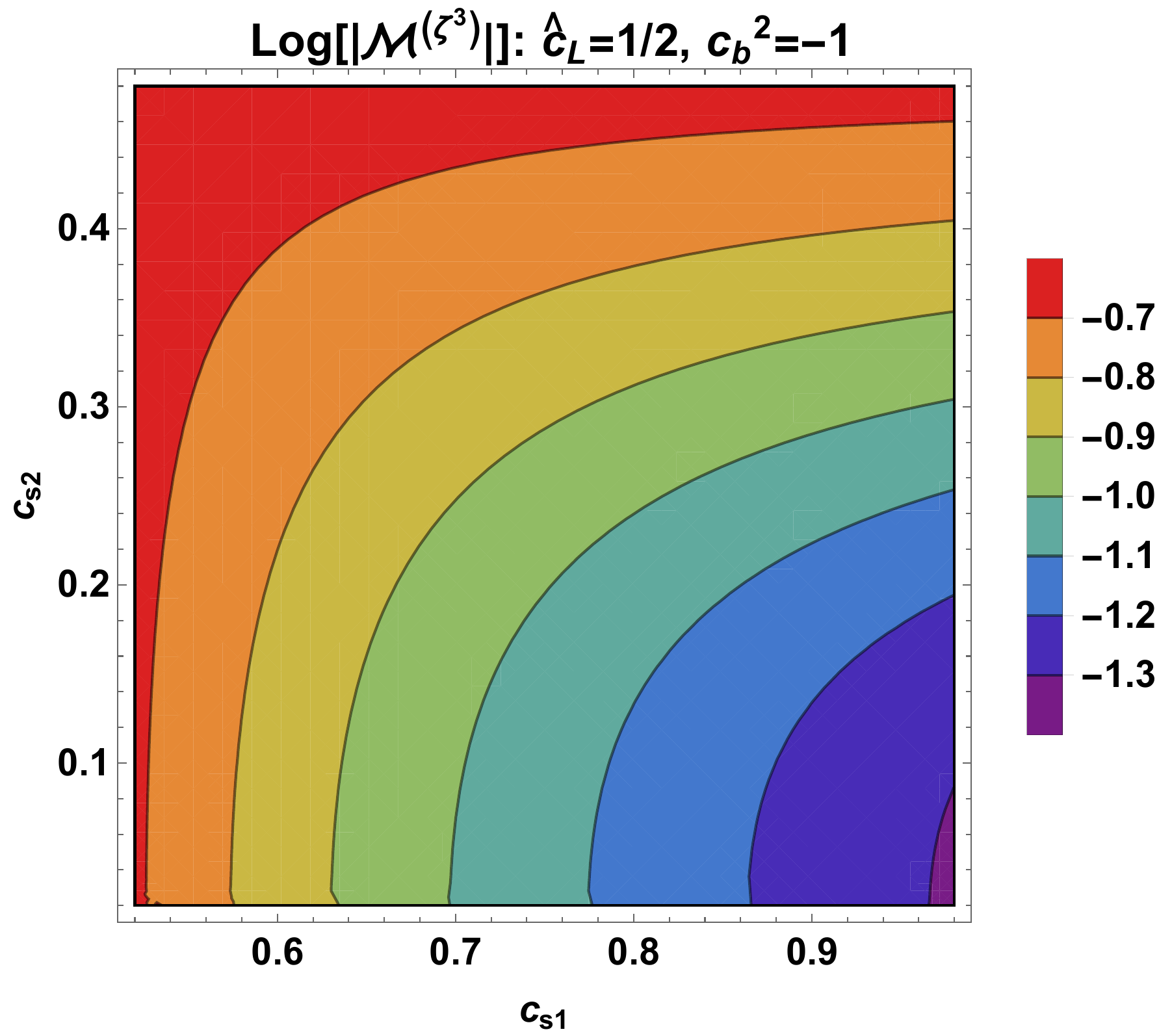}    
  \end{minipage}
  \begin{minipage}[b]{0.45\textwidth}
    \includegraphics[width=\textwidth]{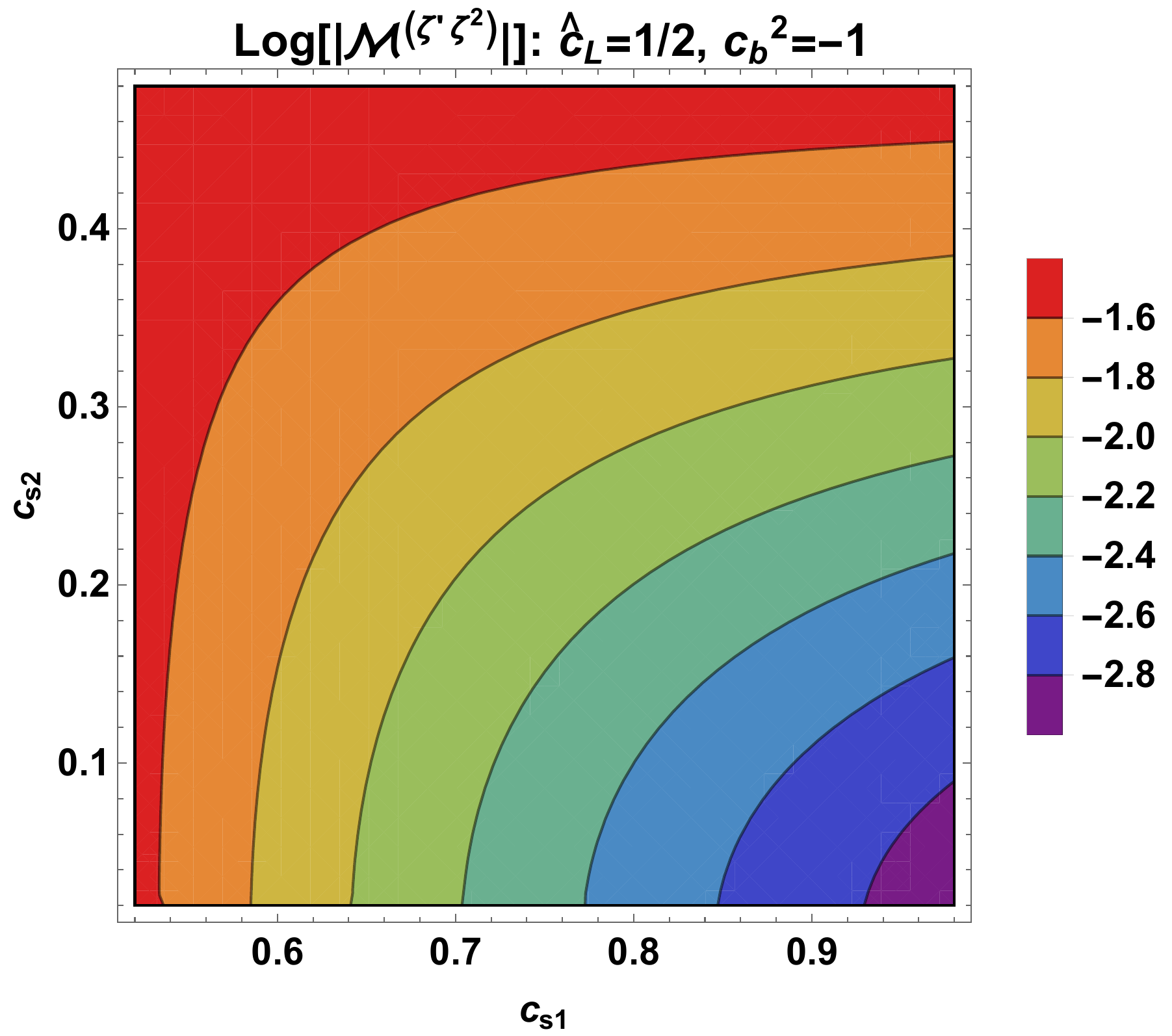}
  \end{minipage}
\caption[Violet vertices: $f_{\text{NL}}$ squeezed]{Squeezed $V$ vertices, $Log_{{}_{10}}\left[\mid {\cal M}^{\zeta_n^3,\,\zeta_n'\,\zeta_n^2} \mid \right]$.}
\label{V_plot}
\end{figure}
\begin{figure}[!tbp]
 \centering
  \begin{minipage}[b]{0.428\textwidth}
    \includegraphics[width=\textwidth]{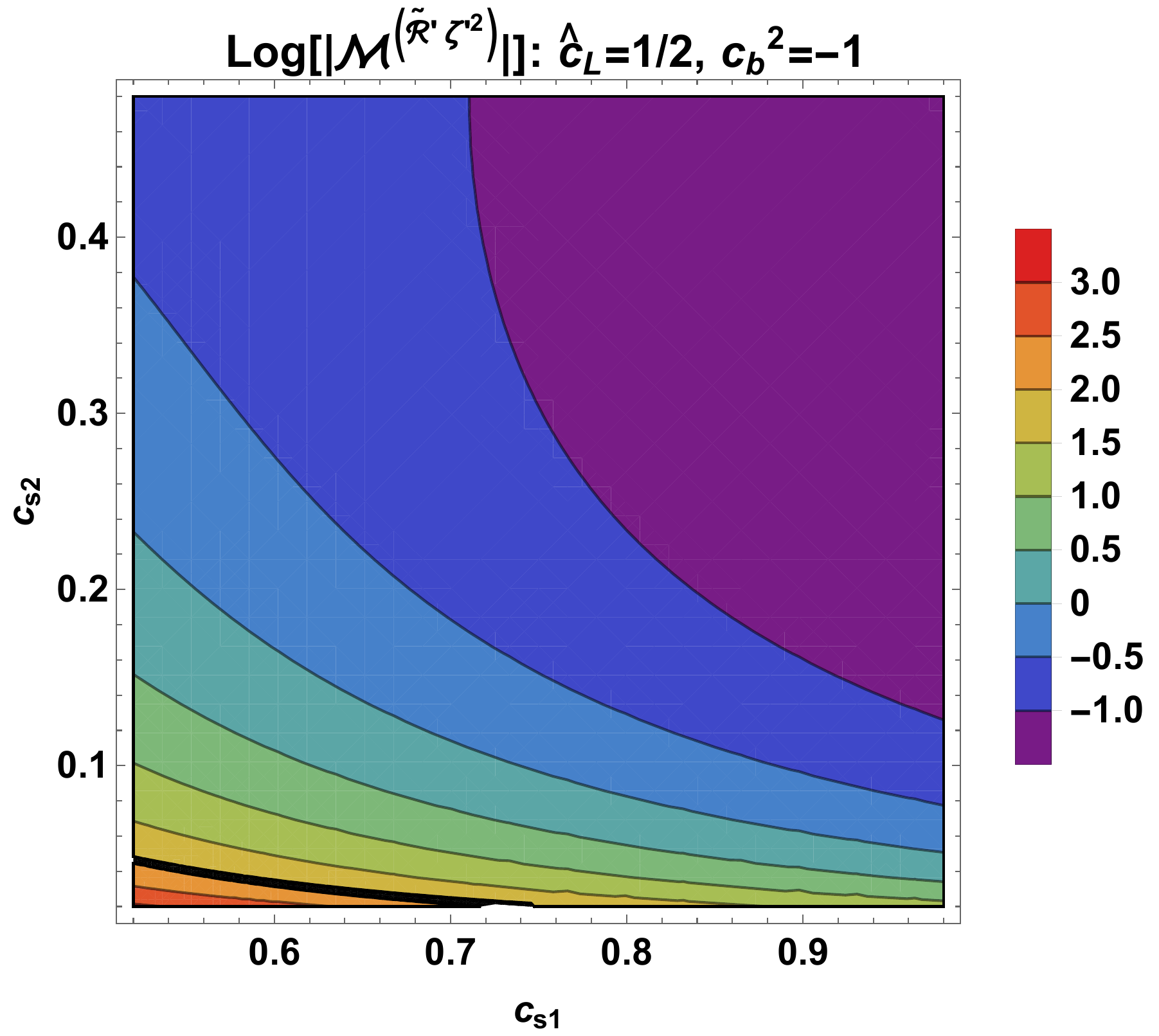}
     \end{minipage}
  \hspace{1cm}
  \begin{minipage}[b]{0.42\textwidth}
    \includegraphics[width=\textwidth]{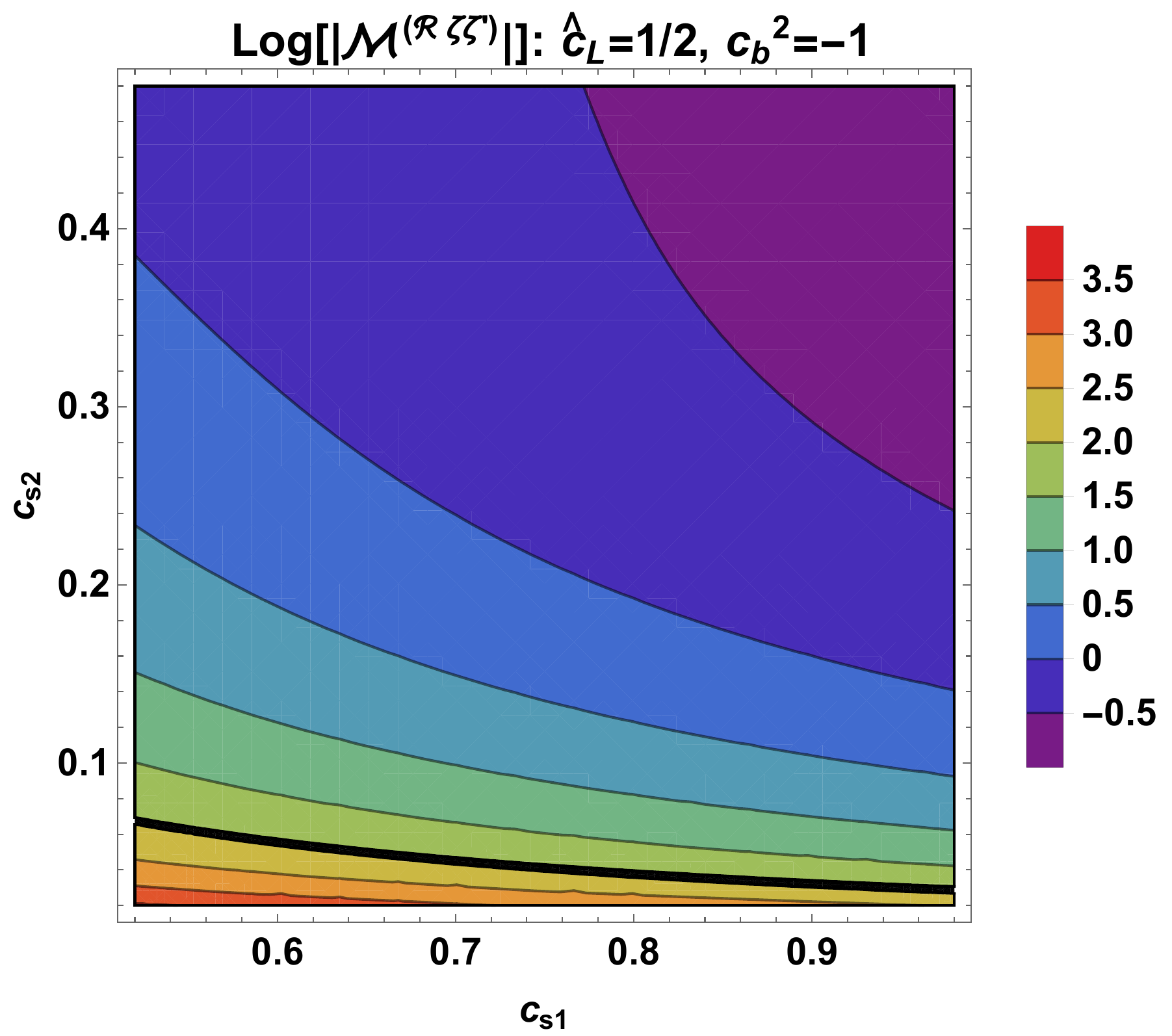}
  \end{minipage}
  \begin{minipage}[b]{0.428\textwidth}
    \includegraphics[width=\textwidth]{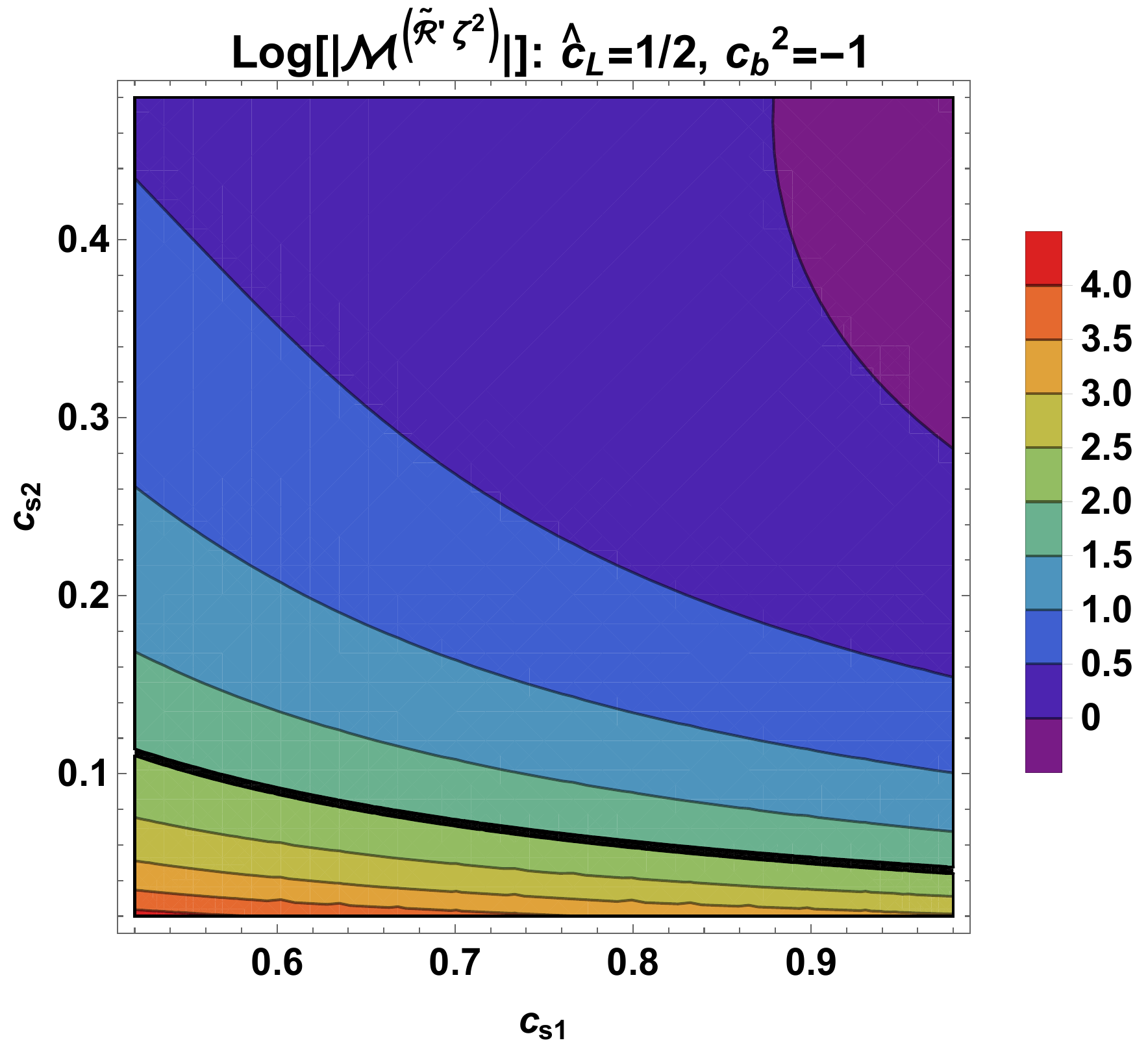}
  \end{minipage}
   \hspace{1cm}
   \begin{minipage}[b]{0.42\textwidth}
    \includegraphics[width=\textwidth]{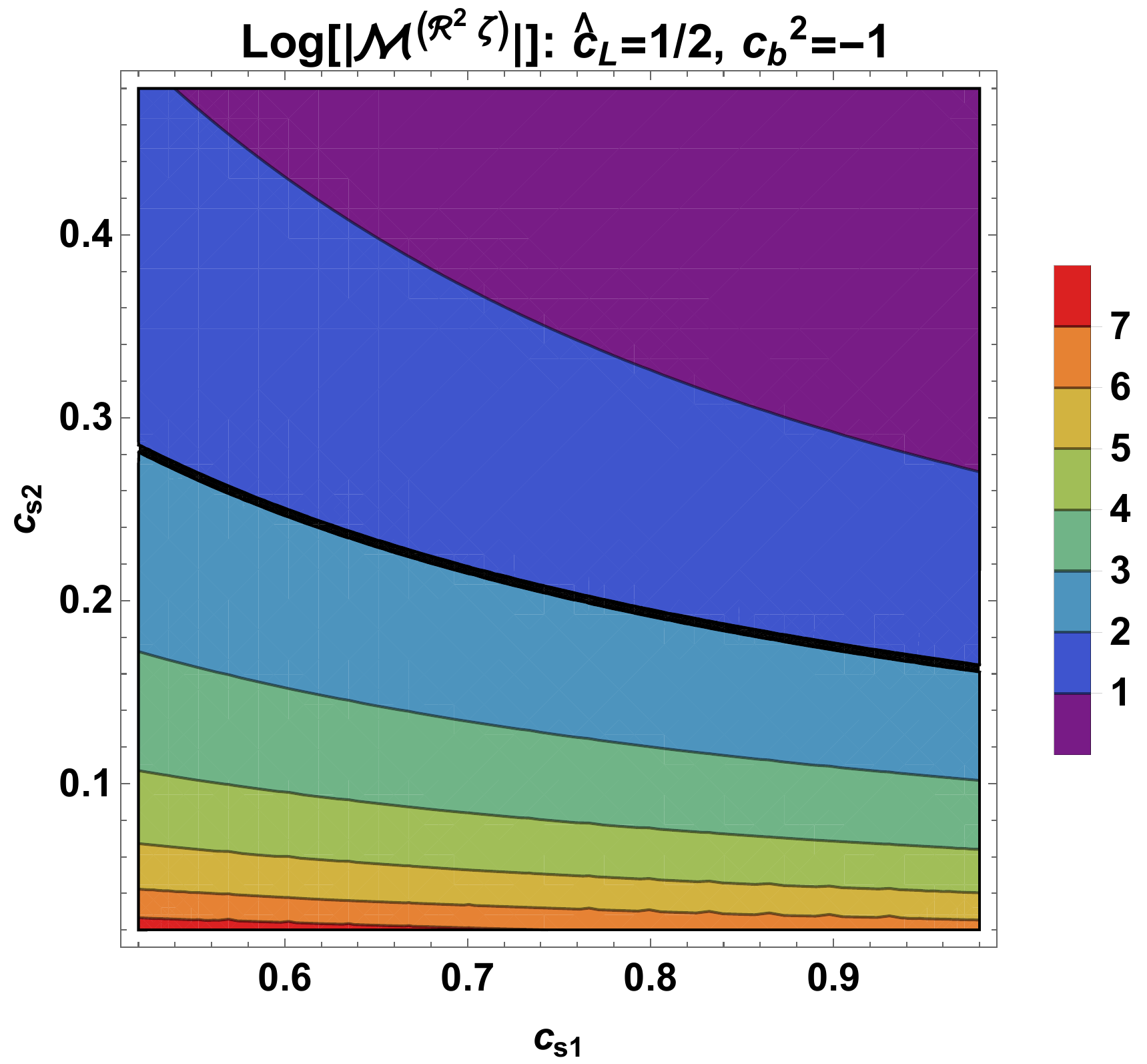}
  \end{minipage}
  \begin{minipage}[b]{0.42\textwidth}
    \includegraphics[width=\textwidth]{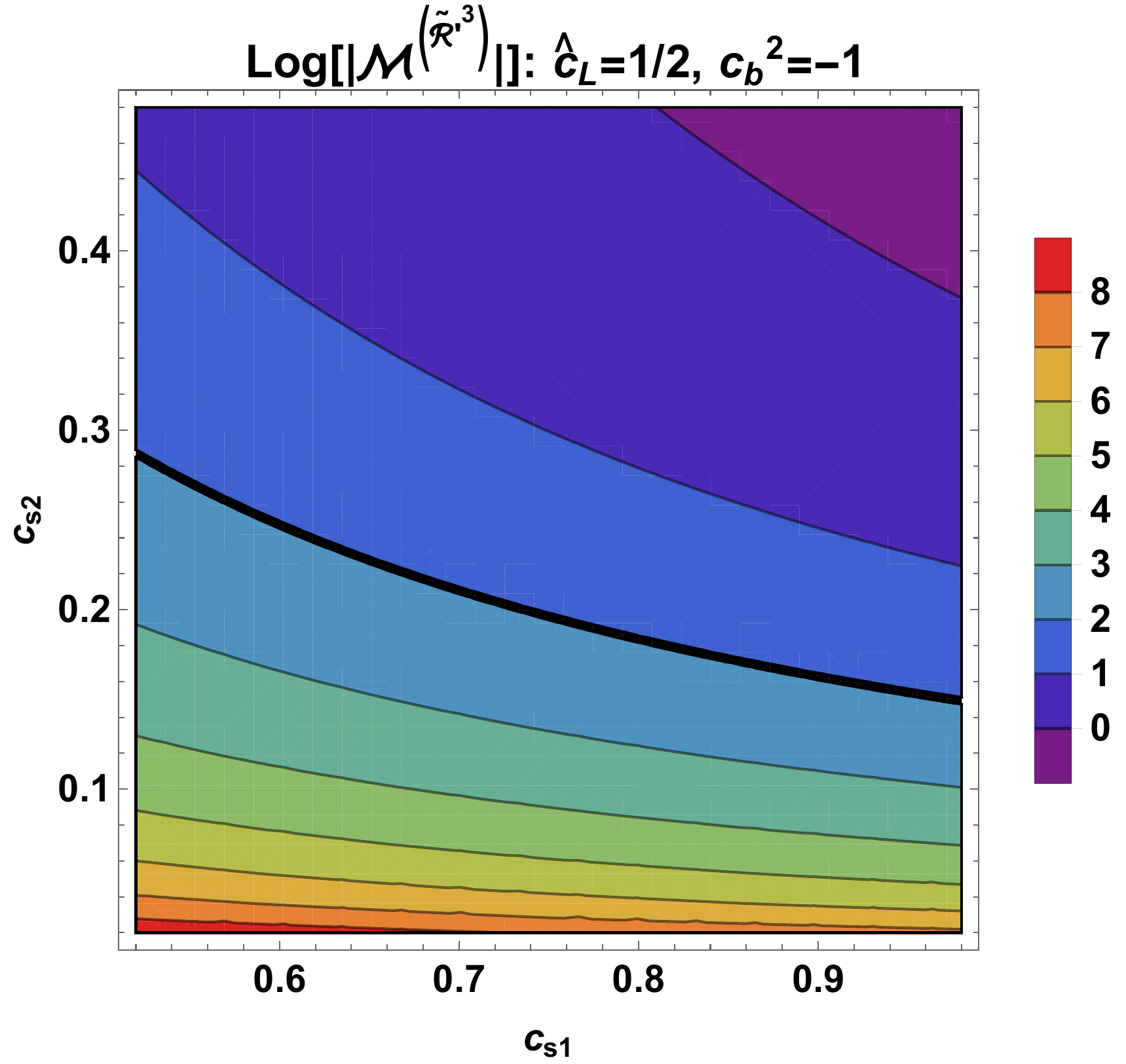}
  \end{minipage}
   \hspace{1cm}
   \begin{minipage}[b]{0.42\textwidth}
    \includegraphics[width=\textwidth]{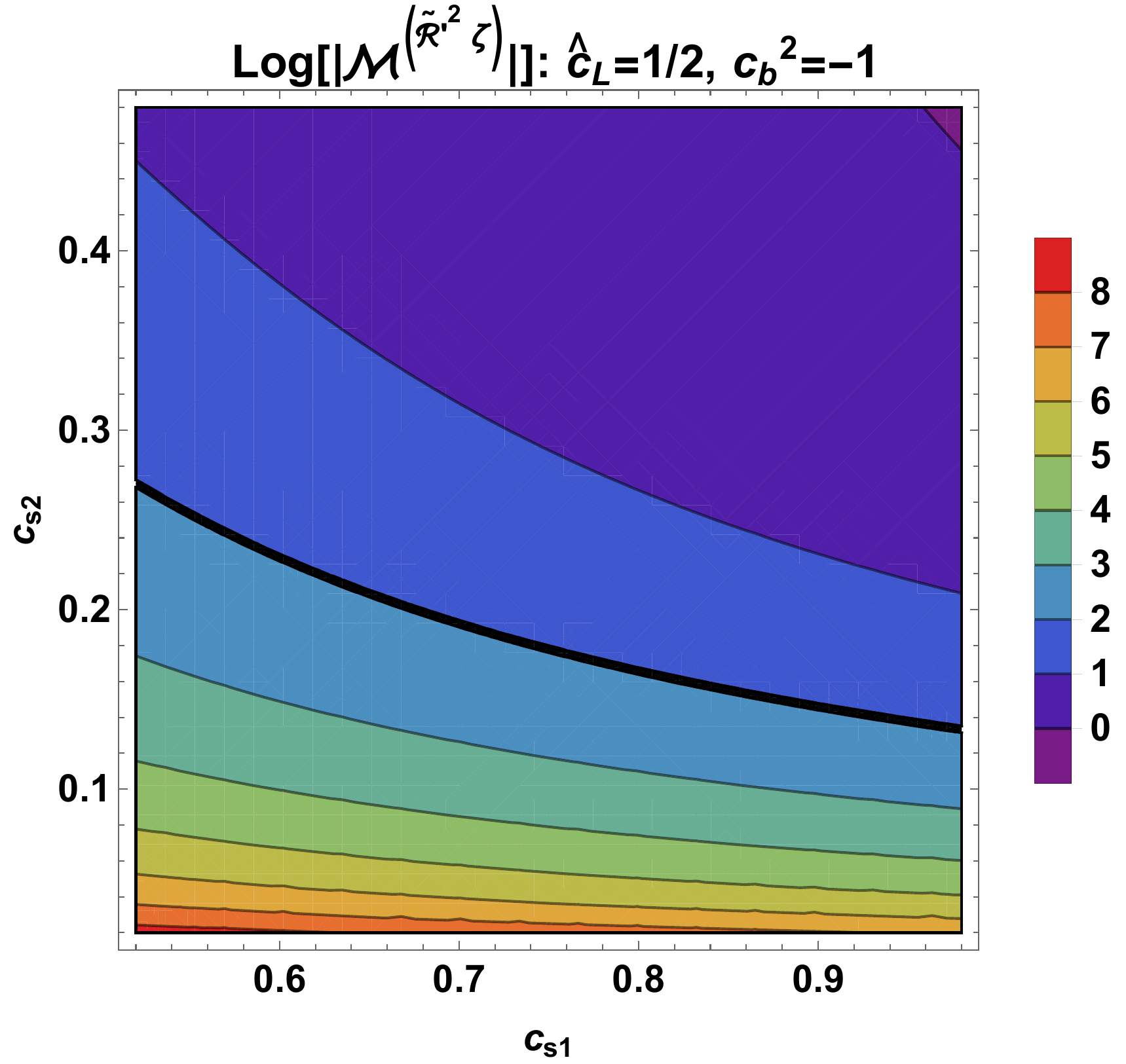}
  \end{minipage}
\caption[Blue, green and orange vertices: $f_{\text{NL}}$ squeezed]{$B_{1,...,5}$, $G_{1,2}$ and $O_{1,2}$ squeezed vertices. The black line delimits the region where ${\cal M}^{{\cal O}} \sim 10^2$}
\label{B_G_O_plot}
\end{figure}
\begin{figure}[!ht]
  \centering
   \begin{minipage}[b]{0.45\textwidth}
    \includegraphics[width=\textwidth]{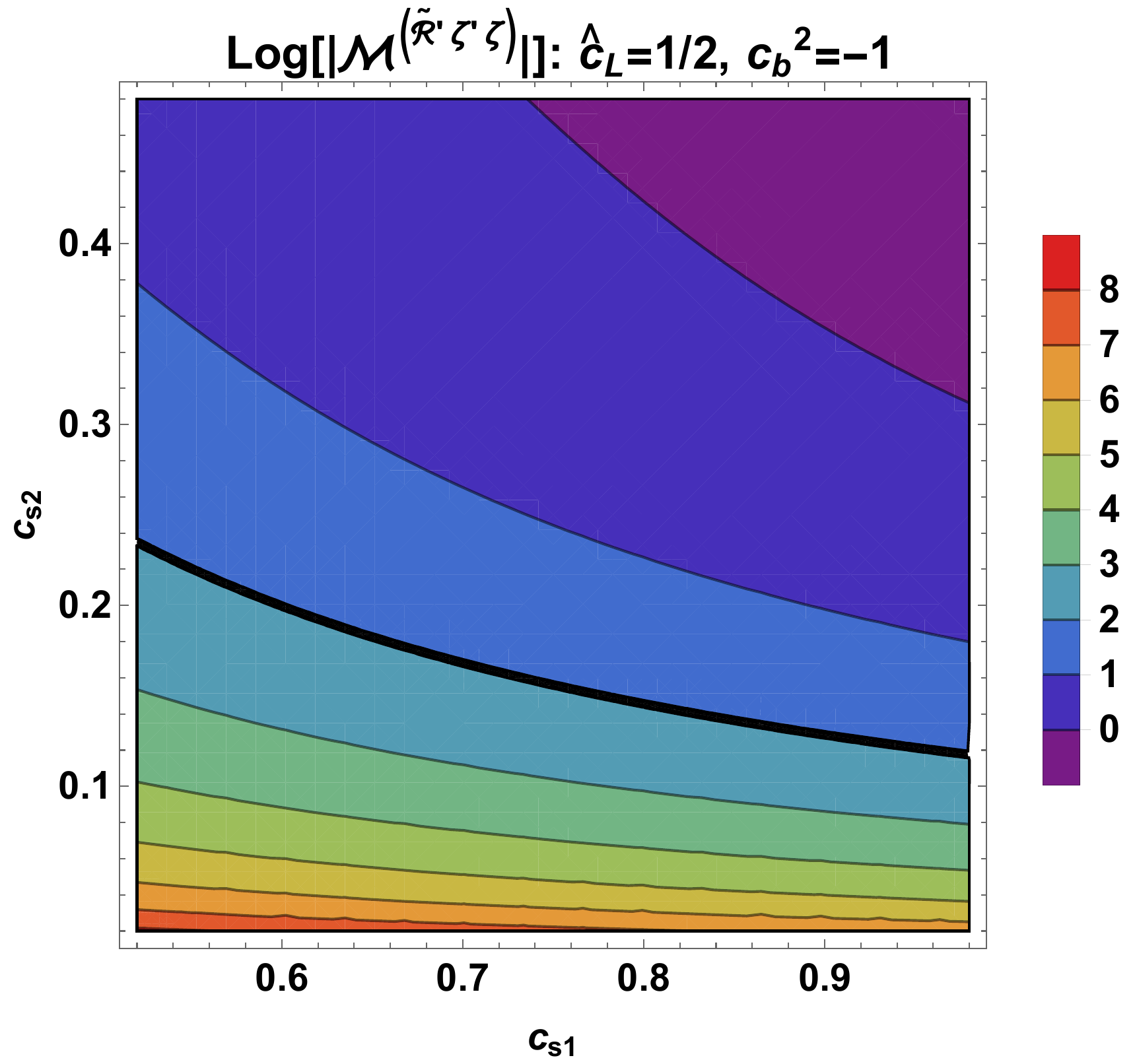}
    \subcaption[$O_{3}$ vertex: $f_{\text{NL}}$ squeezed]{$O_{3}$ squeezed vertex.}
\label{O_plot}    
  \end{minipage}
  \begin{minipage}[b]{0.45\textwidth}
    \includegraphics[width=\textwidth]{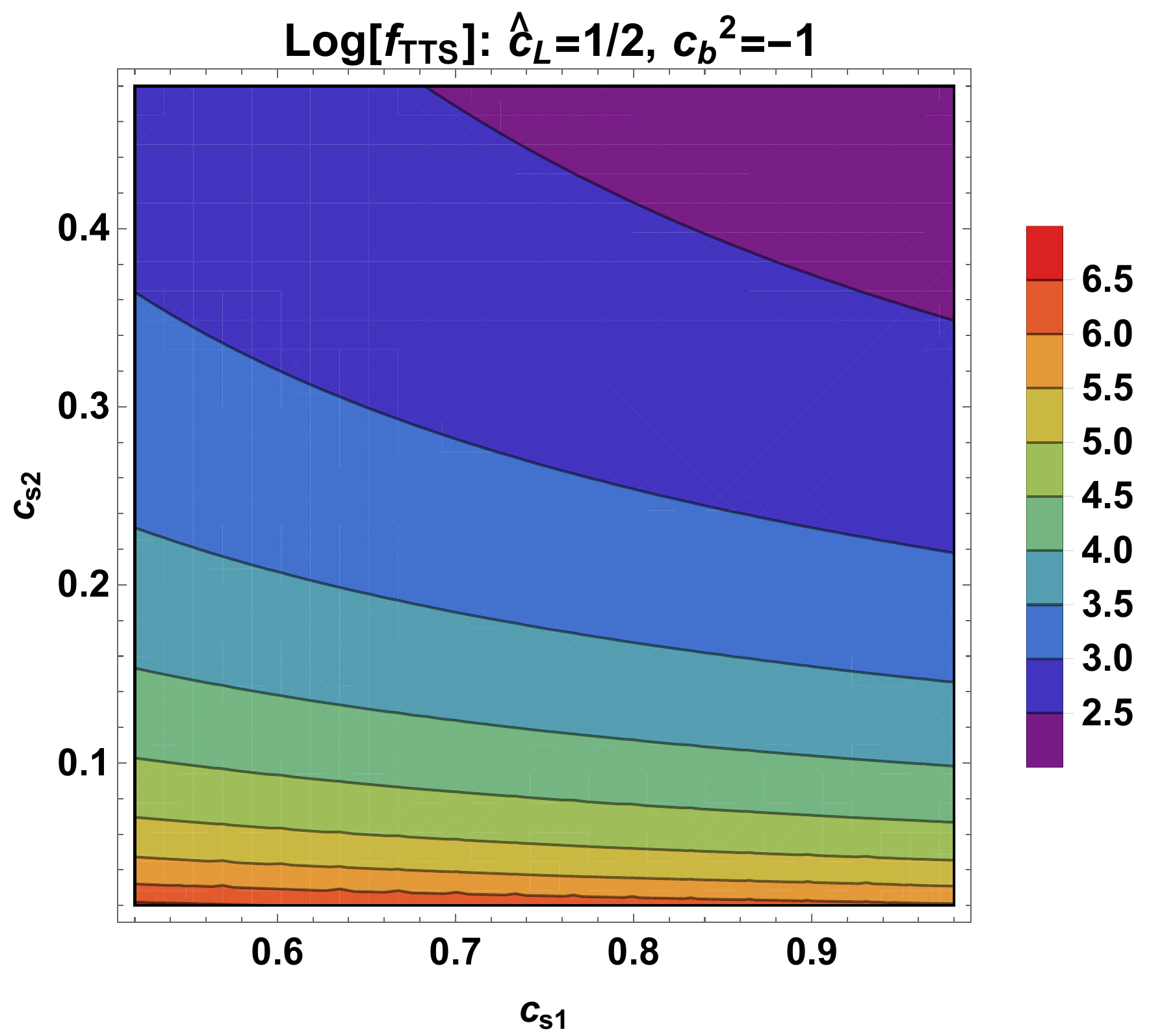}
    \subcaption[Dominant $f_{TTS}$ parameter.]{Dominant $f_{TTS}^{(SQ)}$. $B^{(tts)}\,\epsilon$ set to one.}
\label{Log_TTS}
\end{minipage}
\end{figure}
\begin{figure}[!ht]
  \centering
   \begin{minipage}[b]{0.45\textwidth}
    \includegraphics[width=\textwidth]{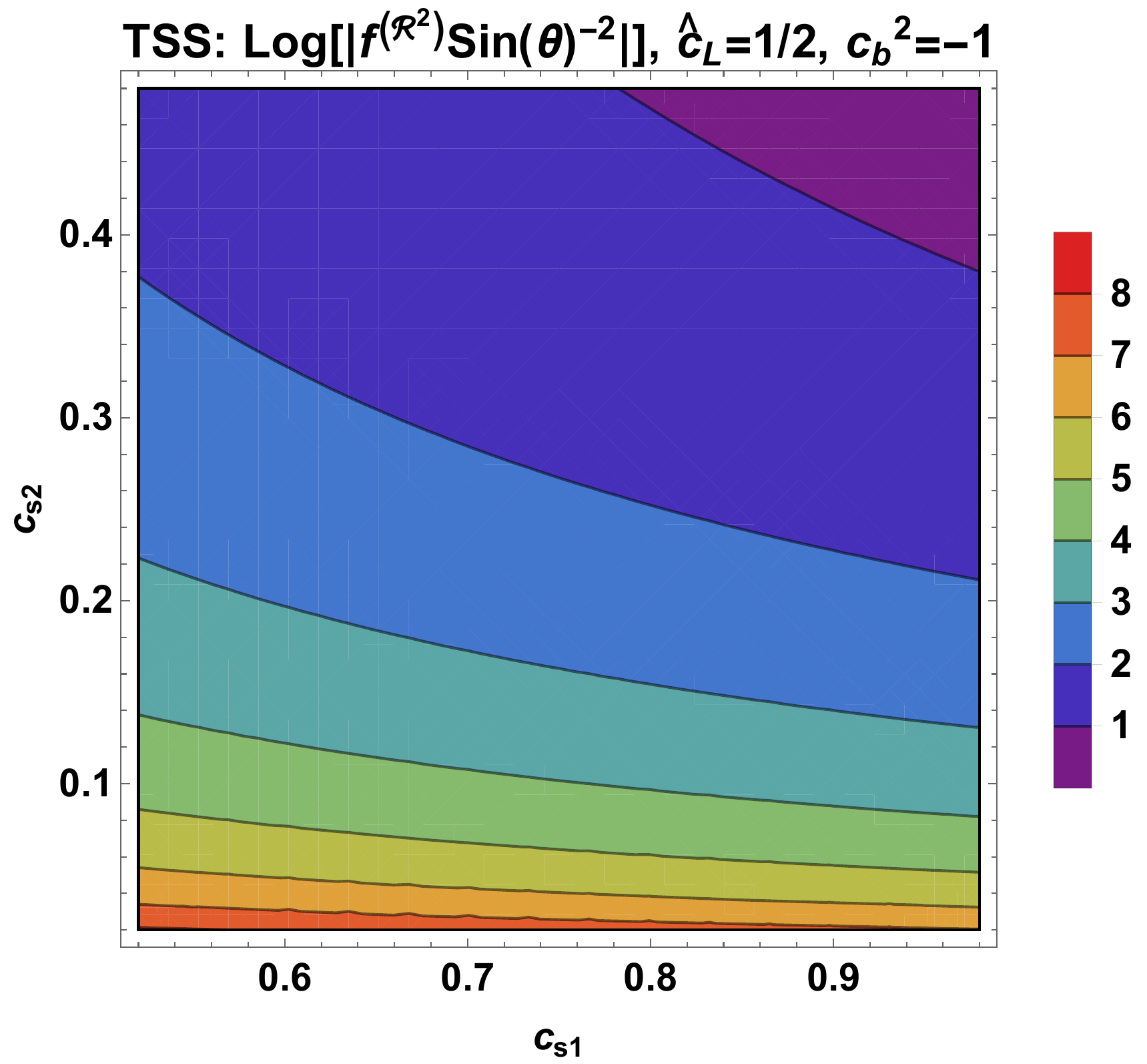}    
  \end{minipage}
  \begin{minipage}[b]{0.45\textwidth}
    \includegraphics[width=\textwidth]{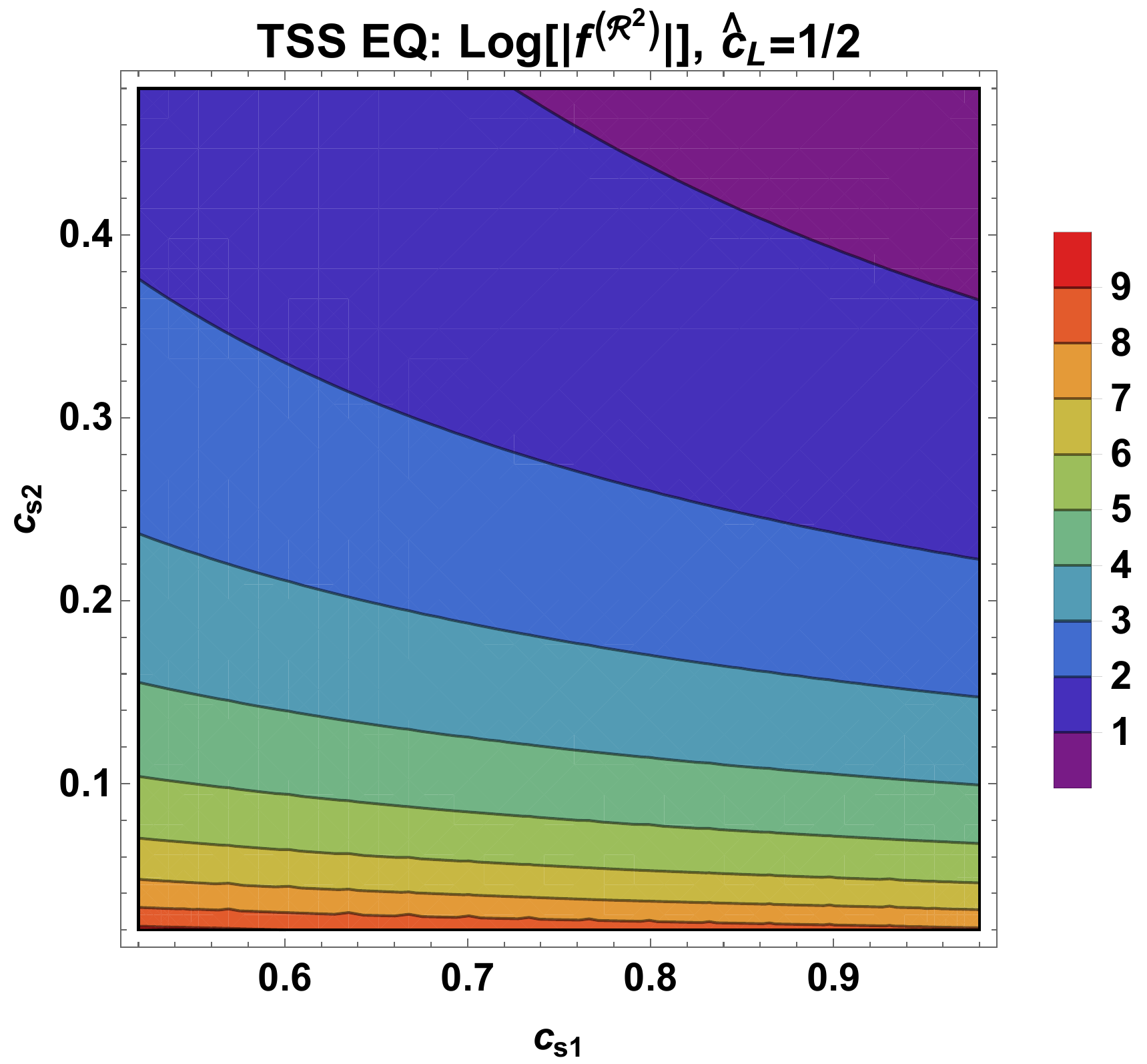}
  \end{minipage}
\caption[TSS dominant vertex: $f_{\text{NL}}$ squeezed and equilateral]{ The dominant $f_{TSS}$ shape $h \,\pi_0^2$ in the $c_b^2=-1$ case. On the left the squeezed $f_{TSS}$, on the right the equilateral one. $G^{(tss)}$ and $\sin(\theta)^2$ set to one.}
\label{TSS_dom}
\end{figure}
\begin{figure}[!ht]
 \centering
   \begin{minipage}[b]{0.45\textwidth}
    \includegraphics[width=\textwidth]{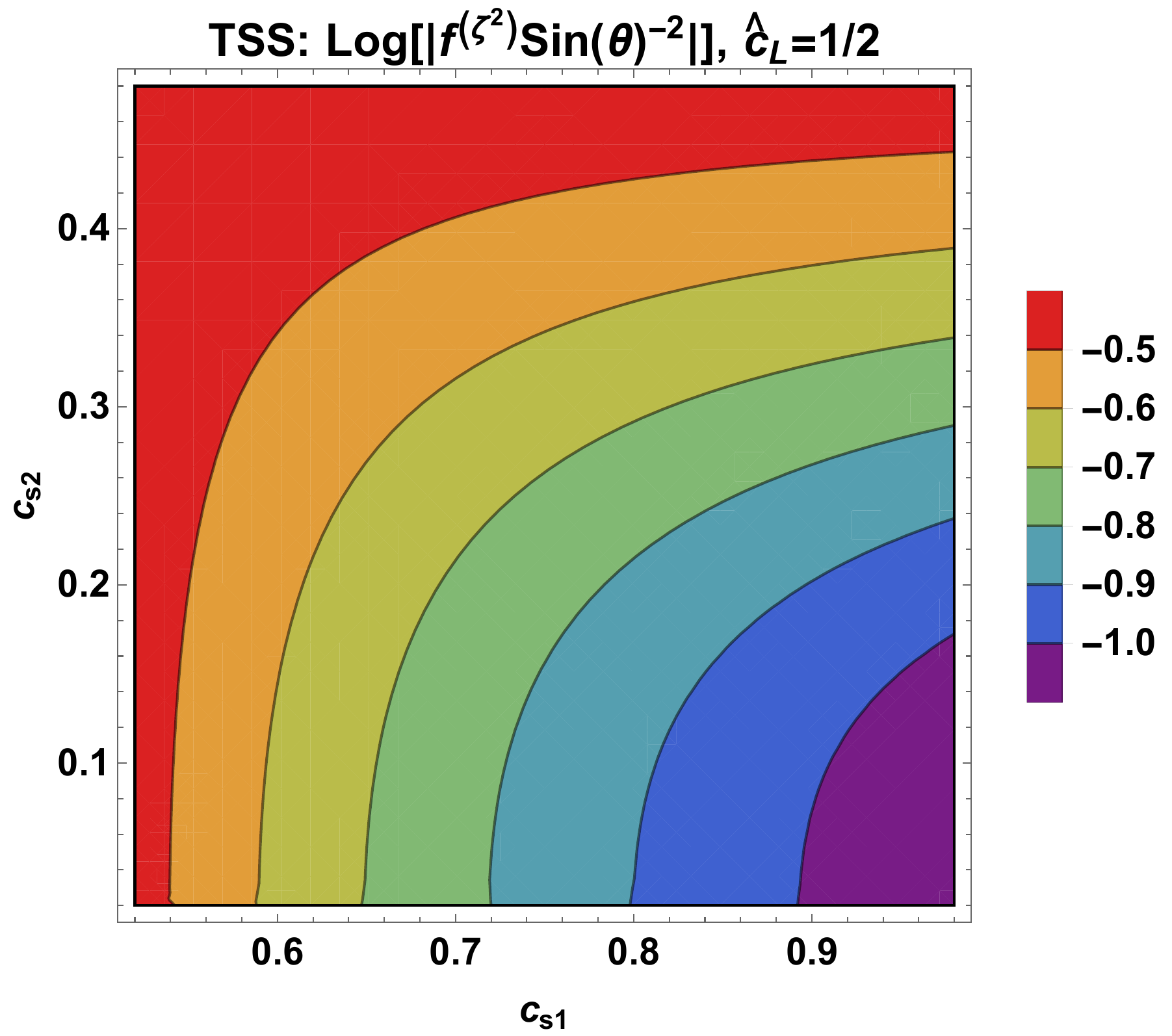}
     \end{minipage}
  \begin{minipage}[b]{0.45\textwidth}
    \includegraphics[width=\textwidth]{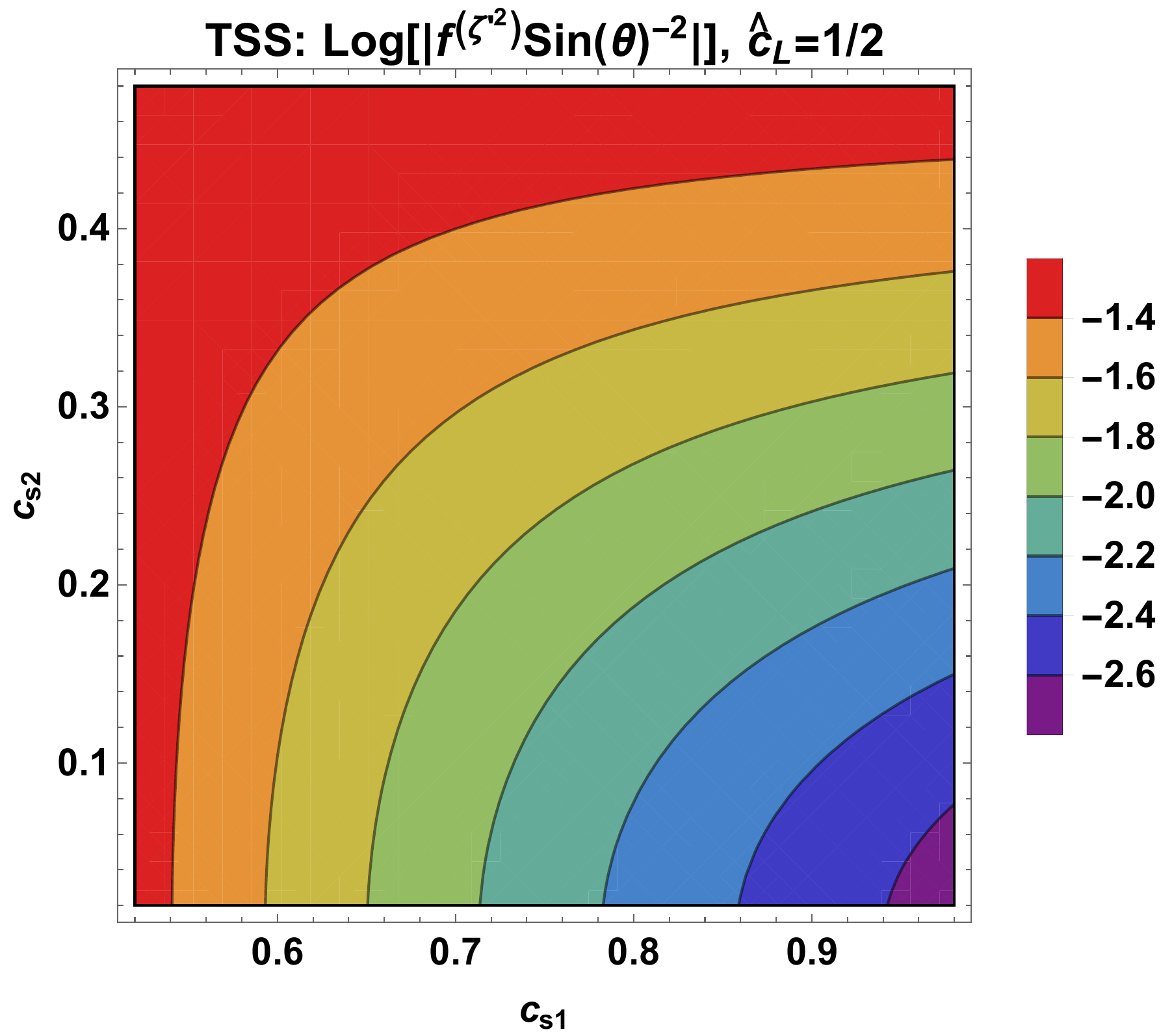}
  \end{minipage}
  \begin{minipage}[b]{0.45\textwidth}
    \includegraphics[width=\textwidth]{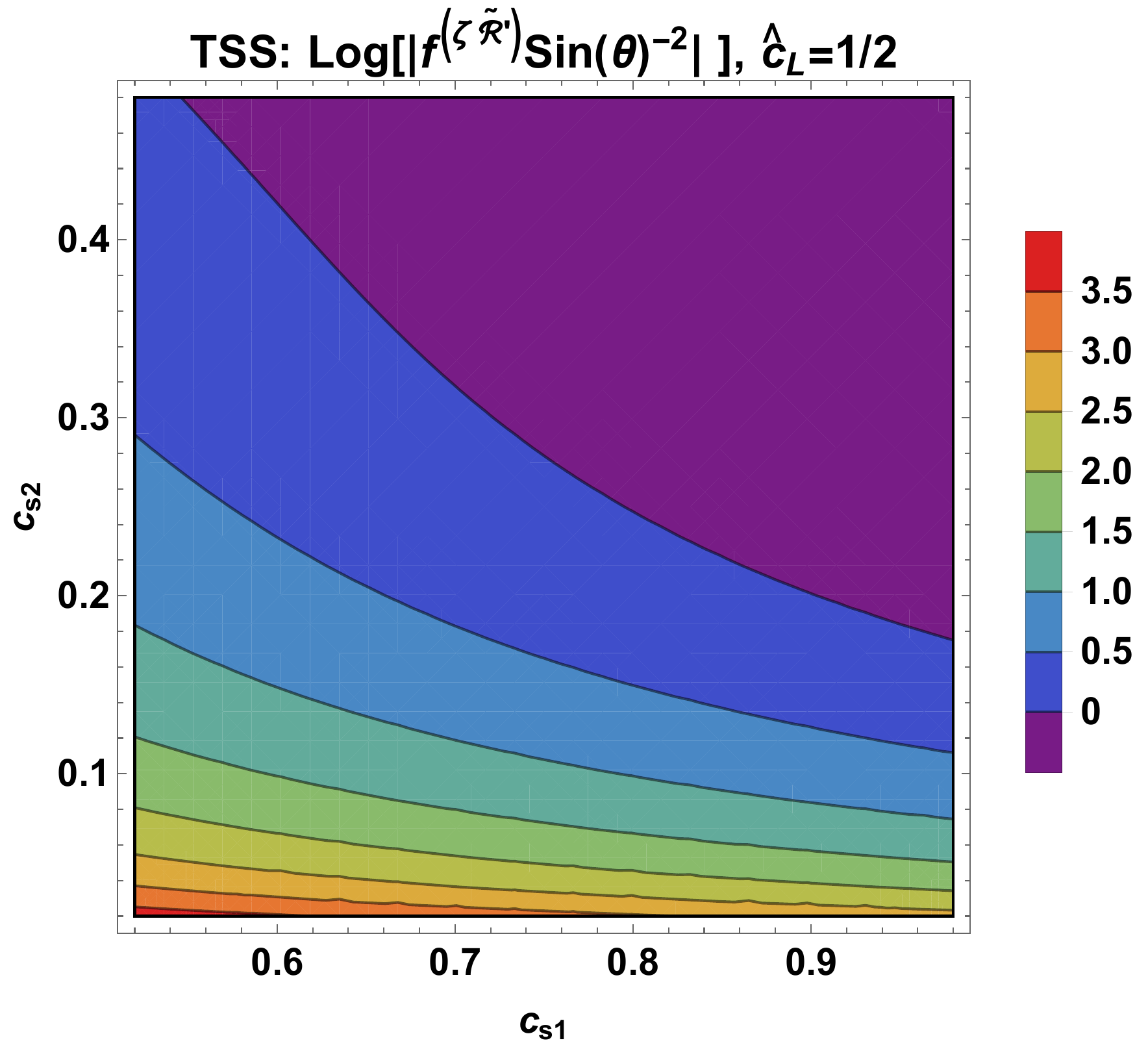}
     \end{minipage}
  \begin{minipage}[b]{0.45\textwidth}
    \includegraphics[width=\textwidth]{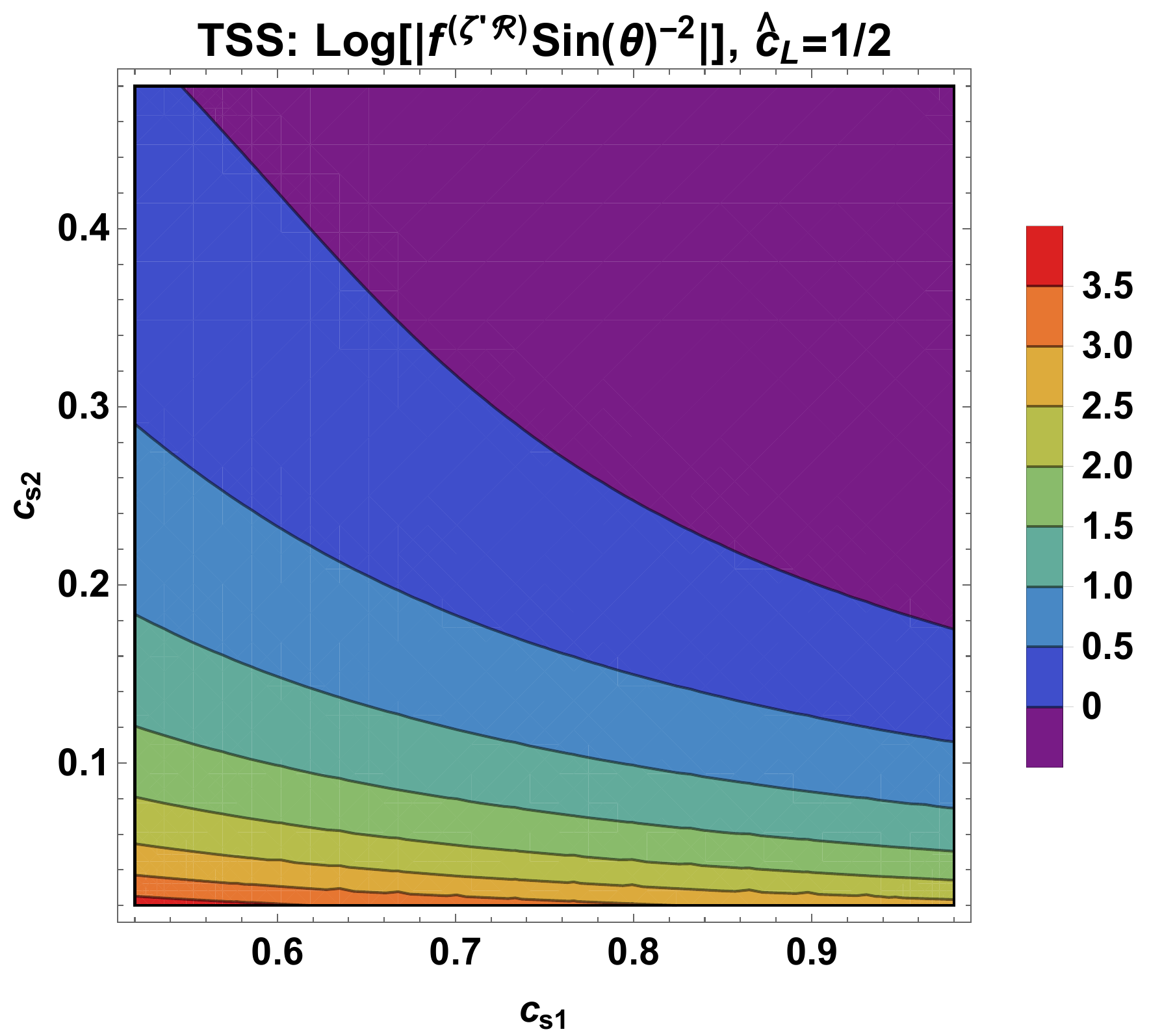}
  \end{minipage}
\caption[TSS violet and blue vertices: $f_{\text{NL}}$ squeezed]{Squeezed TSS plots. Note that the shape $h \,\pi_0^2$ is the dominant one. We set V-B constants and $\sin(\theta)^2$ set to one.}
\label{TSS_cb1_squeezed}
\end{figure}
\begin{figure}[!ht]
 \centering
   \begin{minipage}[b]{0.45\textwidth}
    \includegraphics[width=\textwidth]{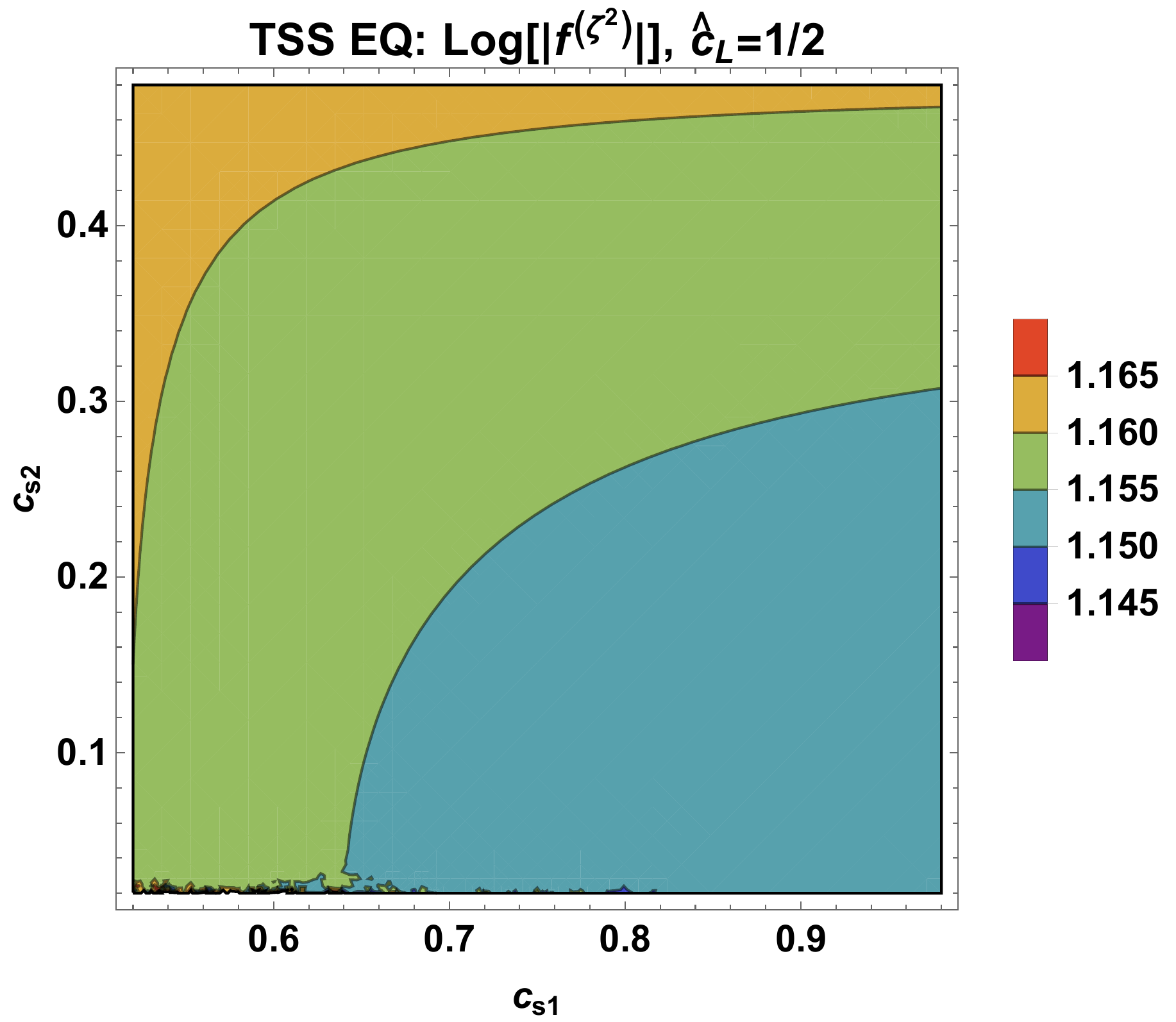}
     \end{minipage}
  \begin{minipage}[b]{0.45\textwidth}
    \includegraphics[width=\textwidth]{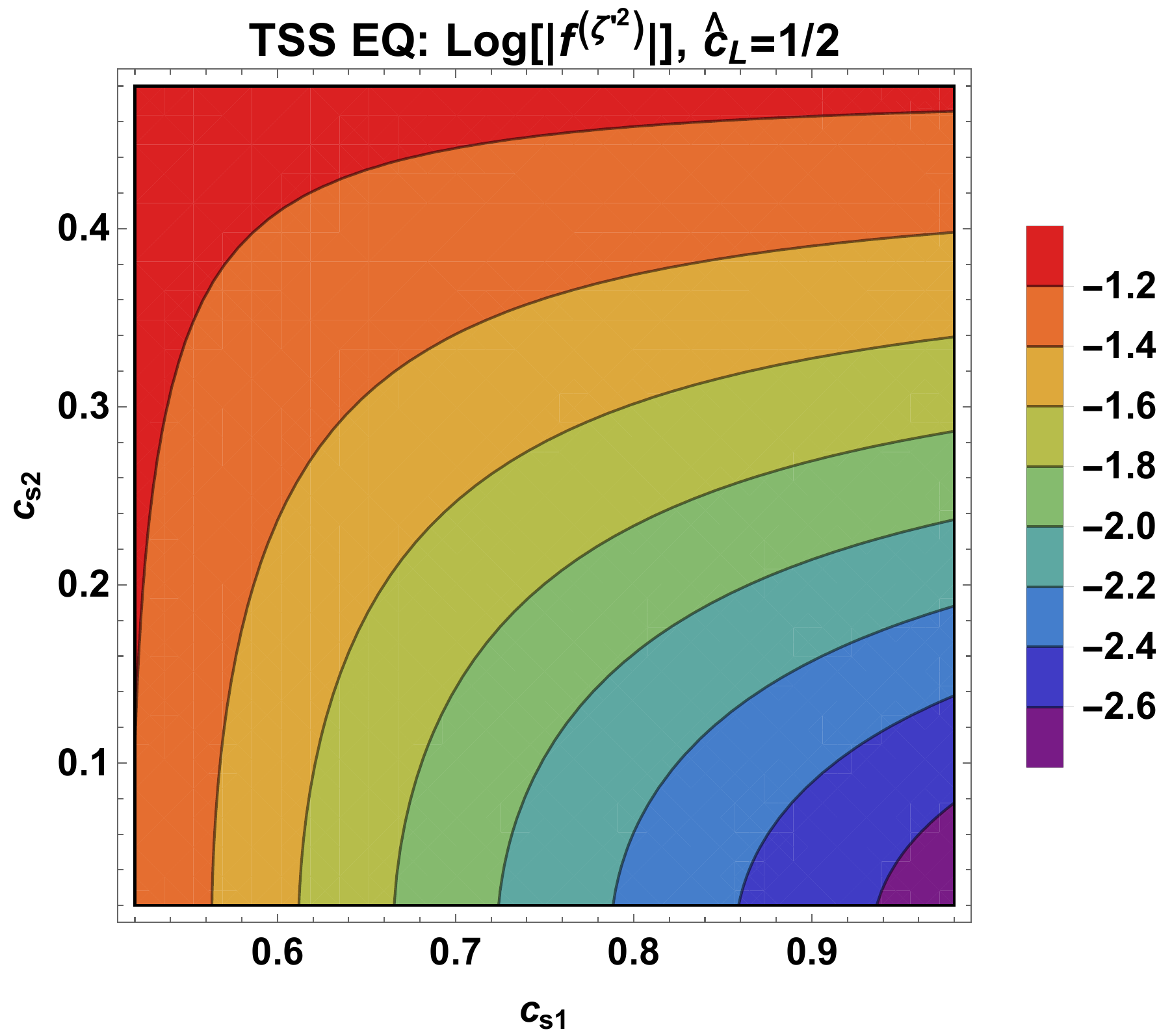}
  \end{minipage}
  \begin{minipage}[b]{0.45\textwidth}
    \includegraphics[width=\textwidth]{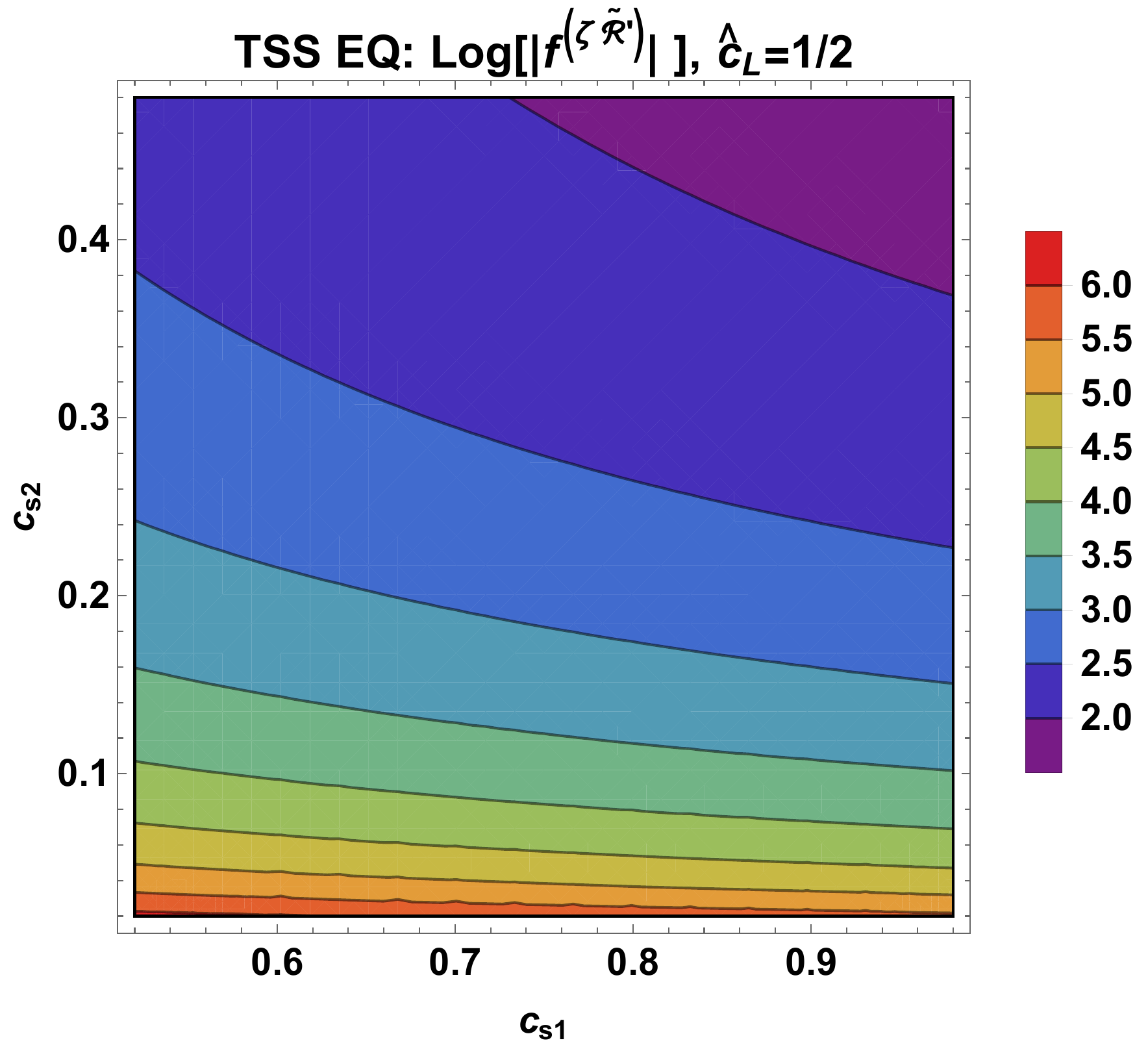}
     \end{minipage}
  \begin{minipage}[b]{0.45\textwidth}
    \includegraphics[width=\textwidth]{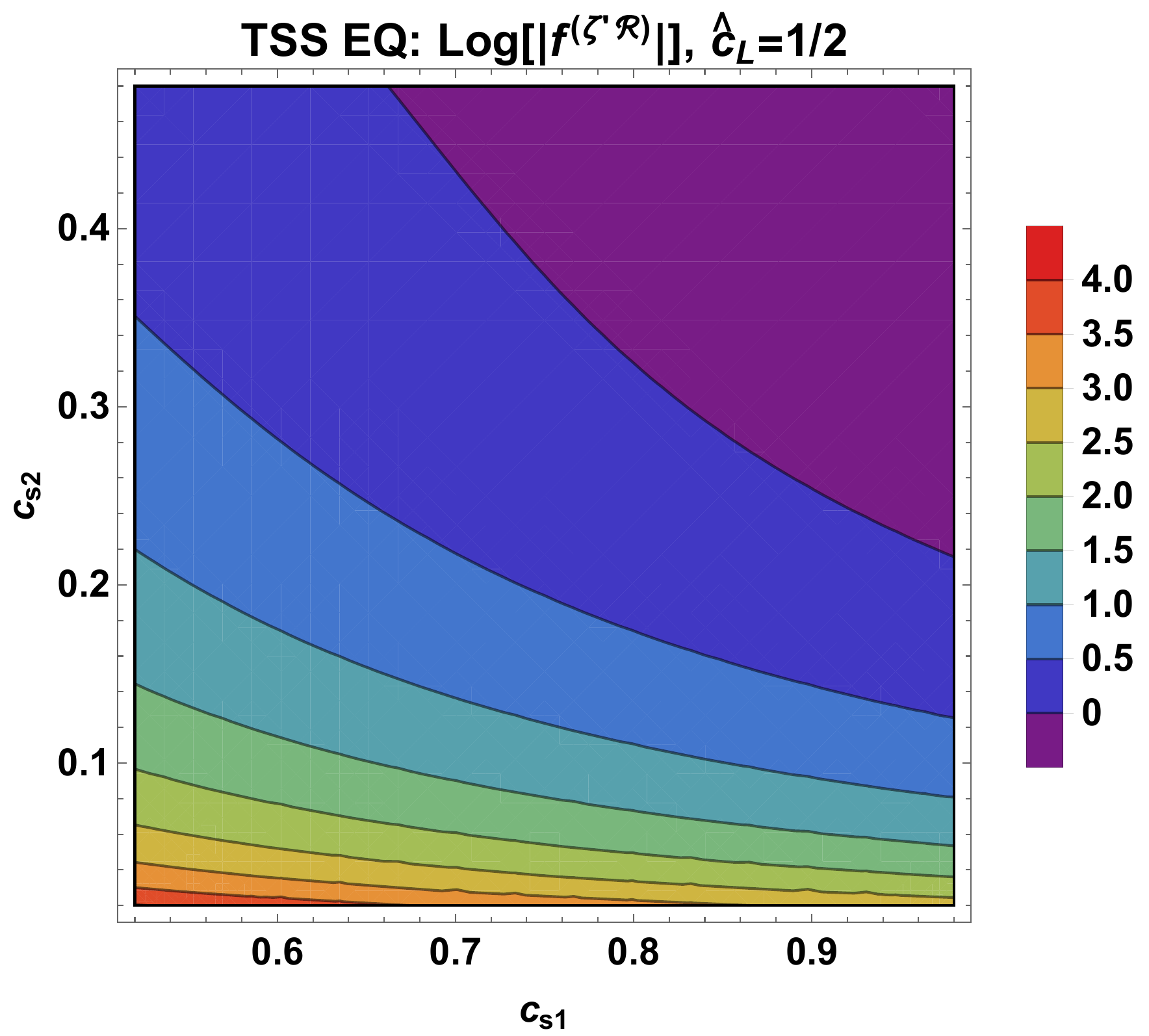}
  \end{minipage}
\caption[TSS violet and blue vertices: $f_{\text{NL}}$ equilateral]{Equilateral TSS plots. We set V-B constants to one and $k\,t=10^{-5}$.}
\label{cb01_EQ}
\end{figure}
\clearpage
\bibliographystyle{unsrt}  
\bibliography{biblio}

\begin{thebibliography}{10}

\bibitem{Cheung:2007st}
Clifford Cheung, Paolo Creminelli, A.~Liam Fitzpatrick, Jared Kaplan, and
  Leonardo Senatore.
\newblock {The Effective Field Theory of Inflation}.
\newblock {\em JHEP}, 03:014, 2008.

\bibitem{Weinberg:2003sw}
S.~Weinberg.
\newblock {Adiabatic modes in cosmology}.
\newblock {\em Phys. Rev.}, D67:123504, 2003.

\bibitem{Maldacena:2002vr}
Juan~Martin Maldacena.
\newblock {Non-Gaussian features of primordial fluctuations in single field
  inflationary models}.
\newblock {\em JHEP}, 05:013, 2003.

\bibitem{Acquaviva:2002ud}
Viviana Acquaviva, Nicola Bartolo, Sabino Matarrese, and Antonio Riotto.
\newblock {Second order cosmological perturbations from inflation}.
\newblock {\em Nucl. Phys. B}, 667:119--148, 2003.

\bibitem{Rubakov:2004eb}
V.~A. Rubakov.
\newblock {Lorentz-violating graviton masses: Getting around ghosts, low strong
  coupling scale and VDVZ discontinuity}.
\newblock {\em hep-th/0407104}, 2004.

\bibitem{Dubovsky:2004sg}
S.~L. Dubovsky.
\newblock {Phases of massive gravity}.
\newblock {\em JHEP}, 10:076, 2004.

\bibitem{Son:2005ak}
D.T. Son.
\newblock {Effective Lagrangian and topological interactions in supersolids}.
\newblock {\em Phys. Rev. Lett.}, 94:175301, 2005.

\bibitem{Celoria:2017bbh}
Marco Celoria, Denis Comelli, and Luigi Pilo.
\newblock {Fluids, Superfluids and Supersolids: Dynamics and Cosmology of Self
  Gravitating Media}.
\newblock {\em JCAP}, 1709(09):036, 2017.

\bibitem{Celoria:2020diz}
Marco Celoria, Denis Comelli, Luigi Pilo, and Rocco Rollo.
\newblock {Boosting GWs in Supersolid Inflation}.
\newblock {\em JHEP}, 01:185, 2021.

\bibitem{Endlich:2012pz}
S.~Endlich, A.~Nicolis, and J.~Wang.
\newblock {Solid Inflation}.
\newblock {\em JCAP}, 1310:011, 2013.

\bibitem{ussgf}
G.~Ballesteros, D.~Comelli, and L.~Pilo.
\newblock {Massive and modified gravity as self-gravitating media}.
\newblock {\em Phys. Rev.}, D94(12):124023, 2016.

\bibitem{Pilo:2017fyg}
Luigi Pilo.
\newblock {Recent results in massive gravity}.
\newblock In {\em {14th Marcel Grossmann Meeting on Recent Developments in
  Theoretical and Experimental General Relativity, Astrophysics, and
  Relativistic Field Theories}}, volume~1, pages 428--437, 2017.

\bibitem{Celoria:2017hfd}
Marco Celoria, Denis Comelli, and Luigi Pilo.
\newblock {Sixth mode in massive gravity}.
\newblock {\em Phys. Rev.}, D98(6):064016, 2018.

\bibitem{Bartolo:2015qvr}
Nicola Bartolo, Dario Cannone, Angelo Ricciardone, and Gianmassimo Tasinato.
\newblock {Distinctive signatures of space-time diffeomorphism breaking in EFT
  of inflation}.
\newblock {\em JCAP}, 1603(03):044, 2016.

\bibitem{Akrami:2018odb}
Y.~Akrami et~al.
\newblock {Planck 2018 results. X. Constraints on inflation}.
\newblock {\em Submitted to A\&A}, 2018.

\bibitem{Bartolo:2013msa}
Nicola Bartolo, Sabino Matarrese, Marco Peloso, and Angelo Ricciardone.
\newblock {Anisotropy in solid inflation}.
\newblock {\em JCAP}, 08:022, 2013.

\bibitem{Bartolo:2014xfa}
Nicola Bartolo, Marco Peloso, Angelo Ricciardone, and Caner Unal.
\newblock {The expected anisotropy in solid inflation}.
\newblock {\em JCAP}, 11:009, 2014.

\bibitem{ade}
P.~A.~R. Ade et~al.
\newblock {Planck 2015 results. XVII. Constraints on primordial
  non-Gaussianity}.
\newblock {\em Astron. Astrophys.}, 594:A17, 2016.

\bibitem{Creminelli:2011rh}
Paolo Creminelli, Guido D'Amico, Marcello Musso, and Jorge Norena.
\newblock {The (not so) squeezed limit of the primordial 3-point function}.
\newblock {\em JCAP}, 1111:038, 2011.

\bibitem{Senatore:2012wy}
Leonardo Senatore and Matias Zaldarriaga.
\newblock {A Note on the Consistency Condition of Primordial Fluctuations}.
\newblock {\em JCAP}, 08:001, 2012.

\bibitem{Creminelli:2012ed}
Paolo Creminelli, Jorge Norena, and Marko Simonovic.
\newblock {Conformal consistency relations for single-field inflation}.
\newblock {\em JCAP}, 07:052, 2012.

\bibitem{Hinterbichler:2012nm}
Kurt Hinterbichler, Lam Hui, and Justin Khoury.
\newblock {Conformal Symmetries of Adiabatic Modes in Cosmology}.
\newblock {\em JCAP}, 08:017, 2012.

\bibitem{Hinterbichler:2013dpa}
Kurt Hinterbichler, Lam Hui, and Justin Khoury.
\newblock {An Infinite Set of Ward Identities for Adiabatic Modes in
  Cosmology}.
\newblock {\em JCAP}, 01:039, 2014.

\bibitem{Hui:2018cag}
Lam Hui, Austin Joyce, and Sam~S.C. Wong.
\newblock {Inflationary soft theorems revisited: A generalized consistency
  relation}.
\newblock {\em JCAP}, 02:060, 2019.

\bibitem{Matarrese:2020why}
Sabino Matarrese, Luigi Pilo, and Rocco Rollo.
\newblock {Resilience of long modes in cosmological observables}.
\newblock {\em JCAP}, 01:062, 2021.

\bibitem{Landry:2019iel}
Michael~J. Landry.
\newblock {The coset construction for non-equilibrium systems}.
\newblock {\em JHEP}, 07:200, 2020.

\bibitem{Celoria:2017idi}
Marco Celoria, Denis Comelli, and Luigi Pilo.
\newblock {Self-gravitating $\Lambda$-media}.
\newblock {\em JCAP}, 1901(01):057, 2019.

\bibitem{Celoria:2019oiu}
Marco Celoria, Denis Comelli, Luigi Pilo, and Rocco Rollo.
\newblock {Adiabatic Media Inflation}.
\newblock {\em JCAP}, 12:018, 2019.

\bibitem{Ferrero:2020jts}
Renata Ferrero and Roberto Percacci.
\newblock {Dynamical diffeomorphisms}.
\newblock {\em arXiv: 2012.04507}, 2020.

\bibitem{Ricciardone_2017}
Angelo Ricciardone and Gianmassimo Tasinato.
\newblock Primordial gravitational waves in supersolid inflation.
\newblock {\em Physical Review D}, 96(2), Jul 2017.

\bibitem{Rubakov:2008nh}
V.~A. Rubakov and P.~G. Tinyakov.
\newblock {Infrared-modified gravities and massive gravitons}.
\newblock {\em Phys. Usp.}, 51:759--792, 2008.

\bibitem{Comelli:2013txa}
Denis Comelli, Fabrizio Nesti, and Luigi Pilo.
\newblock {Massive gravity: a General Analysis}.
\newblock {\em JHEP}, 07:161, 2013.

\bibitem{Akhshik_2015}
Mohammad Akhshik.
\newblock Clustering fossils in solid inflation.
\newblock {\em Journal of Cosmology and Astroparticle Physics},
  2015(05):043–043, May 2015.

\bibitem{Akrami:2019izv}
Y.~Akrami et~al.
\newblock {Planck 2018 results. IX. Constraints on primordial non-Gaussianity}.
\newblock {\em Astron. Astrophys.}, 641:A9, 2020.

\bibitem{Endlich:2013jia}
Solomon Endlich, Bart Horn, Alberto Nicolis, and Junpu Wang.
\newblock {Squeezed limit of the solid inflation three-point function}.
\newblock {\em Phys. Rev. D}, 90(6):063506, 2014.

\bibitem{Malhotra:2020ket}
Ameek Malhotra, Ema Dimastrogiovanni, Matteo Fasiello, and Maresuke Shiraishi.
\newblock {Cross-correlations as a Diagnostic Tool for Primordial Gravitational
  Waves}.
\newblock 12 2020.

\bibitem{Adshead:2020bji}
Peter Adshead, Niayesh Afshordi, Emanuela Dimastrogiovanni, Matteo Fasiello,
  Eugene~A. Lim, and Gianmassimo Tasinato.
\newblock {Multimessenger Cosmology: correlating CMB and SGWB measurements}.
\newblock 4 2020.

\bibitem{Shiraishi_2019}
Maresuke Shiraishi.
\newblock Tensor non-gaussianity search: Current status and future prospects.
\newblock {\em Frontiers in Astronomy and Space Sciences}, 6, Jul 2019.

\bibitem{Bartolo2_2019}
N.~Bartolo, V.~De~Luca, G.~Franciolini, A.~Lewis, M.~Peloso, and A.~Riotto.
\newblock Primordial black hole dark matter: Lisa serendipity.
\newblock {\em Physical Review Letters}, 122(21), May 2019.

\bibitem{Bartolo_2019}
N.~Bartolo, D.~Bertacca, S.~Matarrese, M.~Peloso, A.~Ricciardone, A.~Riotto,
  and G.~Tasinato.
\newblock Anisotropies and non-gaussianity of the cosmological gravitational
  wave background.
\newblock {\em Physical Review D}, 100(12), Dec 2019.

\bibitem{Bartolo:2019yeu}
Nicola Bartolo, Daniele Bertacca, Sabino Matarrese, Marco Peloso, Angelo
  Ricciardone, Antonio Riotto, and Gianmassimo Tasinato.
\newblock {Characterizing the cosmological gravitational wave background:
  Anisotropies and non-Gaussianity}.
\newblock {\em Phys. Rev. D}, 102(2):023527, 2020.

\bibitem{Boyle_2008}
Latham~A. Boyle and Paul~J. Steinhardt.
\newblock Probing the early universe with inflationary gravitational waves.
\newblock {\em Physical Review D}, 77(6), Mar 2008.

\bibitem{Smith_2019}
Tristan~L. Smith and Robert~R. Caldwell.
\newblock Lisa for cosmologists: Calculating the signal-to-noise ratio for
  stochastic and deterministic sources.
\newblock {\em Physical Review D}, 100(10), Nov 2019.

\bibitem{Bartolo:2016ami}
Nicola Bartolo et~al.
\newblock {Science with the space-based interferometer LISA. IV: Probing
  inflation with gravitational waves}.
\newblock {\em JCAP}, 12:026, 2016.

\bibitem{Guzzetti:2016mkm}
M.C. Guzzetti, N.~Bartolo, M.~Liguori, and S.~Matarrese.
\newblock {Gravitational waves from inflation}.
\newblock {\em Riv. Nuovo Cim.}, 39(9):399--495, 2016.

\bibitem{Bordin:2017ozj}
Lorenzo Bordin, Paolo Creminelli, Mehrdad Mirbabayi, and Jorge Norena.
\newblock {Solid Consistency}.
\newblock {\em JCAP}, 1703(03):004, 2017.

\bibitem{Chen_2007}
Xingang Chen, Min-xin Huang, Shamit Kachru, and Gary Shiu.
\newblock Observational signatures and non-gaussianities of general
  single-field inflation.
\newblock {\em Journal of Cosmology and Astroparticle Physics},
  2007(01):002–002, Jan 2007.

\bibitem{Dimastrogiovanni:2014ina}
Emanuela Dimastrogiovanni, Matteo Fasiello, Donghui Jeong, and Marc
  Kamionkowski.
\newblock {Inflationary tensor fossils in large-scale structure}.
\newblock {\em JCAP}, 12:050, 2014.

\bibitem{Dimastrogiovanni_2016}
Emanuela Dimastrogiovanni, Matteo Fasiello, and Marc Kamionkowski.
\newblock Imprints of massive primordial fields on large-scale structure.
\newblock {\em Journal of Cosmology and Astroparticle Physics},
  2016(02):017–017, Feb 2016.

\bibitem{Iacconi:2020yxn}
Laura Iacconi, Matteo Fasiello, Hooshyar Assadullahi, and David Wands.
\newblock {Small-scale Tests of Inflation}.
\newblock {\em JCAP}, 12:005, 2020.

\bibitem{Iacconi:2019vgc}
Laura Iacconi, Matteo Fasiello, Hooshyar Assadullahi, Emanuela
  Dimastrogiovanni, and David Wands.
\newblock {Interferometer Constraints on the Inflationary Field Content}.
\newblock {\em JCAP}, 03:031, 2020.

\bibitem{Cabass:2021iii}
Giovanni Cabass.
\newblock {Zoology of Graviton non-Gaussianities}.
\newblock {\em arXiv:2103.09816}, 2021.

\bibitem{Ballesteros:2016kdx}
G.~Ballesteros, D.~Comelli, and L.~Pilo.
\newblock {Thermodynamics of perfect fluids from scalar field theory}.
\newblock {\em Phys. Rev.}, D94(2):025034, 2016.

\bibitem{Garcia_Saenz_2020}
Sebastian Garcia-Saenz, Lucas Pinol, and Sébastien Renaux-Petel.
\newblock Revisiting non-gaussianity in multifield inflation with curved field
  space.
\newblock {\em Journal of High Energy Physics}, 2020(1), Jan 2020.

\bibitem{Arroja_2011}
Frederico Arroja and Takahiro Tanaka.
\newblock {A note on the role of the boundary terms for the non-Gaussianity in
  general k-inflation}.
\newblock {\em JCAP}, 05:005, 2011.

\bibitem{Rigopoulos_2011}
Gerasimos Rigopoulos.
\newblock Gauge invariance and non-gaussianity in inflation.
\newblock {\em Physical Review D}, 84(2), Jul 2011.

\bibitem{Burrage_2011}
Clare Burrage, Raquel~H. Ribeiro, and David Seery.
\newblock {Large slow-roll corrections to the bispectrum of noncanonical
  inflation}.
\newblock {\em JCAP}, 07:032, 2011.

\bibitem{Seery_2005}
David Seery and James~E. Lidsey.
\newblock {Primordial non-Gaussianities in single field inflation}.
\newblock {\em JCAP}, 06:003, 2005.

\end{thebibliography}

\end{document}